\documentclass[11pt, a4paper]{article} 
\usepackage{jheppub}\makeatletter\gdef\@fpheader{~}\makeatother
\usepackage{todonotes}
\usepackage{dynkin-diagrams}
\usepackage{amsmath, amsthm, amssymb, amsbsy, amsfonts, mathrsfs, enumerate, float, wrapfig, mathtools, graphbox, framed, tcolorbox, microtype, adjustbox}
\allowdisplaybreaks 
\usepackage[section]{placeins}\makeatletter\AtBeginDocument{\expandafter\renewcommand\expandafter\subsection\expandafter{\expandafter\@fb@secFB\subsection}}\makeatother
\usepackage{booktabs}
\usepackage{multirow}
\usepackage[dvipsnames]{xcolor}  
\usepackage{tikz}\usetikzlibrary{calc,
  decorations.markings,
  decorations.pathreplacing,
  arrows,arrows.meta}  
\usepackage{orcidlink}\makeatletter\gdef\@OrigHeightRecip{0.005}\makeatother
\hypersetup{linkcolor=blue, urlcolor=magenta, citecolor=magenta}

\newcommand{\nn}{\nonumber}  
\newcommand{\Z}{\ensuremath{\mathbb{Z}}}
\newcommand{\D}{\ensuremath{D}}  
\newcommand{\E}{\ensuremath{E}}    
\newcommand{\M}{{\sf M}}    
  
\newcommand{\F}{\textbf{F}}
  
\newcommand{\SO}{\text{SO}}
\newcommand{\U}{\text{U}}
\newcommand{\SU}{\text{SU}}
\title{5d \texorpdfstring{$\mathbb{Z}_{n}$}{Zn} S-folds, quotient curves, and isogenies}
\author[a]{Sung-Soo Kim,\,\orcidlink{0000-0001-7977-193X}}
\author[a]{Song Liu,\,\orcidlink{0009-0000-9824-7850}}
\author[a]{Mudassar Sabir,\,\orcidlink{0000-0002-8551-2608}}
\author[b]{Futoshi Yagi\,\orcidlink{0000-0002-6217-5151}}

\affiliation[a]{School of Physics, University of Electronic Science and Technology of China, \\
No.~2006 Xiyuan Ave, West Hi-Tech Zone, Chengdu, Sichuan 611731, China}
\affiliation[b]{School of Science, Huzhou Normal University, \\
No. 759, East 2nd Road, Huzhou, Zhejiang 313000, China}

\emailAdd{sungsoo.kim@uestc.edu.cn}
\emailAdd{sliu.susy@std.uestc.edu.cn} 
\emailAdd{mudassar.sabir@uestc.edu.cn} 
\emailAdd{60714@huznu.edu.cn}

\abstract{
We construct Seiberg--Witten curves for the $\mathbb{Z}_3$, $\mathbb{Z}_4$, and $\mathbb{Z}_6$ S-folds of 5d rank-1 superconformal theories. Compared with the $\mathbb{Z}_2$ case \cite{Kim:2026yeh},
$\mathbb{Z}_{3,4,6}$ S-folds require stronger constraints on the mass parameters. We provide a systematic procedure of realizing the S-foldings at the mirror curve level. We obtain the same elliptic curve families by specializing the complexified mass parameters of the rank-1 $E_6$, $E_7$, and $E_8$ theories, with holonomies motivated by the 7-brane insertions required by the S-fold prescription. The two constructions agree in Weierstrass form. The families with continuous mass deformations organize into characters of $\mathfrak{su}(2)$ and $\mathfrak{g}_2$, consistently with their singular fibers. Finally, we discuss that the Seiberg--Witten curves for $\Z_n$ S-folds can be obtained through cyclic $n$-isogenies, which may provide a uniform quotient curve description of rank-1 S-folds.
}

\begin{document}
\maketitle

\section{Introduction and discussion}\label{sec:intro}
Type IIB $(p,q)$ 5-brane webs provide a powerful description of five-dimensional superconformal field theories (SCFTs) with eight supercharges translating Coulomb branch data into the geometry of the web~\cite{Aharony:1997ju}. For example, a compact face in a 5-brane web supports a monopole string whose tension is its area, and integrating these tensions leads to the cubic prepotential $\mathcal{F}$~\cite{Aharony:1997bh, Intriligator:1997pq}. Although this prepotential is exact at one loop within a gauge theory chamber~\cite{Seiberg:1996bd, Intriligator:1997pq}, compactification introduces additional dependence on the circle radius and instanton effects~\cite{Nekrasov:1996cz, Eguchi:2000fv}. The latter are incorporated in the holomorphic curve wrapped by the M5-brane obtained after T-duality and the M-theory lift~\cite{Witten:1997sc, Kol:1997fv, Brandhuber:1997ua}, also known as the Seiberg--Witten curve. Toric and generalized toric diagrams supply Laurent polynomial equations for these Seiberg--Witten curves in coordinates $(t,w)\in(\mathbb{C}^{*})^2$~\cite{Leung:1997tw, Benini:2009gi, Kim:2014nqa}. This presentation retains the connection to the 5-brane web while allowing its discrete symmetries to be implemented algebraically~\cite{Acharya:2021jsp,Kim:2021fxx,Arias-Tamargo:2026rrf}.

S-folds use a rotational symmetry of the web together with an $\mathrm{SL}(2,\Z)$ identification of 5-brane charges. Their geometric and brane realizations have been investigated in~\cite{Acharya:2021jsp, Kim:2021fxx, Furrer:2024zzu}, and extended to generalized quotients and holographic descriptions in~\cite{Apruzzi:2022nax}. For a rotation of order $n$ with its fixed point inside a compact face, the brane prescription keeps a sector of angle $2\pi/n$ with a suitable value of the axio-dilaton and supplies the missing monodromy through the particular 7-brane insertion associated with $\Z_n$ S-folds~\cite{Kim:2021fxx}. The inserted 7-brane configurations for $n=2,3,4,6$ are of type $D_4$, $E_6$, $E_7$, and $E_8$, with deficit angles $\pi$, $\frac{4\pi}{3}$, $\frac{3\pi}{2}$, and $\frac{5\pi}{3}$, respectively. The reduction in face areas accounts for the prepotential relation $\mathcal{F}_{\mathcal{T}/\mathbb{Z}_n}=\frac{1}{n}\mathcal{F}_{\mathcal{T}}$ between the parent theory $\cal T$ and the S-folded theory $\mathcal{T}/\Z_n$~\cite{Kim:2021fxx}.

For $\Z_2$ S-folds for rank-1 SCFTs, the Seiberg--Witten curves are computed from two distinct methods~\cite{Kim:2026yeh}. One is to directly implement the $\Z_2$ quotient by introducing the two-to-one map. The other is to impose special holonomies for the complexified mass parameters associated with the $D_4$ 7-brane configuration inserted at the fixed point of the $\Z_2$ action on the 5-brane web. These two curves, obtained from two different physical origins, perfectly agree with each other when the curves are expressed in terms of the Weierstrass form.

Inserting 7-branes for the $\Z_2$ S-fold, namely four D7-branes and an O7$^-$-plane, changes the resulting SCFT. A notable example is the $\SU(2)$ theory with six hypermultiplets in the fundamental representation ($\SU(2)+6\F$). To preserve the $\Z_2$ symmetry, the six masses are identified pairwise, so only three masses remain deformable. The $\Z_2$ S-fold action then adds four D7-branes whose positions are not deformable. As a result, the $\Z_2$ S-fold of $\SU(2)+6\F$ yields $\SU(2)+7\F$. We stress that only four mass parameters, including the instanton mass, remain, and they form the \textit{apparent $F_4$ flavor symmetry}. This is realized by assigning a particular holonomy to the inserted D7-branes, which we call $D_4$-tuning where four flavor fugacities are tuned to $(1,\!-1,1,\!-1)$, starting from the 8 mass parameters of $\SU(2)+7\F$~\cite{Kim:2026yeh}.

In this paper, we extend and generalize the analysis of the $\Z_2$ S-folds to higher-order $\Z_n$ ($n=3,4,6$) quotients. The extension from $\Z_2$ to $\Z_{3,4,6}$ is not merely a replacement of the order of the quotient. Several qualitatively new features arise for which the $\Z_2$ analysis does not carry over directly. First, while $\Z_2$ is a simultaneous inversion of two torus coordinates $(t,w)\mapsto(t^{-1}, w^{-1})$, admitting a direct construction in terms of two invariant coordinates, the higher-order $\Z_n$ S-fold actions mix the two torus coordinates through finite monomial transformations, which impose additional relations among the coefficients of the Seiberg--Witten curves for the parent theories and restrict mass deformations. It is hence more non-trivial to find the invariant coordinates. To find such invariant coordinates and an $n$-to-$1$ map systematically, we embed the algebraic curves in $(\mathbb{C}^*)^2 / \Z_n$ into $(\mathbb{C}^*)^n / \Z_n$. It is then easier to find an $n$-to-$1$ map such that $n$ points in $(\mathbb{C}^*)^n$ on which $\Z_n$ acts are mapped to one point in another copy of $(\mathbb{C}^*)^n$, satisfying a set of constraint equations. There, the Seiberg--Witten curves for the $\Z_n$ S-folds can be readily obtained by partially solving the set of such constraints. Second, the 7-branes inserted depend on $n$. They are  $E_6, E_7,$ or $E_8$ 7-branes for $n=3,4,6$, respectively. Accordingly, the mass tuning prescription involves holonomies in different exceptional group embeddings, rather than only the alternating signs of the $\Z_2$ construction. As we discuss in later sections, each $\Z_n$ S-fold with $n=3,4,6$ requires a special tuning, which we call the $E_{6,7,8}$-tunings, in which $6,7$, and $8$ mass parameters are tuned, respectively. Again, the Seiberg--Witten curves obtained in two different ways agree.

The resulting apparent flavor symmetries can be identified from the quotient geometries and are summarized in table \ref{tab:summary}, together with the generic Kodaira fiber configurations and the $E_{6,7,8}$-tuning for the resulting S-folded theories of $E_N$ flavor symmetries.
\begin{table}[t]
\centering
\small
\setlength{\tabcolsep}{4pt}
\renewcommand{\arraystretch}{1.18}
\begin{tabular}{clcccc}
\toprule
$\Z_n$
& \shortstack{Parent theory \\$\mathcal{T}$}
& \shortstack{S-folded theory \\$\mathcal{T}/ \Z_n$}
& \shortstack{Kodaira fiber\\
              $(F_\infty;F_{\mathrm{bulk}})$}
& \shortstack{Mass parameter\\$ \mathrm{rank}_{F}(\mathcal{T}/\Z_n) $}
& \shortstack{Flavor\\ $\mathfrak{g}_F$}
\\
\midrule
$\Z_2$
& $\SU(2)_0$ 
& $E_5$
& $(I_4;I_2^4)$
& $1$
& $\mathfrak{su}(2)$
\\
&
$\SU(2)+2\F$
& $E_6$
& $(I_3;I_2^4I_1)$
& $2$
& $\mathfrak{su}(3)$
\\
&
$\SU(2)+4\F$
& $E_7$
& $(I_2;I_2^4I_1^2)$
& $3$
& $\mathfrak{so}(7)$
\\
&
$\SU(2)+6\F$
& $E_8$
& $(I_1;I_2^4I_1^3)$
& $4$
& $\mathfrak{f}_4$
\\
\midrule
$\Z_3$
& $E_0$ (local $\mathbb{P}^2$)
& $E_6$
& $(I_3;I_3^3)$
& $0$
& $.$
\\
&
$\SU(2)+2\F$
& $E_7$
& $(I_2;I_3^3 I_1)$
& $1$
& $\mathfrak{su}(2)$
\\
&
$\SU(2)+5\F$
& $E_8$
& $(I_1;I_3^3 I_1^2)$
& $2$
& $\mathfrak{g}_2$
\\
\midrule
$\Z_4$
& $\SU(2)_0$
& $E_7$
& $(I_2;I_4^2 I_2)$
& $0$
& $.$
\\
&
$\SU(2)+4\F$
& $E_8$
& $(I_1;I_4^2 I_2 I_1)$
& $1$
& $\mathfrak{su}(2)$
\\
\midrule
$\Z_6$
& $\SU(2)+2\F$
& $E_8$
& $(I_1;I_6 I_3 I_2)$
& $0$
& $.$
\\
\bottomrule
\end{tabular}
\caption{Summary of $\Z_n$ S-folds and the corresponding Kodaira fibers and global symmetry. The Kodaira fiber for the parent theory $\SU(2)+N_f\mathbf{F}$ takes the form $(F_\infty, F_{\rm bulk})=(I_{8-N_f};I_1^{4+N_f})$. 
The $\text{rank}_{F}(\mathcal{T}/\Z_n)$ is the $\text{rank}(\text{MW})$ for the associated $n$-isogeny.}
\label{tab:summary}
\end{table}

It is worth noting that the $n$-to-1 map used to realize the $\mathbb{Z}_n$ S-folds is nothing but a cyclic fiberwise $n$-isogeny acting on the curves in Weierstrass form. The quotient curves can further be characterized through torsion points. The arithmetic of elliptic fibrations, namely the group structure of rational sections and their torsion, has already provided important information about discrete quotients and the global structures of compactified theories~\cite{Xie:2022lcm, Closset:2021lhd, Closset:2023pmc, Furrer:2024zzu}. Our observation is that $n$-isogenies provide yet another use of this arithmetic: an explicit realization of the $\mathbb{Z}_n$ S-folds.

Concretely, we identify rational sections of exact order $n$ on the $\mathbb{Z}_n$-symmetric parent curves and construct the corresponding cyclic fiberwise $n$-isogenies using V\'elu's formulas~\cite{Velu1971, Silverman2009arithmetic}. The resulting curves coincide with the Weierstrass forms obtained from the 5-brane web coordinate quotients. The same method applied to the $\mathbb{Z}_2$ S-folds gives their $2$-isogeny realization. The $2$-, $3$-, $4$-, and $6$-isogeny computations, therefore, place all the rank-1 curves for the $\mathbb{Z}_n$ S-folds in a common algebraic framework. We also discuss a $\Z_5$ quotient curve  from the 5-isogeny applied to the Seiberg--Witten curve of the $\SU(2)+3\F$ theory. Although it is not an S-fold, this obtained curve can be reproduced from a specialization of the mass parameters of the $E_8$ Seiberg--Witten curve.

The paper is organized as follows. In section~\ref{sec:sec:S-foldings}, we review 5-brane webs for 5d rank-1 theories and the S-fold prescription, emphasizing the restrictions on the mass parameters. Section~\ref{sec:SW-for-zn} develops a systematic method for constructing the Seiberg--Witten curves of $\Z_n$ S-folds by embedding the web coordinates into an ambient space and then taking the corresponding quotient. We show that the resulting curves agree with those obtained via the
$E_{6,7,8}$-tunings. In section~\ref{sec:346-isogeny}, we reconstruct the Seiberg--Witten curves of the S-fold theories using cyclic $n$-isogenies. In section~\ref{sec:Otherquotient}, we present a $\Z_5$ quotient curve from 5-isogeny, and discuss a 5-brane web interpretation. Appendix~\ref{app:n-to-1-map} provides explicit maps for the $\mathbb{Z}_3$ and $\mathbb{Z}_4$ quotients. Appendix~\ref{app:SW-geometry} discusses the Kodaira fiber configurations and Mordell--Weil groups. Appendix~\ref{app:Z236-curve-web} implements the $E_{6,7,8}$-tunings in curves derived from 5-brane webs. Appendix~\ref{app:rel-fug} relates fugacities associated with different exceptional group embeddings and collects the relevant character identities. 

\bigskip

\section{S-foldings of 5d rank-1 SCFTs}\label{sec:sec:S-foldings}
We review the S-folds of 5d $\mathcal{N}=1$ superconformal theories (SCFTs) of rank-1 based on type IIB 5-brane constructions~\cite{Kim:2021fxx}.

\subsection{5-brane constructions for 5d rank-1 SCFTs}\label{sec:rank1review}

$\mathcal{N}=1$ superconformal theories in five dimensions (5d) are supersymmetric gauge/non-Lagrangian theories with eight supercharges and $\SU(2)_R$ R-symmetry~\cite{Seiberg:1996bd, Intriligator:1997pq}. When admitting a gauge theory description of a gauge group $G$ at low energies, $\mathcal{N}=1$ supersymmetric multiplets have the vector multiplet and charged matter hypermultiplets. The vector multiplet consists of a vector field $A_{\mu}$ and a real scalar field $\phi$. The hypermultiplets consist of complex scalar fields forming a doublet under $\SU(2)_R$ R-symmetry, which transform in a representation $R$ of $G$. Fermionic fields are the superpartners in each multiplet. 

Its moduli space has two branches. The Coulomb branch is parameterized by the vacuum expectation values (VEV) of the scalar in the vector multiplet $\phi_{i=1,\dots,\mathrm{rank}(G)}$, taking values in the Cartan of the gauge group $G$, while the Higgs branch is parameterized by the VEV of the scalars in the hypermultiplet. In this work, we will be interested in the Coulomb  branch. The corresponding low energy theory where the gauge group $G$ is broken to $\U(1)^{\mathrm{rank}(G)}$ is then described by an effective theory of the Abelian gauge groups, which is characterized by a cubic prepotential~\cite{Intriligator:1997pq}, 
\begin{align}\label{eq:IMSprep}
    \mathcal{F}(\phi) = \frac{m_0}{2}h_{ij}\phi_i\phi_j+\frac{\kappa}{6}d_{ijk}\phi_i\phi_j\phi_k +\frac{1}{12}\bigg(\sum_{\mathsf{r}\in\mathbf{R}}|\mathsf{r}\cdot\phi|^3 - \sum_f\sum_{\mathsf{w}\in \mathbf{W}_f} |\mathsf{w}\cdot\phi+m_f|^3\bigg)\ , 
\end{align}
where $m_0$ is the instanton mass $1/g_0^2$, $\kappa$ is the classical Chern-Simons level, and $m_f$ is the mass of the $f$-th hypermultiplet. Here, $h_{ij}=\mathrm{Tr}(T_iT_j)$ with the Cartan generators $T_i$ and $d_{ijk}=\frac12\mathrm{Tr}\big(T_i\{T_j,T_k\}\big)$ is non-zero only when $G=\SU(N\ge 3)$. 
The last term in \eqref{eq:IMSprep} comes from the one-loop contribution, where $\mathbf{R}$ is the root of the Lie algebra $\mathfrak{g}$ of $G$, and $\mathbf{W}_f$ is the weight associated with the matter. For example, the prepotential for the $\SU(2)$ theory with $N_f$ hypermultiplet in the fundamental representation ($\SU(2)+N_f\F$) takes the following simple form,  
\begin{align}\label{eq:FforSU2+NfF}
    \mathcal{F}_{\SU(2)+N_f\F} (\phi) = m_0 \phi^2 +\bigg(\,\frac{8}{6} |\phi|^3 -\frac1{12}\sum_{i=1}^{N_f} \big(|\phi-m_i|^3+|\phi+m_i|^3 \big)\bigg)\ .
\end{align}

A large class of theories has a non-trivial fixed point at the UV, giving rise to superconformal theories (SCFTs) in the infinite coupling limit, where $m_0\to 0$. At the fixed point, various interesting phenomena arise, one of which is global symmetry enhancement, where the flavor symmetry of the hypermultiplets and the conserved instanton $\U(1)_I$ symmetry merge together to form a bigger symmetry. For $\SU(2)+N_f\F$, the $N_f$ hypermultiplets in the fundamental representation  (flavor) of the $\SU(2)$ gauge group form an $\SO(2N_f)$ flavor symmetry, and together with the instanton $\U(1)_I$ symmetry, its global symmetry is then enhanced to $\E_{N_f+1}\supset \SO(2N_f)\times \U(1)_I$ for $N_f\le 7$ at the fixed point~\cite{Seiberg:1996bd}.

These 5d $\mathcal{N}=1$ SCFTs can be constructed from 5-brane webs in Type IIB string theory~\cite{Aharony:1997ju, Aharony:1997bh}, which are also known as $(p,q)$ 5-brane webs, where the coprime integers $p$ and $q$ are the brane-charges of D5-brane $(1,0)$ and NS5-branes $(0,1)$. A $(p,q)$ 5-brane is a bound state of $p$ charges of D5-brane and $q$ charges of NS5-brane. Their charge conserving configuration makes a 5-brane web~\cite{Aharony:1997bh} that describes BPS configurations of 5d supersymmetric gauge theories. The D5-branes and NS5-branes share the 5d worldvolume directions but occupy different directions in the transverse directions, as shown in Table~\ref{tab:brane-config}. The $x^5$-$x^6$ plane, known as $(p,q)$-plane, effectively describes the Coulomb branch physics of the theory, where the separation of D5-branes represents the different VEV of the scalar field in the vector multiplet, while the separation of two NS5-branes captures the strength of coupling of the theory; the larger the separation, the weaker the coupling.  
\begin{table}[ht]
\centering
\begin{tabular}{c|ccccc|cc|ccc}
\hline
5-/7-brane & 0 & 1 & 2 & 3 & 4 & 5 & 6 & 7 & 8 & 9 \\
\hline
NS5~(0,1)  & $\times$ & $\times$ & $\times$ & $\times$ & $\times$ & $\times$ &  &  &  &  \\
D5~(1,0)   & $\times$ & $\times$ & $\times$ & $\times$ & $\times$ &  & $\times$ &  &  &  \\
$(p,q)$    & $\times$ & $\times$ & $\times$ & $\times$ & $\times$ & $\theta$ & $\theta$ &  &  &  \\
7-brane    & $\times$ & $\times$ & $\times$ & $\times$ & $\times$ &  &  & $\times$ & $\times$ & $\times$ \\
\hline
\end{tabular}
\caption{Type IIB brane configuration for 5d $\mathcal{N}=1$ supersymmetric gauge theories. The symbol `$\times$' indicates the directions along which the brane extends. A $(p,q)$ 5-brane is represented by a line making an angle $\theta = \tan^{-1}(q/p)$.}
\label{tab:brane-config}
\end{table}

The inclusion of 7-branes in 5-brane webs enriches our understanding of the symmetries of 5d SCFTs. The 7-branes extend along the $x^{7,8,9}$ directions and appear as points in the $(p,q)$-plane, on which external $(p,q)$ 5-branes terminate. Each 7-brane carries a monodromy cut determined by its charge. When an $[r,s]$ 7-brane crosses the monodromy cut of another $[p, q]$ 7-brane, its charge is transformed by the corresponding monodromy, \begin{align}\label{eq:monodromy}
\Big[\begin{smallmatrix} r' \\ s' \end{smallmatrix}\Big]
=K_{[p,q]}\Big[\begin{smallmatrix} r \\ s \end{smallmatrix}\Big]\ , 
\quad \mathrm{where}\quad 
K_{[p,q]} =\begin{pmatrix}
    1+pq & -p^2\\
     q^2 & 1- p q
    \end{pmatrix}\ , 
\end{align}
where $K_{[p,q]}$ is the monodromy matrix associated with the $[p, q]$ 7-brane. The enhanced global symmetry $\E_{N_f+1}$ of the $\SU(2)+N_f\F$ theories can be understood based on 5-brane webs with 7-branes as follows. First, one brings all the external 7-branes for the $\SU(2)+N_f\F$ theory inside the compact face, and then rearranges the 7-branes using the monodromy so that they are organized to reveal the $ADE$-type symmetries (up to $\mathrm{SL}(2,\Z)$ transformation),  
\begin{align}\label{eq:7-brane ADE}
    \SU(n):& \quad \mathbf{A}^n \cr
    \mathrm{SO}(2n):& \quad \mathbf{A}^n\mathbf{B}\mathbf{C} \qquad (n\ge4) \cr
    \E_{n}:&\quad \mathbf{A}^{n-1}\mathbf{B}\mathbf{C}\mathbf{B} \quad  ( n\ge 6)\ , 
\end{align}
where boldface letters denote the 7-branes with the following charges
\begin{align}
\mathbf{A}=[1,0],\qquad  \mathbf{B}=[1,-1],\qquad \mathbf{C}=[1,1]\ .
\end{align}
\begin{figure}[htbp]
  \centering

  \tikzset{
    box line/.style={
      draw=black,
      line width=1.5pt,
      line cap=round,
      line join=round
    },
    red dashed line/.style={
      draw=red,
      line width=1.5pt,
      dash pattern=on 3.5pt off 4.5pt,
      line cap=round
    },
    group brace/.style={
      draw=black,
      line width=0.75pt,
      decorate,
      decoration={
        brace,
        mirror,
        amplitude=5pt
      }
    },
    diagram label/.style={
      inner sep=1pt,
      font=\normalsize
    }
  }

  \begin{tikzpicture}[x=1cm,y=1cm, scale=0.6,
  transform shape]

    \draw[red dashed line]
      (-1.8,4.8) -- (0.55,2.45);

    \draw[red dashed line]
      (-1.8,-1.8) -- (0.55,0.55);

    \draw[red dashed line]
      (9.8,4.8) -- (7.45,2.45);

    \draw[red dashed line]
      (9.8,-1.8) -- (7.45,0.55);

    \foreach \x in {2.5,3.5,5.5}{
      \draw[red dashed line]
        (\x,0.55) -- (\x,-3);
    }

    \draw[box line]
      (0,0) rectangle (8,3);

    \node[diagram label,font=\Large] at (0.85,2.15) {$\mathbf B$};
    \node[diagram label,font=\Large] at (0.85,0.85) {$\mathbf C$};
    \node[diagram label,font=\Large] at (7.15,2.15) {$\mathbf C$};
    \node[diagram label,font=\Large] at (7.15,0.85) {$\mathbf B$};

    \node[diagram label,font=\Large] at (2.5,0.9) {$\mathbf A$};
    \node[diagram label,font=\Large] at (3.5,0.9) {$\mathbf A$};
    \node[diagram label,font=\Large] at (4.5,0.9) {$\boldsymbol{\cdots}$};
    \node[diagram label,font=\Large, color=red] at (4.5,-1.5) {$\boldsymbol{\cdots}$};
    \node[diagram label,font=\Large] at (5.5,0.9) {$\mathbf A$};

    \draw[group brace,line width=1.25pt]
      (2.5,-3.25) -- (5.5,-3.25)
      node[midway,below=13pt,diagram label,font=\large] {$\boldsymbol{N_f}$};

  \end{tikzpicture}
 \caption{7-brane configuration for the $\SU(2)+N_f\F$ theories, counterclockwise arranged $\mathbf{A}^{N_f}\mathbf{BCBC}$, which shows the $E_{N_f+1}$ flavor symmetry. Red dotted lines denote the monodromy cuts of the 7-branes.}
    \label{fig:7-braneEn}
\end{figure}
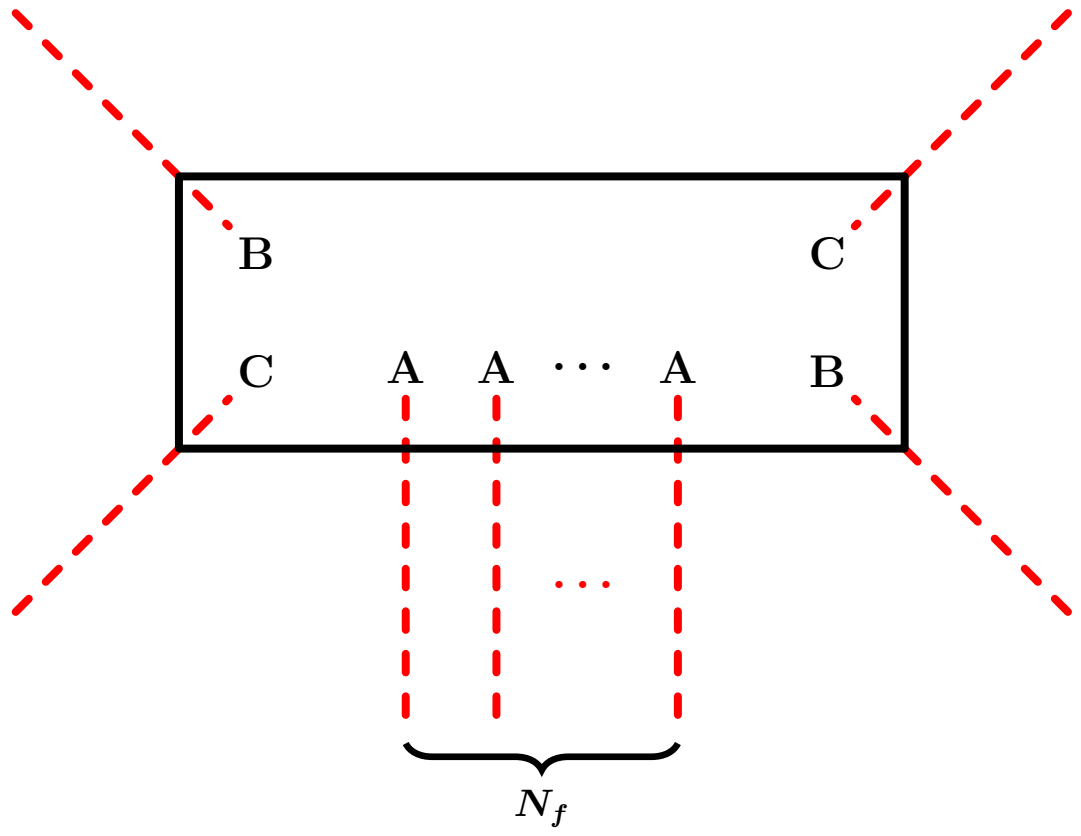
Figure~\ref{fig:7-braneEn} illustrates a 5-brane configuration for the $\SU(2)+N_f\F$ theories in which all external 7-branes have been moved into the compact face associated with the Coulomb branch. Reading the 7-branes counterclockwise yields $\mathbf{A}^{N_f}\mathbf{BCBC}$, which, by \eqref{eq:7-brane ADE}, exhibits the $\E_{N_f+1}$ flavor symmetry of these theories.

\tikzset{
    web/.style={thick, black, line cap=round, line join=round},
    dualweb/.style={densely dashed, blue!70!cyan, thick},
    labeldash/.style={dashed, black, thick},
    poly/.style={thick, black},
    pt/.style={circle, draw=black, thick, inner sep=2.2pt},
    redpt/.style={circle, draw=red, very thick, inner sep=3.5pt},
    bluept/.style={circle, draw=blue!70!cyan, very thick, inner sep=3.5pt},
    greenpt/.style={circle, draw=green!70!black, very thick, inner sep=3.5pt}
}

\subsection{\texorpdfstring{$\Z_n$}{Zn} S-folding of 5d rank-1 SCFTs}\label{sec:Zn-S-folds}
In this subsection, we review the $S$-folds of 5d rank-1 theories realized in Type IIB 5-brane web \cite{Kim:2021fxx}. Certain 5-brane webs admit discrete $\Z_n$ symmetries generated by a rotation through an angle $2\pi/n$ with the proper value of the axio-dilaton $\tau$, accompanied by an appropriate SL(2,\Z) transformation of the $(p,q)$ charges. The combined action, rather than the spatial rotation alone, preserves the charged brane configuration. For the $n=2$ case, it was extensively discussed \cite{Kim:2026yeh}. Here, we focus on the cases $n=3,4,6$. 
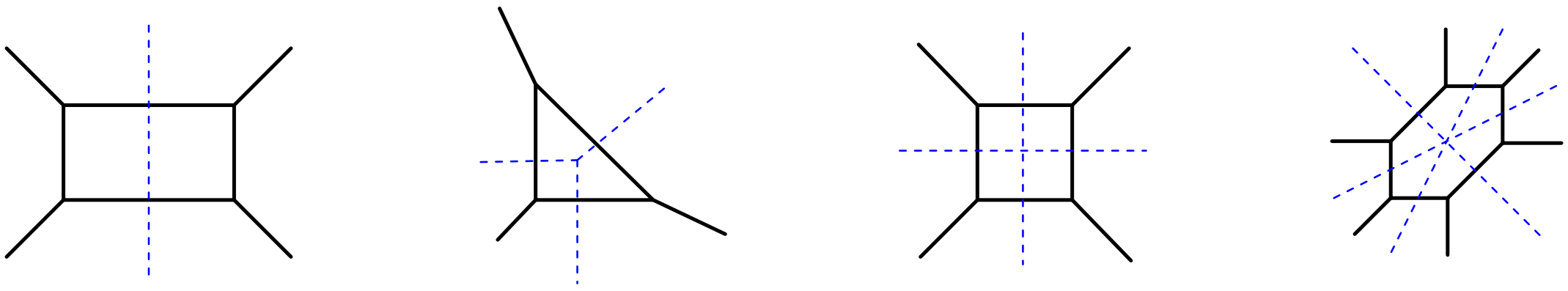
\begin{figure}[htbp]
  \centering
  \tikzset{
    solid line/.style={
      draw=black,
      line width=1.45pt,
      line cap=round,
      line join=round
    },
    guide line/.style={
      draw=blue,
      line width=0.75pt,
      dash pattern=on 2.5pt off 3.5pt,
      line cap=round
    }
  }

  \resizebox{0.93\linewidth}{!}{%
    \begin{tikzpicture}[
      x=0.75pt,
      y=0.75pt,
      yscale=-1
    ]

\begin{scope}
  \draw[solid line]
    (80,90) rectangle (170,140);

  \draw[solid line] (50,60) -- (80,90);
  \draw[solid line] (170,90) -- (200,60);

  \draw[solid line] (50,170) -- (80,140);
  \draw[solid line] (170,140) -- (200,170);

  \draw[guide line]
    (125,48) -- (125,182);
\end{scope}

      \begin{scope}
        \draw[solid line]
          (329,79) -- (391,140) -- (329,140) -- cycle;

        \draw[solid line] (310,39) -- (329,79);
        \draw[solid line] (391,140) -- (429,158);
        \draw[solid line] (329,140) -- (309,161);

        \draw[guide line] (300,120) -- (351,119);
        \draw[guide line] (351,119) -- (397,81);
        \draw[guide line] (351,119) -- (351,184);
      \end{scope}

      \begin{scope}
        \draw[solid line]
          (562,90) rectangle (612,140);

        \draw[solid line] (531,58) -- (562,90);
        \draw[solid line] (612,90) -- (642,60);
        \draw[solid line] (612,140) -- (643,172);
        \draw[solid line] (562,140) -- (532,170);

        \draw[guide line] (521,114) -- (651,114);
        \draw[guide line] (586,52) -- (586,174);
      \end{scope}

      \begin{scope}
        \draw[solid line]
          (809,50) -- (809,80) -- (839,80) -- (839,110) -- (870,110);

        \draw[solid line]
          (839,80) -- (858,61);

        \draw[solid line]
          (839,110) -- (810,139) -- (810,169);

        \draw[solid line]
          (809,80) -- (780,109) -- (780,139) -- (810,139);

        \draw[solid line] (780,109) -- (749,109);
        \draw[solid line] (780,139) -- (761,158);

        \draw[guide line] (760,60) -- (860,160);
        \draw[guide line] (839,50) -- (779,170);
        \draw[guide line] (750,139) -- (869,79);
      \end{scope}

    \end{tikzpicture}%
  }

\caption{Examples of $\Z_2$, $\Z_3$, $\Z_4$, $\Z_6$ symmetries of rank-1 theories. Here we chose the axio-dilaton to be~$\tau=i$. From the left, $\SU(2)_0$ with a generic $m_0$, $\E_0$, $\SU(2)_0$ with $m_0=0$, and $\SU(2)+2\F$.}
    \label{fig:z2z346web}
\end{figure}

As shown in figure \ref{fig:z2z346web}, simple rank-1 examples exhibit all three types of symmetry. The web of the $\E_0$ theory admits a $\Z_3$ symmetry, while that of the $\SU(2)_0$ theory admits a $\Z_4$ symmetry in the infinite-coupling limit. Similarly, the $\SU(2)+2\F$ theory admits a $\Z_6$-symmetric web at the appropriate symmetric locus. These symmetries can be gauged through the S-folds~\cite{Acharya:2021jsp, Kim:2021fxx}. In the brane-web description, this construction combines a quotient by the spatial rotation with the corresponding SL(2,\Z) identification of brane charges and the insertion of a suitable 7-brane configuration at the fixed point.

\paragraph{Prescription.} Explicit constructions of $\Z_n$ S-folds using Type IIB 5-brane webs were discussed in~\cite{Kim:2021fxx}. For a fixed point lying in the interior of a compact face, the prescription of~\cite{Kim:2021fxx} can be summarized as follows:
\begin{enumerate}
    \item[(I)]
    Given a 5-brane web invariant under the combined $\Z_n$ action, retain a fundamental domain consisting of a sector of opening angle $2\pi/n$. Its images under the S-fold action reconstruct the original web.

    \item[(II)]
    Insert the 7-brane configuration $\mathcal{S}_{\D_4}$, $\mathcal{S}_{\E_6}$, $\mathcal{S}_{\E_7}$, or $\mathcal{S}_{\E_8}$ at the fixed point for $n=2,3,4,6$, respectively. The associated monodromy implements the identification of the boundary 5-branes of the fundamental domain.
\end{enumerate}
Here, the relevant 7-brane configurations are
\begin{align}\label{eq:7-branes}
    \mathcal{S}_{\D_4}
    &= \mathbf{A}^{4}\mathbf{BC},
    &
    \mathcal{S}_{\E_6}
    &= \mathbf{A}^{5}\mathbf{BCB},
    &
    \mathcal{S}_{\E_7}
    &= \mathbf{A}^{6}\mathbf{BCB},
    &
    \mathcal{S}_{\E_8}
    &= \mathbf{A}^{7}\mathbf{BCB}.
\end{align}
Their deficit angles are
\begin{align}
    \delta_n
    = 2\pi\left(1-\frac{1}{n}\right),
    \qquad n=2, 3,4,6,
\end{align}
namely $\pi$, $4\pi/3$, $3\pi/2$, and $5\pi/3$, respectively. Together with the corresponding monodromies, these deficit angles allow the boundary 5-branes of the retained sector to be consistently identified.

The 5-brane descriptions of the $\Z_n$ S-folds yield that the area of the compact faces is reduced by $1/n$, and it was proposed~\cite{Kim:2021fxx} that for the parent theory $\mathcal{T}$, the prepotential of the S-folded theory $\mathcal{T}/\Z_n$ is given by 
\begin{align}\label{eq:1/nF}
    \mathcal{F}(\mathcal{T}/\Z_n) = \frac1n\mathcal{F}(\mathcal{T})\Big|_{\Z_n\text{-identification}}\ .
\end{align}
As demonstrated in \cite{Kim:2021fxx,Kim:2026yeh}, the prepotential can be reinterpreted as that of the resulting theory where the 7-branes \eqref{eq:7-branes} are inserted. Such inserted 7-branes supply additional degrees of freedom localized at the S-fold fixed point. The global symmetry of the resulting SCFT must therefore be determined through the total 7-brane configuration, including both the inserted stack and the 7-branes inherited from the original web.

As demonstrated via explicit computations of Seiberg–Witten curves in~\cite{Kim:2026yeh}, the flavor symmetries of $\Z_2$ S-folds of rank-1 theories are determined by the $D_4$-tuning associated with the inserted 7-brane $\mathcal{S}_{\D_4}$. We claim that this reasoning extends to the flavor symmetry for $\Z_n$ S-folds. Namely, although the inserted 7-branes provide additional degrees of freedom, the flavor symmetry visible in physical observables, say the Seiberg--Witten curve in our case, is not the flavor symmetry of the resulting SCFT due to the insertion of 7-branes. We refer to this manifest symmetry as the \emph{apparent flavor symmetry}. It follows then that the apparent flavor symmetry for the S-folded theory is not the full $E_n$ symmetry but rather a subgroup of $E_n$ restricted by the holonomies for the fixed position of the inserted 7-branes $\mathcal{S}_{D_4, E_{6,7,8}}$. Hence, the rank of apparent flavor symmetry is given by
\begin{align}\label{eq:1/nrank}
    \mathrm{rank}_{F}(\mathcal{T}/\Z_n) 
    = \frac1n\Big(\mathrm{rank}_{F}(\mathcal{T})+3\Big)-1 \ . 
\end{align}
Here, the inserted 7-branes do not contribute to the apparent flavor rank, since they are localized at the fixed point of the $\Z_n$ symmetry.\footnote{Unlike Eq.~(3.4) of~\cite{Kim:2021fxx}, \eqref{eq:1/nrank} contains no contribution $r_F(\mathcal{S})$ from the inserted 7-branes. This is because \eqref{eq:1/nrank} gives the rank of the \emph{apparent} flavor symmetry, which counts only the independently deformable mass parameters, excluding those associated with the 7-branes localized at the fixed point.}
This $ \mathrm{rank}_{F}(\mathcal{T}/\Z_n)$ then tells us the number of independent mass parameters, \#(masses), that are deformable after the $\Z_n$ S-folding.

The complete list of rank-1 theories of the $\Z_n$ S-folds is summarized as follows.
\begin{align}
    \mathcal{T}\quad \xrightarrow{\ \Z_n\ } \quad \mathcal{T}/\Z_n~~\text{with~\#(masses)~parameters}\ .
\end{align}

\subsection*{\underline{\texorpdfstring{$\Z_2$}{Z2} S-folds}}

    \begin{table}[H]
    \centering
    \small
    \renewcommand{\arraystretch}{1.3}
    \begin{tabular}{c||c|c}
        \hline
        Parent theory $\mathcal{T}$ &  S-fold $\mathcal{T}/\Z_2$ & \#(masses) \\
        \hline
        $\SU(2)_0$ & $\E_5$ & 1 \\
        $\SU(2)+2\F$ & $\E_6$ & 2 \\
        $\SU(2)+4\F$ & $\E_7$ & 3 \\
        $\SU(2)+6\F$ & $\E_8$ & 4 \\
        \hline
    \end{tabular}
\label{tab:list-z2}
\end{table}
Here, $\#\mathrm{(masses)}$ is the number of mass parameters that appear in the Seiberg–Witten curves for the S-folded theories. It is the rank of the apparent flavor symmetry \eqref{eq:1/nrank}.

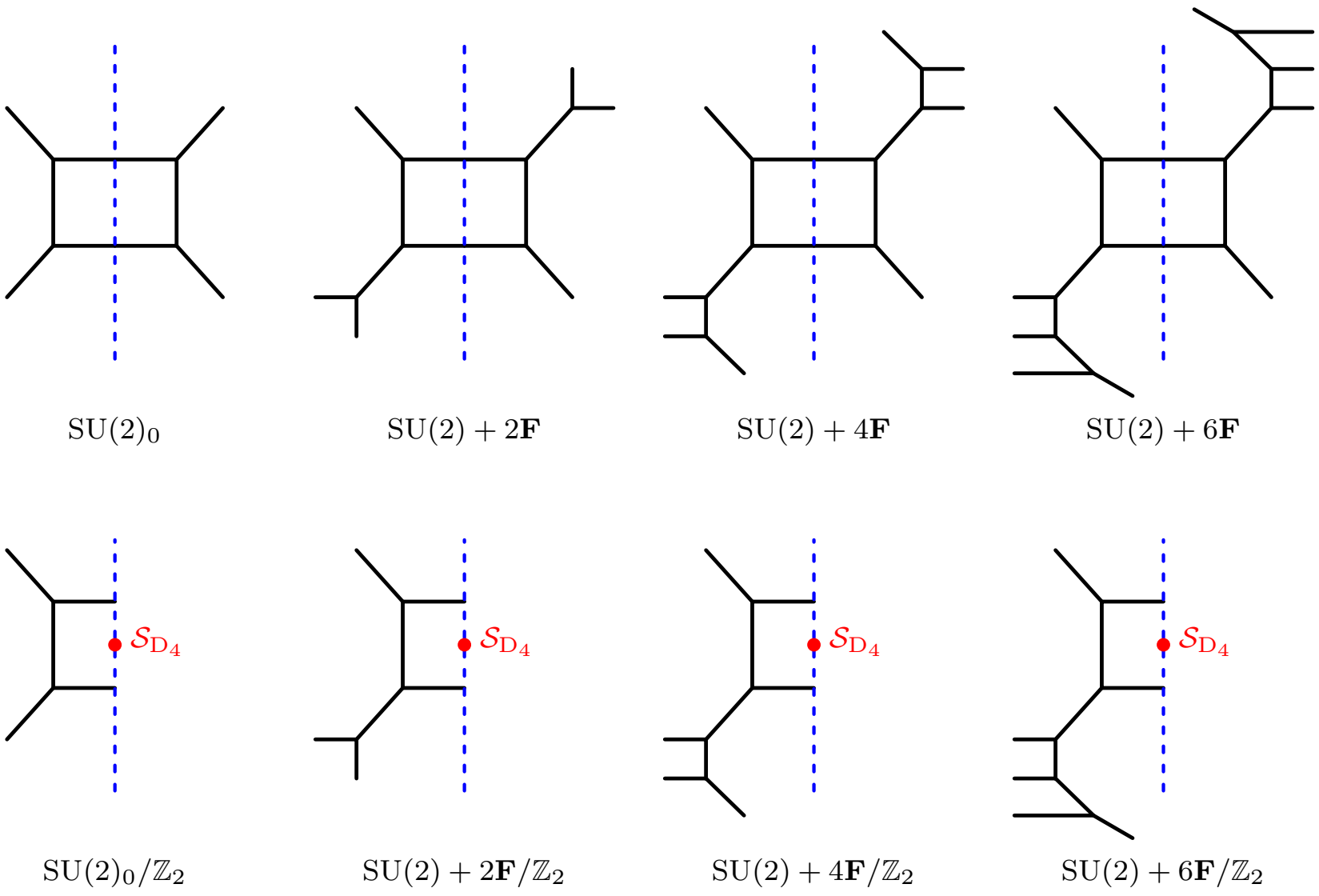
\begin{figure}[H]
    \centering
    \newcommand{\tikzfigscale}{0.85}

    \begin{tikzpicture}[
        scale=\tikzfigscale,
        transform shape,
        line cap=round,
        line join=round,
        line width=0.9pt,
        font=\small,
        guide/.style={
            draw=blue,
            line width=0.8pt,
            dash pattern=on 1.5pt off 3pt
        },
        label/.style={
            font=\scriptsize,
            inner sep=1pt
        },
        dot/.style={
            draw=red,
            fill=red,
            line width=0.5pt
        }
    ]

    \def\figsep{3.40}
    \def\rowsep{4.30}

    \newcommand{\halfweb}[1]{%
        \draw (0,0.42)
            -- (-0.6,0.42)
            -- (-0.6,-0.42)
            -- (0,-0.42);

        \draw (-1.05,0.92) -- (-0.6,0.42);
        \draw (-0.6,-0.42) -- (-1.05,-0.92);

        \ifnum#1>5\relax
            \draw (-1.45,-0.92)
                -- (-1.05,-0.92)
                -- (-1.05,-1.30);
        \fi

        \ifnum#1>6\relax
            \draw (-1.45,-1.30)
                -- (-1.05,-1.30)
                -- (-0.68,-1.66);
        \fi

        \ifnum#1>7\relax
            \draw (-1.45,-1.66)
                -- (-0.68,-1.66)
                -- (-0.30,-1.88);
        \fi
    }

     \foreach \n/\theory [count=\i from 0] in {
        5/{\SU(2)_0},
        6/{\SU(2)+2\F},
        7/{\SU(2)+4\F},
        8/{\SU(2)+6 \F}
    }{
        \begin{scope}[shift={({\i*\figsep},0)}]

            \begin{scope}[shift={(0,\rowsep)}]
                \halfweb{\n}

                \begin{scope}[rotate=180]
                    \halfweb{\n}
                \end{scope}

                \draw[guide] (0,-1.52) -- (0,1.52);

                                \node[label] at (0,-2.22)
                    {$\theory$};
            \end{scope}

            \halfweb{\n}

            \draw[guide] (0,-1.42) -- (0,1.02);

            \filldraw[dot] (0,0) circle[radius=0.055];

            \node[label,red,anchor=west]
                at (0.10,0.03)
                {$\mathcal S_{\mathrm D_4}$};

 \node[label] at (0,-2.22)
                {$\theory/\Z_2$};

        \end{scope}
    }

    \end{tikzpicture}
    \caption{The $\Z_2$ S-folds of rank-1 theories: the parent 5-brane webs for $\SU(2)_0$, $\SU(2)+2\F$, $\SU(2)+4\F$, $\SU(2)+6\F$ (upper row) and their quotient configurations (lower row) with the insertion of the $\mathcal{S}_{D_{4}}$ 7-brane at
the fixed point.}\label{fig:Z2-S-fold-fla}
\end{figure}

From the point of view of the insertion of  $\mathcal{S}_{D_4}=\mathbf{A^4BC}$, 4 mass parameters of the inserted 7-branes are tuned, while the  instanton mass is rescaled as $q\to\sqrt{q}$. See figure \ref{fig:Z2-S-fold-fla}.

\subsection*{\underline{\texorpdfstring{$\Z_3$}{Z3} S-folds}}
\begin{figure}[H]
    \centering
    \begin{tikzpicture}[
        x=0.65cm,
        y=0.65cm,
        line width=0.9pt,
        line cap=round,
        line join=round,
        every node/.style={
            font=\small,
            inner sep=2pt
        },
        branch cut/.style={
            draw=blue,
            line width=0.8pt,
            dash pattern=on 1.5pt off 3pt
        },
        sfold point/.style={
            draw=red,
            fill=red,
            line width=0.5pt
        }
    ]

    \foreach \k/\xshift in {0/0,1/7,2/14} {
        \begin{scope}[shift={(\xshift,0)}]

            \coordinate (A) at (0,2);
            \coordinate (B) at (0,0);
            \coordinate (C) at (2,0);

            \coordinate (P) at (-0.6,3.2);
            \coordinate (Q) at (-1,-1);
            \coordinate (R) at (3.2,-0.6);

            \draw (A) -- (B) -- (C) -- cycle;
            \draw (A) -- (P);
            \draw (B) -- (Q);
            \draw (C) -- (R);

            \ifnum\k>0\relax
                \draw (P) ++(-0.7,0.7) -- (P) -- ++(0,0.7);
                \draw (Q) ++(-0.7,0)   -- (Q) -- ++(0,-0.7);
                \draw (R) ++(0.7,0)    -- (R) -- ++(0.7,-0.7);
            \fi

            \ifnum\k>1\relax
                \coordinate (Ptwo) at (-1.3,3.9);
                \coordinate (Qtwo) at (-1,-1.7);
                \coordinate (Rtwo) at (3.9,-0.6);

                \draw (Ptwo) ++(-0.7,0)
                    -- (Ptwo) -- ++(0,0.7);
                \draw (Qtwo) ++(-0.7,0)
                    -- (Qtwo) -- ++(0.7,-0.7);
                \draw (Rtwo) ++(0,0.7)
                    -- (Rtwo) -- ++(0.7,-0.7);
            \fi

            \coordinate (O) at ({2/3},{2/3});

            \draw[branch cut] (O) -- ++(-1.5,0);
            \draw[branch cut] (O) -- ++(0,-1.5);
            \draw[branch cut] (O) -- ++(1.5,1.5);

        \end{scope}
    }

    \node at (0.8,-2.8)  {\text{local $\mathbb P^2$}};
    \node at (7.8,-2.8)  {$\SU(2)+2\F$};
    \node at (14.8,-2.8) {$\SU(2)+5\F$};

    \foreach \k/\xshift in {0/0,1/7,2/14} {
        \begin{scope}[shift={(\xshift,-5.2)}]

            \coordinate (O) at (0.8,0.8);
            \coordinate (V) at (0.1,0.1);
            \coordinate (Q) at (-0.6,-0.6);

            \draw[branch cut] (O) -- ++(-1.7,0);
            \draw[branch cut] (O) -- ++(0,-1.7);

            \draw (0.1,0.8) -- (V) -- (0.8,0.1);
            \draw (V) -- (Q);

            \ifnum\k>0\relax
                \draw (Q) ++(-0.7,0) -- (Q) -- ++(0,-0.7);
            \fi

            \ifnum\k>1\relax
                \coordinate (Qtwo) at (-0.6,-1.3);
                \draw (Qtwo) ++(-0.7,0)
                    -- (Qtwo) -- ++(0.7,-0.7);
            \fi

            \draw[sfold point] (O) circle[radius=2pt];
            \node[red,anchor=west] at (1.12,0.8) {$\mathcal S_{\E_6}$};

        \end{scope}
    }

    \node at (0.8,-7.7)
        {$\mathbb P^2/\Z_3$};
    \node at (7.8,-7.7)
        {$\SU(2)+2\F/\Z_3$};
    \node at (14.8,-7.7)
        {$\SU(2)+5\F/\Z_3$};

    \end{tikzpicture}
    \caption{The $Z_3$ S-folds of rank-1 theories: the parent 5-brane webs for local $\mathbb P^2$, $\SU(2) + 2\F$, and $\SU(2) + 5\F$ (upper row) and the corresponding $\Z_3$ quotient configurations (lower row) with the insertion of the $S_{E_6}$ 7-brane at the fixed point.}\label{fig:Z3-S-fold-fla}
\end{figure}
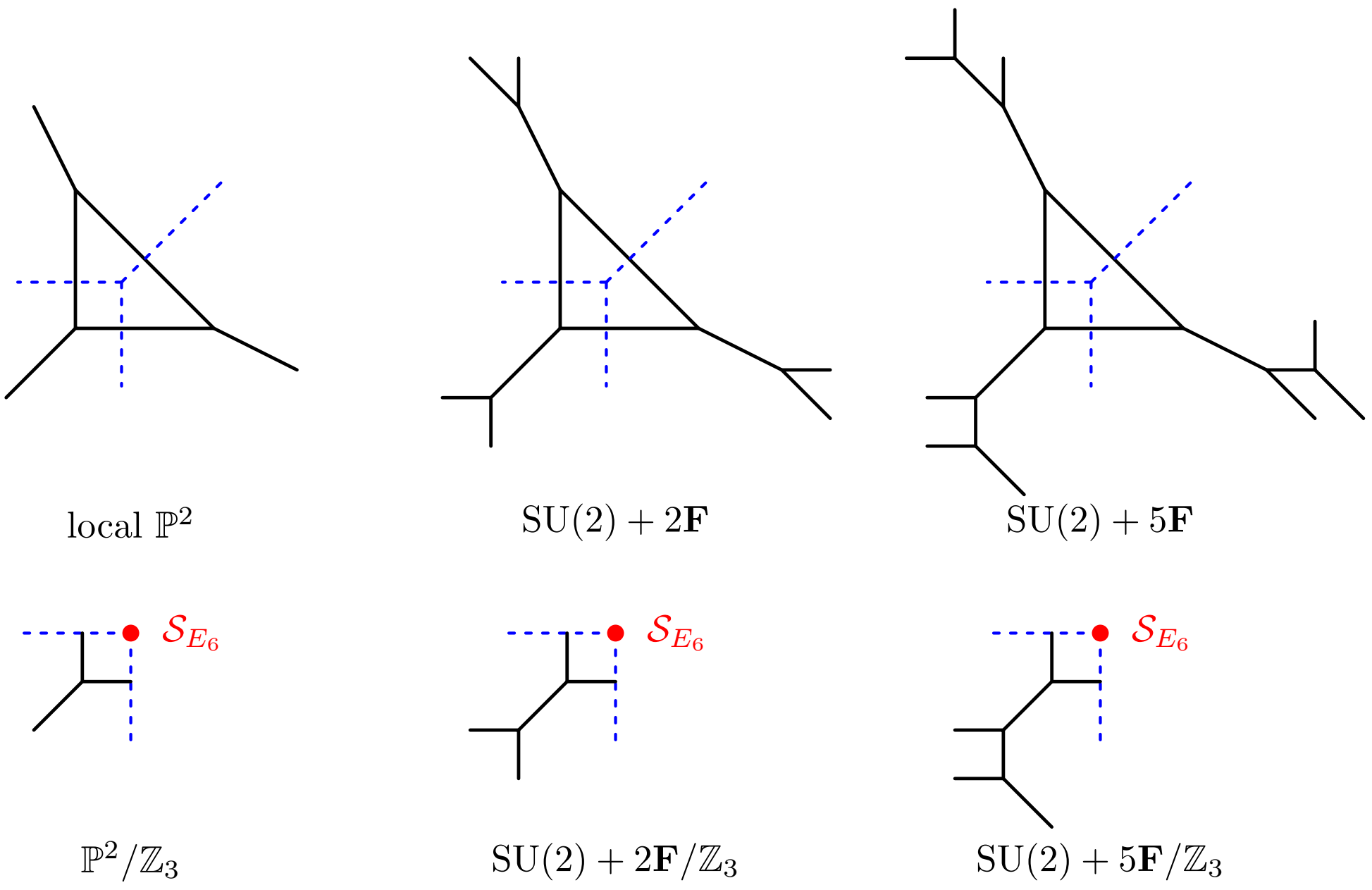

\begin{table}[H]
    \centering
    \small
    \renewcommand{\arraystretch}{1.3}
    \begin{tabular}{c||c|c}
        \hline
        Parent theory $\mathcal{T}$ & S-fold $\mathcal{T}/\Z_3$ & \#(masses) \\
        \hline
        $\text{local $\mathbb P^2$}$ & $\E_6$ & 0 \\
        $\SU(2)+2\F$ & $\E_7$ & 1 \\
        $\SU(2)+5\F$ & $\E_8$ & 2 \\
        \hline
    \end{tabular}
\label{tab:list-z3}
\end{table}
From the point of view of the insertion of $\mathcal{S}_{\E_6}=\mathbf{A^5BCB}$, 6 mass parameters of the inserted 7-branes are tuned.   See figure \ref{fig:Z3-S-fold-fla}.

\subsection*{\underline{\texorpdfstring{$\Z_4$}{Z4} S-folds}}
\begin{table}[H]
    \centering
    \small
    \renewcommand{\arraystretch}{1.3}
    \begin{tabular}{c||c|c}
        \hline
        Parent theory $\mathcal{T}$ & S-fold $\mathcal{T}/\Z_4$ & \#(masses) \\
        \hline
        $\SU(2)_0$ & $\E_7$ & 0 \\
        $\SU(2)+4\F$ & $\E_8$ & 1 \\
        \hline
    \end{tabular}
\label{tab:list-z4}
\end{table} 

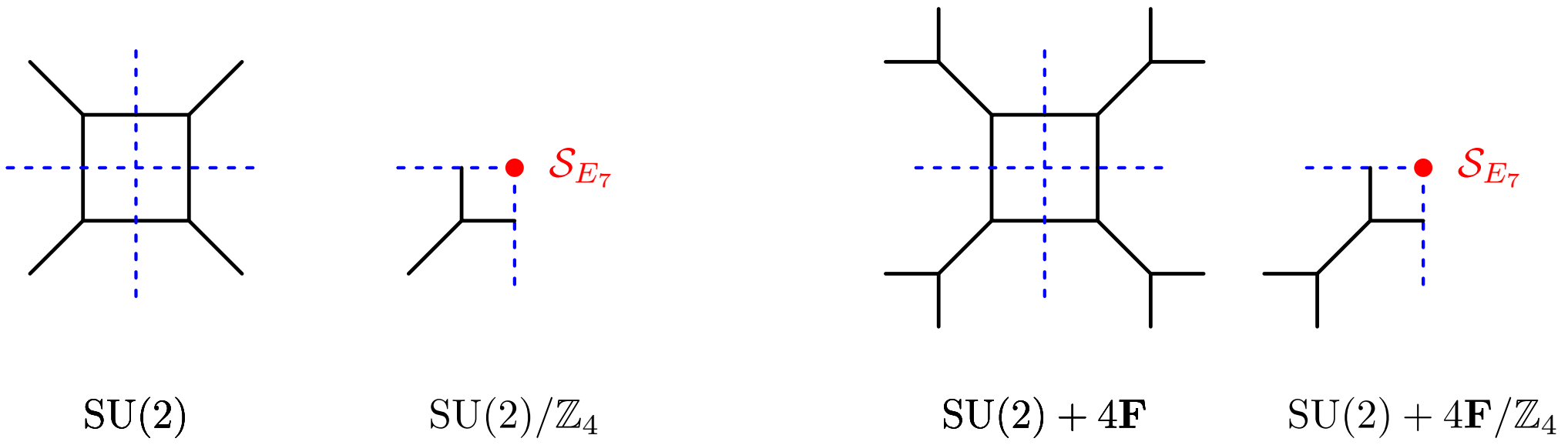
\begin{figure}[H]
    \centering
    \begin{tikzpicture}[
        x=0.65cm,
        y=0.65cm,
        line width=0.9pt,
        line cap=round,
        line join=round,
        every node/.style={
            font=\small,
            inner sep=2pt
        },
        branch cut/.style={
            draw=blue,
            line width=0.8pt,
            dash pattern=on 1.5pt off 3pt
        },
        sfold point/.style={
            draw=red,
            fill=red,
            line width=0.5pt
        }
    ]

    \def\innersep{4.0}

    \newcommand{\quarterweb}[1]{%
        \draw (-0.7,0) -- (-0.7,-0.7) -- (0,-0.7);
        \draw (-0.7,-0.7) -- (-1.4,-1.4);

        \ifnum#1>0\relax
            \draw (-2.1,-1.4)
                -- (-1.4,-1.4)
                -- (-1.4,-2.1);
        \fi
    }

    \foreach \k in {0,1} {
        \begin{scope}[shift={({3*\k*\innersep},0)}]

            \foreach \rotangle in {0,90,180,270} {
                \begin{scope}[rotate=\rotangle]
                    \quarterweb{\k}
                \end{scope}
     \node at (0,-3.3)
    {$\ifnum\k=0 \SU(2)\else \SU(2)+4\F\fi$};
            }

            \draw[branch cut] (-1.7,0) -- (1.7,0);
            \draw[branch cut] (0,-1.7) -- (0,1.7);

           \begin{scope}[shift={(5,0)}]

                \draw[branch cut] (0,0) -- (-1.7,0);
                \draw[branch cut] (0,0) -- (0,-1.7);
                \node at (0,-3.3)
    {$\ifnum\k=0 \SU(2)\else \SU(2)+4\F\fi/\Z_4$};

                \quarterweb{\k}

                \draw[sfold point] (0,0) circle[radius=2pt];
                \node[red,anchor=west] at (0.32,0)
                    {$\mathcal S_{\E_7}$};

            \end{scope}

        \end{scope}
    }

    \end{tikzpicture}
    \caption{The $\Z_4$ S-folds of rank-1 theories: the parent 5-brane webs for pure $\SU(2)_0$ and $\SU(2)+4\F$ (left) and the corresponding $\Z_4$ quotient quarter-webs (right) with the insertion of the $S_{E_7}$ 7-brane at the fixed point.}\label{fig:Z4-S-fold-fla}
\end{figure}

From the point of view of the insertion of $\mathcal{S}_{\E_7}=\mathbf{A^6BCB}$, 7 mass parameters of the inserted 7-branes are tuned.  See figure \ref{fig:Z4-S-fold-fla}.

\subsection*{\underline{\texorpdfstring{$\Z_6$}{Z6} S-fold}}
\begin{figure}[H]
    \centering
    \begin{tikzpicture}[
        x=0.75pt,
        y=0.75pt,
        yscale=-1,
        line cap=round,
        line join=round,
        solid line/.style={
            draw=black,
            line width=0.9pt
        },
        guide line/.style={
            draw=blue,
            line width=0.8pt,
            dash pattern=on 1.5pt off 3pt
        },
        sfold point/.style={
            draw=red,
            fill=red,
            line width=0.5pt
        },
        every node/.style={
            font=\small,
            inner sep=2pt
        }
    ]

    \begin{scope}

        \draw[solid line]
            (0,-60) -- (0,-30)
            -- (30,-30) -- (30,0) -- (60,0);

        \draw[solid line]
            (30,-30) -- (49,-49);

        \draw[solid line]
            (30,0) -- (0,30) -- (0,60);

        \draw[solid line]
            (0,-30) -- (-30,0)
            -- (-30,30) -- (0,30);

        \draw[solid line]
            (-30,0) -- (-60,0);

        \draw[solid line]
            (-30,30) -- (-49,49);

        \draw[guide line] (-50,-50) -- (50,50);
        \draw[guide line] (30,-60) -- (-30,60);
        \draw[guide line] (-60,30) -- (60,-30);

        \node at (0,90)
            {$\SU(2)+2\bf F$};

    \end{scope}

    \begin{scope}[shift={(200,0)}]

        \draw[guide line] (0,0) -- (-60,30);
        \draw[guide line] (0,0) -- (-30,60);

        \draw[solid line]
            (-30,15) -- (-30,30) -- (-15,30);

        \draw[solid line]
            (-30,30) -- (-49,49);

        \filldraw[sfold point]
            (0,0) circle[radius=2pt];

        \node[red,anchor=west] at (8,0)
            {$\mathcal S_{E_8}$};

        \node at (-20,90)
            {$\SU(2)+2\mathbf F/\Z_6$};

    \end{scope}

    \end{tikzpicture}
    \caption{The $\Z_6$ S-fold of the $\SU(2) + 2\F$ theory: the parent 5-brane web with $\Z_6$ symmetry (left) and the quotient web (right) with the insertion of the $\mathcal{S}_{E_8}$ 7-brane at the fixed point.}\label{fig:Z6-S-fold-fla}
\end{figure}
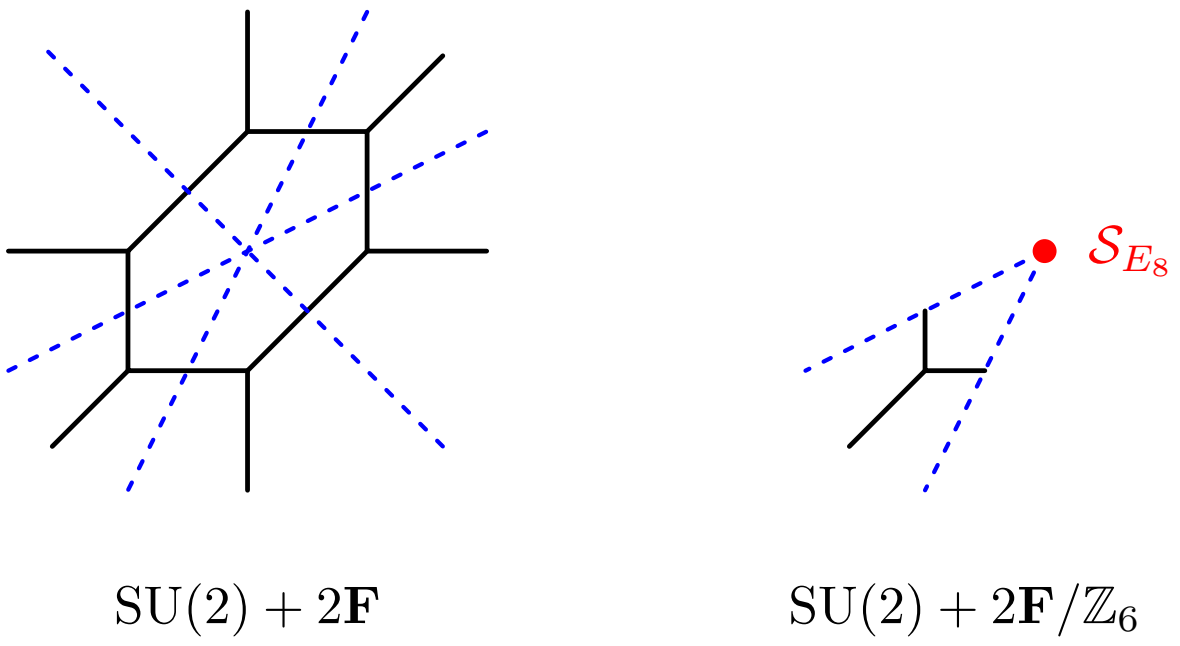

\begin{table}[H]
    \centering
    \small
    \renewcommand{\arraystretch}{1.3}
    \begin{tabular}{c||c|c}
        \hline
        Parent theory $\mathcal{T}$ &  S-fold $\mathcal{T}/\Z_6$ & \#(masses) \\
        \hline
        $\SU(2)+2\F$ & $\E_8$ & 0 \\
        \hline
    \end{tabular}
\label{tab:list-z6}
\end{table}
From the point of view of the insertion of $\mathcal{S}_{\E_8}=\mathbf{A^7BCB}$, all 8 mass parameters of the inserted 7-branes are tuned.  See figure \ref{fig:Z6-S-fold-fla}.

\bigskip

\section{Seiberg–Witten curves for 5d  \texorpdfstring{$\Z_{3,4,6}$}{Z346} S-folds}\label{sec:SW-for-zn}

In the previous section, we discussed that $\Z_n$ S-folds can be realized by a certain set of 7-branes, and as a consequence, the S-fold of  rank-1 theories can be understood as another rank-1 SCFT with higher rank global symmetry with a special tuning of mass parameters. In this section, we investigate this statement at the level of Seiberg–Witten curves. Since the case $n=2$ is already studied in \cite{Kim:2026yeh}, we focus on the cases $n=3,4,6$ in this paper. We construct Seiberg--Witten curves for $\Z_{3,4,6}$ S-folds of rank-1 SCFTs and compare them with the curves for theories with rank-1 SCFTs with higher rank global symmetry with the $E_{6,7,8}$-tuning of the mass parameters.

In order to construct the Seiberg--Witten curve for a $\Z_{n}$ S-fold theory, we start with the Seiberg–Witten curve for corresponding rank-1 SCFTs without S-folding, which is given as a generic algebraic curve of the form
\begin{align}\label{eq:SW-before-Sfold}
F_\text{generic} (t,w \,; \{ C_{\ell m} \}) := \sum_{\ell,m} C_{\ell m} t^{\ell} w^m = 0.
\end{align}
The summation range for the indices $\ell, m$ can be read from the dual graph of the corresponding 5-brane web diagram. This is a conventional expression for a Seiberg–Witten curve embedded in $\mathbb{(C^*)}^2$ parametrized by $(t,w)$.

Interpreting this curve as an M5-brane configuration, the coordinates $t, w$ are related to the coordinates $x_{4,5,6,11}$ parameterizing a 4-dimensional subspace of the 11-dimensional spacetime of M-theory as
\begin{align}
t = e^{-\frac1{R_M}(x_6+i x_{11})}, \quad w = e^{-\frac{1}{R_A}(x_5 + i x_4)},
\end{align}
where the directions labeled by $x_4$ and $x_{11}$ are compactified. Here, $R_M$ is the radius of the M-theory circle along $x_{11}$, while $R_A$ is the radius of the circle compactified in the $x_4$ direction.
In the Type IIB picture, the S-fold is understood as a simultaneous $\Z_n$ orbifold action and S-duality transformation, which is compatible with supersymmetry. The $\Z_n$ orbifold action on the $(x_5,x_6)$-plane in the Type IIB picture is realized in the M-theory picture as the same rotation acting on the same coordinates. On the other hand, the S-duality transformation in the Type IIB picture is identified with a modular transformation of the torus parametrized by $(x_4,x_{11})$ in the M-theory picture. Thus, in the M-theory picture, the S-fold action acts on both $(x_5, x_6)$-space and $(x_4, x_{11})$-space simultaneously, which can be combined into the orbifold $\Z_n$ action on the space parametrized by $(t,w)$. Compatibility with supersymmetry translates into the requirement that the transformation on $(t,w)$ be holomorphic.  

Regarding such a $\Z_n$ action $\sigma$ as a transformation that acts on the coordinates $(t,w)$ as
\begin{align}\label{eq:Zn-sigma}
\Z_n \ni \sigma: \quad (t,w) \to (t'(t,w), w'(t,w)),
\end{align}
we impose that \eqref{eq:SW-before-Sfold} is invariant under this $\Z_{n}$ action as
\begin{align}\label{eq:Zn-constraint}
F_\text{generic} ( t'(t,w), w'(t,w) \,; \{ C_{\ell m} \} ) = 
F_\text{generic} ( t , w \,; \{ C_{\ell m} \} ).
\end{align}
This leads to the constraints on the coefficients $\{ C_{\ell m} \}$. We parametrize these coefficients in terms of mass parameters $M_i$ and Coulomb modulus $U$ so as to satisfy such constraints, 
\begin{align}
C_{\ell m} = C^{(0)}_{\ell m} (U,\{ M_i \}),
\end{align}
where $C^{(0)}_{\ell m}$ is a particular solution to \eqref{eq:Zn-constraint}. 

Since the curve 
\begin{align}\label{eq:F=0}
F(t,w):= F_\text{generic} ( t , w \, ; \{ C^{(0)}_{\ell m} (U, \{ M_i \} ) \} ) = 0
\end{align}
obtained by tuning the coefficients as above is invariant under the $\Z_{n}$ action \eqref{eq:Zn-sigma} by construction, we can consider the $\Z_n$ quotient of this algebraic curve. This quotient curve gives the Seiberg–Witten curve for the $\Z_{n}$ S-fold theory, analogous to the $\Z_{2}$ case. This is understood as a curve embedded in $\mathbb{(C^*)}^2/\Z_n$.

To compare this Seiberg–Witten curve with that of the corresponding rank-1 $E_N$ SCFT with the $E_{6,7,8}$-tuning of the mass parameters, it would be convenient to rewrite the curves in the Weierstrass normal form. An important intermediate step is to rewrite the curve \eqref{eq:F=0} in $\mathbb{(C^*)}^2/\Z_n$ as a curve in $\mathbb{(C^*)}^2$. In principle, this would be possible if we could find a map from $\mathbb{(C^*)}^2/\Z_n$ to $\mathbb{(C^*)}^2$ explicitly. However, unlike the case of $\Z_{2}$ S-fold, finding such a map turns out to be not straightforward, in particular, for the case of $\Z_{6}$ S-fold\footnote{We found explicit maps for the $\Z_{3,4}$ case by trial and error, and we demonstrated the comparison using these maps in the Appendix \ref{app:n-to-1-map}. We could not find an explicit map for the $\Z_{6}$ case.}, due to the complicated manner that $\Z_n$ acts on $\mathbb{(C^*)}^2$. 
In this section, we investigate an alternative approach, which we explain below.

Repeatedly applying the $\Z_{n}$ action  \eqref{eq:Zn-sigma} to the coordinates $(t,w)$, we obtain a sequence of combinations of $(t,w)$, which we denote as $(a_k(t,w), b_k(t,w))$, $(k=0,1,\dots, n-1)$ with $(a_0, b_0):=(t,w)$. That is, if we apply the $\Z_n$ action \eqref{eq:Zn-sigma} $k$ times on $(t,w)$, we obtain
\begin{align}
\Z_n \ni \sigma^k: (t,w) \to (a_k(t,w), b_k(t,w)).
\end{align}

After performing the $\Z_{n}$ action $n$ times, it returns to the original one as $(a_n(t,w), b_n(t,w)) \\= (t,w)$. In all the cases $n=3,4,6$ considered below, the generator is chosen so that $a_1(t,w)=w=b_0(t,w)$, i.e., the $a$-sequence already contains $b_0$. It is therefore enough to consider only the sequences $a_k(t,w)$, $(k=0,1,\dots,n-1)$.

The source of the difficulty for finding a map  $\mathbb{(C^*)}^2/\Z_n \to \mathbb{(C^*)}^2$ mentioned previously was the non-triviality of the expression of $a_k(t,w)$ in $(t,w)$. In order to resolve this difficulty, we embed the $\mathbb{(C^*)}^2$ space parametrized by $(t,w)$ into a $\mathbb{(C^*)}^n$ space parametrized by a set of $n$ coordinates $a_k$ ($k=0,1,\dots, n-1$), where the embedding is given by the combinations mentioned above as 
\begin{align}\label{eq:embed-def}
a_k=a_k(t,w), \quad \quad (k=0,1,2,\dots, n-1).
\end{align}
By doing so, we can reinterpret that the $\Z_{n}$ acts on the $n$ coordinates of the $\mathbb{(C^*)}^n$ space simply as 
\begin{align}\label{eq:Zn-on-a}
\Z_n \ni \sigma: \,\, a_k \to a_{k+1} \quad (k=0,1,2,\dots, n-1), \qquad a_n:=a_0.
\end{align}
That is, we can naturally uplift the $\Z_{n}$ action on $\mathbb{(C^*)}^2$ to act on the $\mathbb{(C^*)}^n$ space in a simple manner. By construction, $\mathbb{(C^*)}^2/\Z_n$ is embedded into $\mathbb{(C^*)}^n / \Z_n$ in a consistent way.
Our strategy is to embed the Seiberg--Witten curve into this $\mathbb{(C^*)}^n/\Z_n$ space and to find a map from $\mathbb{(C^*)}^n/\Z_n$ to $\mathbb{(C^*)}^n$.
Due to the simple manner that $\Z_n$ acts on $\mathbb{(C^*)}^n$, finding such a map turns out to be much easier than finding a map from $\mathbb{(C^*)}^2/\Z_n$ to $\mathbb{(C^*)}^2$ as mentioned above.  An analogous approach has been discussed also in \cite{Acharya:2021jsp} for the case of $n=3$.

At the cost of simplifying the $\Z_n$ action as in \eqref{eq:Zn-on-a}, we need to take into account the non-trivial embedding of $\mathbb{(C^*)}^2/\Z_n$ into $\mathbb{(C^*)}^n/\Z_n$ discussed above. This embedding can also be rewritten by imposing a set of constraints $f_{\ell}( \{ a_k \}) = 0$ ($\ell=1,2,\dots, L_n$ with $L_3=1, L_4=2, L_6=5$) in such a way that its solution uniquely reproduces the original expressions for the embedding \eqref{eq:embed-def}. 
Rewriting the Seiberg--Witten curve given in \eqref{eq:F=0} in terms of the new variables $\{a_k\}$, we define
\begin{align}
f_0(\{a_k\})=F(t,w)\ .
\end{align}
Together with the relation derived above, the equation $f_\ell=0$ defines the Seiberg--Witten curve through the following system:
\begin{align}\label{eq:SW-gen-Fla}
f_{\ell}( \{ a_k \}) = 0\ 
\quad 
\left\{
\begin{array}{ll}
\ell = 0,1 & \text{ for~} n=3\\
\ell = 0,1,2 & \text{ for~} n=4\\
\ell = 0,1,2,3,4,5 & \text{ for~} n=6
\end{array}
\right. .
\end{align}
In this way, we can embed the Seiberg–Witten curve, which is originally defined as \eqref{eq:F=0}, into the $\mathbb{(C^*)}^n/\Z_n$ space.

In order to find an explicit expression for the map from $\mathbb{(C^*)}^n/\Z_n$ to $\mathbb{(C^*)}^n$, it would be convenient to introduce a new set of $n$ coordinates 
\begin{align}\label{eq:atox}
x_j := \sum_{k=0}^{n-1} \omega_n^{jk} a_k \ , \quad (j=0,1,2,\dots,n-1),
\qquad
\omega_n := e^{\frac{2 \pi i}{n}},
\end{align}
on which $\Z_n$ acts in an even simpler manner as
\begin{align}\label{eq:Zn-on-x}
\Z_n: x_j \to \omega_n^{n-j} x_j, \quad (j=0,1,2,\dots n-1).
\end{align}
Using $\omega_n^n=1$ and $\sum_{k=0}^{n-1} \omega_n^{k} = 0$, 
we find that the inverse of the coordinate transformation \eqref{eq:atox} is given by 
\begin{align}\label{eq:xtoa}
a_k = \frac{1}{n} \sum_{j=0}^{n-1} \omega_n^{-jk} x_j,
\qquad
(k=0,1,2,\dots n-1).
\end{align}

Candidates for the desired map $\mathbb{(C^*)}^n/\Z_n \to \mathbb{(C^*)}^n$ are obtained by finding $n$ independent $\Z_n$-invariant quantities constructed by combining $\{ x_j \}$.
Since acting $\Z_n$ on $x_j$ just generates simple phases $\omega_n^{-j}$ as in \eqref{eq:Zn-on-x} without mixing with other coordinates $x_k$ ($k\neq j$), it would be enough to restrict the candidate of such invariant combinations to be a set of monomials in $\{x_j\}$ such that the phases cancel. That is,
\begin{align}\label{eq:nby1}
T_k = \prod_{j=0}^{n-1} x_j^{N_{k,j}}, 
\quad
N_{k,j} \in \Z, \quad \sum_{j=1}^{n-1} j N_{k,j} = 0 \text{ mod } n, 
\quad (k=1,\dots, n).
\end{align}
Among such candidates, we look for the ones that give an $n:1$ map. That is, if we solve the set of equations \eqref{eq:nby1} with given $\{T_k\}$, we need to obtain $n$ sets of solutions for $\{ x_j \}$
\begin{align}\label{eq:n-sol-for-x}
x_j = x_j^{(i)} (\{ T_k \}), \qquad (j=0, 1,2, \dots, n-1),
\end{align}
where $i=1,2,\dots, n$ labels $n$ different solutions.

Once we find such a map, we can rewrite the Seiberg--Witten curve given in \eqref{eq:SW-gen-Fla} in terms of the invariant coordinates $\{ T_k \}$ in the form
\begin{align}\label{eq:SW-gen-Gla}
h_{\ell}( \{ T_k \}) = 0 \ 
\quad 
\left\{
\begin{array}{ll}
\ell = 0,1 & \text{ for~} n=3\\
\ell = 0,1,2 & \text{ for~} n=4\\
\ell = 0,1,2,3,4,5 & \text{ for~} n=6
\end{array}
\right.,
\end{align}
by combining the coordinate transformation \eqref{eq:xtoa} and \eqref{eq:n-sol-for-x}. 
By construction, the choice of the solution out of $n$ in \eqref{eq:n-sol-for-x} does not affect the result. 
The Seiberg--Witten curve expressed in the form \eqref{eq:SW-gen-Gla} is now embedded in $\mathbb{(C^*)}^n$ that is parametrized by $\{ T_k \}$ rather than in a quotient space.

\begin{figure}
    \centering
   \begin{tikzpicture}[
    box/.style={
        draw,
        rounded corners=3pt,
        line width=0.8pt,
        align=center,
        text width=5.6cm,
        minimum height=2cm,
        inner xsep=10pt,
        inner ysep=9pt,
        font=\small
    },
    arr/.style={->, >=stealth, line width=0.8pt},
    darr/.style={->, >=stealth, dashed, line width=0.8pt}
]

    \node[box] (A) at (0,0) {
        SW curve in $(\mathbb C^*)^n/\Z_n$ \eqref{eq:SW-gen-Fla} \\[6pt]
        $f_{\ell}(\{a_k\})\!=\!0$ \\[3pt]
    };

    \node[box] (B) at (8.4,0) {
        SW curve in $(\mathbb C^*)^n$ \eqref{eq:SW-gen-Gla} \\[6pt]
        $h_{\ell}(\{T_k\})\!=\!0$
    };

    \node[box] (C) at (8.4,-4) {
        SW curve in $(\mathbb C^*)^2$ \eqref{eq:SWinC2}, \eqref{eq:Weierstrass} \\[6pt]
        $H(T_{k_1},T_{k_2})\!=\!0$ \\[4pt]
    };

    \node[box] (D) at (0,-4) {
        SW curve in $(\mathbb C^*)^2/\Z_n$ \eqref{eq:F=0} \\[6pt]
        $F(t,w)=0$
    };

    \draw[arr]
        (A.east) -- node[above=4pt, fill=white, inner sep=2pt] {$n$-to-$1$ map}
        (B.west);

    \draw[arr]
        (B.south) -- node[right=4pt, fill=white, inner sep=2pt] {Partially solve}
        (C.north);

    \draw[arr]
        (D.north) -- node[right=4pt, fill=white, inner sep=2pt, align=left] {
            Embedding\\[-1pt]
            $(\mathbb C^*)^2/\Z_n\subset (\mathbb{C}^*)^n/\Z_n$
        }
        (A.south);

    \draw[darr]
        (D.east) -- node[above=4pt, fill=white, inner sep=2pt] {?}
        (C.west);

\end{tikzpicture}
\caption{Our process of obtaining an algebraic curve in the Weierstrass normal form from the quotient algebraic curve. Since it is difficult to obtain an algebraic curve in $\mathbb{(C^*)}^2$ directly, we first uplift to $\mathbb{(C^*)}^n /\Z_n$. This makes it easier to find an $n$-to-$1$ map, which maps $n$ points in $(\mathbb{C}^*)^n$ on which $\Z_n$ acts to one point in another copy of $\mathbb{(C^*)}^n$. After obtaining a curve embedded in $\mathbb{(C^*)}^n$, we can rewrite it as a curve in $\mathbb{(C^*)}^2$ by partially solving the set of equations. Finally, we obtain Weierstrass normal form by a proper coordinate transformation.}
\label{fig:process}
\end{figure}
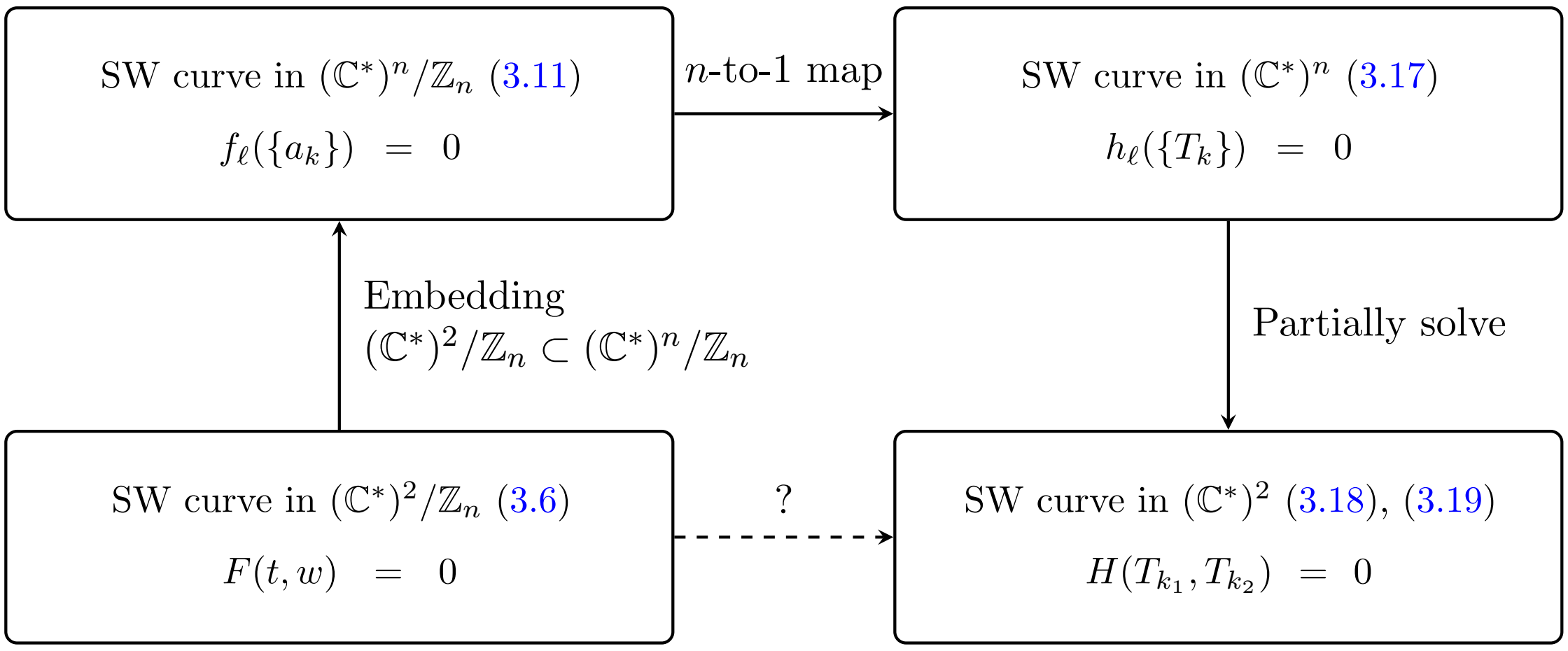

In order to obtain a standard expression for the Seiberg--Witten curve from \eqref{eq:SW-gen-Gla}, we solve $(n-2)$ equations in \eqref{eq:SW-gen-Gla} for $(n-2)$ variables out of $n$ so that they are expressed in terms of the remaining two variables $T_{k_1}, T_{k_2}$. By substituting them into the remaining equation(s), we obtain an equation for two variables 
\begin{align}\label{eq:SWinC2}
H(T_{k_1}, T_{k_2}) = 0.
\end{align}
This is an algebraic curve embedded into $\mathbb{(C^*)}^2$ similar to the usual case. With a proper coordinate transformation, we can finally express it in Weierstrass normal form
\begin{align}\label{eq:Weierstrass}
E:\quad y^2 = 4x^3 - g_2\, x - g_3,
\end{align}
by following the procedure presented in \cite{Huang:2013yta}.\footnote{
In our convention, $U$ denotes the Coulomb branch modulus of the Seiberg--Witten curve obtained from a 5-brane web and $u$ denotes the corresponding modulus in the Weierstrass normal form.} When expressing the curve $E$, we only list $g_2$ and $g_3$ with a superscript indicating the corresponding theory.
We summarize the procedure explained above in figure  \ref{fig:process}.

This procedure enables us to compare the Seiberg–Witten curve for S-fold theories with that for $\E_N$ SCFT with the $E_{6,7,8}$-tuning of the mass parameters. In the following, we perform this process more explicitly for $\Z_{3}$, $\Z_{4}$, and $\Z_{6}$ S-fold theories, respectively. In each case, we find an explicit tuning of the mass parameters of the corresponding rank-1 $\E_N$ SCFT for which its Seiberg–Witten curve agrees with that of the S-fold theories.

\subsection{\texorpdfstring{Seiberg–Witten curves for $\Z_{3}$}{Z3} S-folds}\label{sec:Z3Sfold}

Due to the presence of the $\Z_3$ S-fold, the theory exhibits $\Z_3$ symmetry under the following transformation   
\begin{align}\label{eq:twz3}
    t\to w\to\frac1{tw}\to t\ .
\end{align}
We define
\begin{align}\label{eq:Z3-defa}
a_0=t, \quad a_1=w, \quad a_2=\frac1{tw}
\end{align}
so that the $\Z_3$ transformation is written as
\begin{align}\label{eq:z3atrans}
    a_0\to a_1\to a_2\to a_0\ .
\end{align}
The embedding \eqref{eq:Z3-defa} of $\mathbb{(C^*)}^2$ parametrized by the coordinates $t,w$ into $\mathbb{(C^*)}^3$ parametrized by the coordinates $a_0,a_1,a_2$ is equivalently given by a constraint%
\footnote{
Our treatment of the $\Z_3$ S-fold closely parallels, with a minor distinction, the construction of \cite{Acharya:2021jsp}, where the Seiberg–Witten curve is embedded in $\mathbb{P}^2$. The homogeneous coordinates $X,Y,Z$ of the $\mathbb{P}^2$ are related to our coordinates $a_0, a_1, a_2$ of $\mathbb{(C^*)}^3$ as 
$$
a_0 = \frac{X}{Z}, \quad
a_1 = \frac{Z}{Y}, \quad
a_2 = \frac{Y}{X}\ .
$$
The cyclic action $X\to Z\to Y\to X$ induces the $\Z_3$ transformation in \eqref{eq:z3atrans}, allowing the corresponding invariant coordinates to be constructed in the same manner.
The essential distinction lies in the underlying constraints: our coordinates satisfy \eqref{eq:consZ3}, whereas the homogeneous coordinates obey the projective identification 
$$(X,Y,Z) \sim (\lambda X, \lambda Y,\lambda Z).$$ 
Consequently, although the two descriptions are closely related for the $\Z_3$ S-fold, extending the construction of \cite{Acharya:2021jsp} to the $\Z_4$ and $\Z_6$ cases does not yield an equally direct correspondence with the approach adopted here.}
\begin{align}\label{eq:consZ3}
f_1(\{a_i\})
:= a_0a_1a_2-1=0\ .
\end{align}

We further introduce a new set of coordinates
\begin{align}\label{eq:xkai}
&x_0 = a_0+a_1+a_2
\cr
&x_1 = a_0 + \omega_3 a_1 + \omega_3^2 a_2 
\cr
&x_2 = a_0 + \omega_3^2 a_1 + \omega_3 a_2 
\end{align}
with $\omega_3= \exp (\frac{2\pi i}3)$ so that $\Z_3$ acts
on these coordinates as
\begin{align}\label{eq:Z3-on-x}
x_0 \to x_0,
\qquad
x_1 \to \omega_3^{2} x_1,
\qquad
x_2 \to \omega_3 x_2.
\end{align}
Inverting \eqref{eq:xkai}, we obtain
\begin{align}\label{eq:Z3aixi}
    a_0&= \frac{1}{3} (x_0+x_1+x_2)\ , \nonumber\\
    a_1&=\frac{1}{3} (x_0+\omega_3^2 x_1 + \omega_3 x_2)\ , \nonumber\\
    a_2&=\frac{1}{3} (x_0+\omega_3 x_1+\omega_3^2 x_2)\ . 
\end{align}
Substituting \eqref{eq:Z3aixi} into \eqref{eq:consZ3}, we find the constraint can be rewritten in terms of $x_i$ as
\begin{align}\label{eq:consZ3-x}
\tilde{f}_1(\{ x_i \}) := x_0^3 + x_1^3 + x_2^3 - 3 x_0 x_1 x_2 - 27 = 0\ ,
\end{align}
up to an overall constant.

Then, we introduce three functions $T_i$, $i=1,2,3$, as  monomials of $x_k$:
\begin{align}\label{eq:Z3-defT}
    T_1=x_0\ ,\quad T_2=x_1x_2\ ,\quad T_3=x_2^3\ .
\end{align}
These are invariant under the transformation \eqref{eq:Z3-on-x} and have three solutions for $\{ x_i \}$ when $\{T_i\}$ are given.
Without loss of generality, we choose one of them
\begin{align}\label{eq:sol3}
    x_0=T_1\ ,\quad x_1=\frac{T_2}{T_3^{1/3}}\ ,\quad x_2=T_3^{1/3}\ .
\end{align}

Inserting \eqref{eq:sol3} into \eqref{eq:consZ3-x} yields
\begin{align}\label{eq:constZ3Xi}
h_1(\{ T_k \}) := T_3 T_1^3+T_2^3+T_3^2-3 T_2 T_3 T_1-27 T_3=0\ .
\end{align}
This is one of the constraints \eqref{eq:SW-gen-Gla} to define the Seiberg--Witten curve in $\mathbb{(C^*)}^3$. This constraint is common for all the $\Z_3$ S-fold theories.
In addition, we need one more constraint $h_0(\{ T_k \}) = 0$
coming from the original expression \eqref{eq:F=0} for the Seiberg--Witten curve to be given as in \eqref{eq:SW-gen-Gla}. 
In the following, we compute it for each $\Z_3$ S-fold theory to obtain the Seiberg--Witten curve in the Weierstrass normal form.
Then we show that the curve for each theory is reproduced by the Seiberg–Witten curve for the rank-1 $\E_N$ SCFT with a certain tuning of the mass parameters.

\subsubsection{\texorpdfstring{Local $\mathbb{P}^2$ with $\Z_3$ S-fold}{Local P2 with Z3 S-fold}}

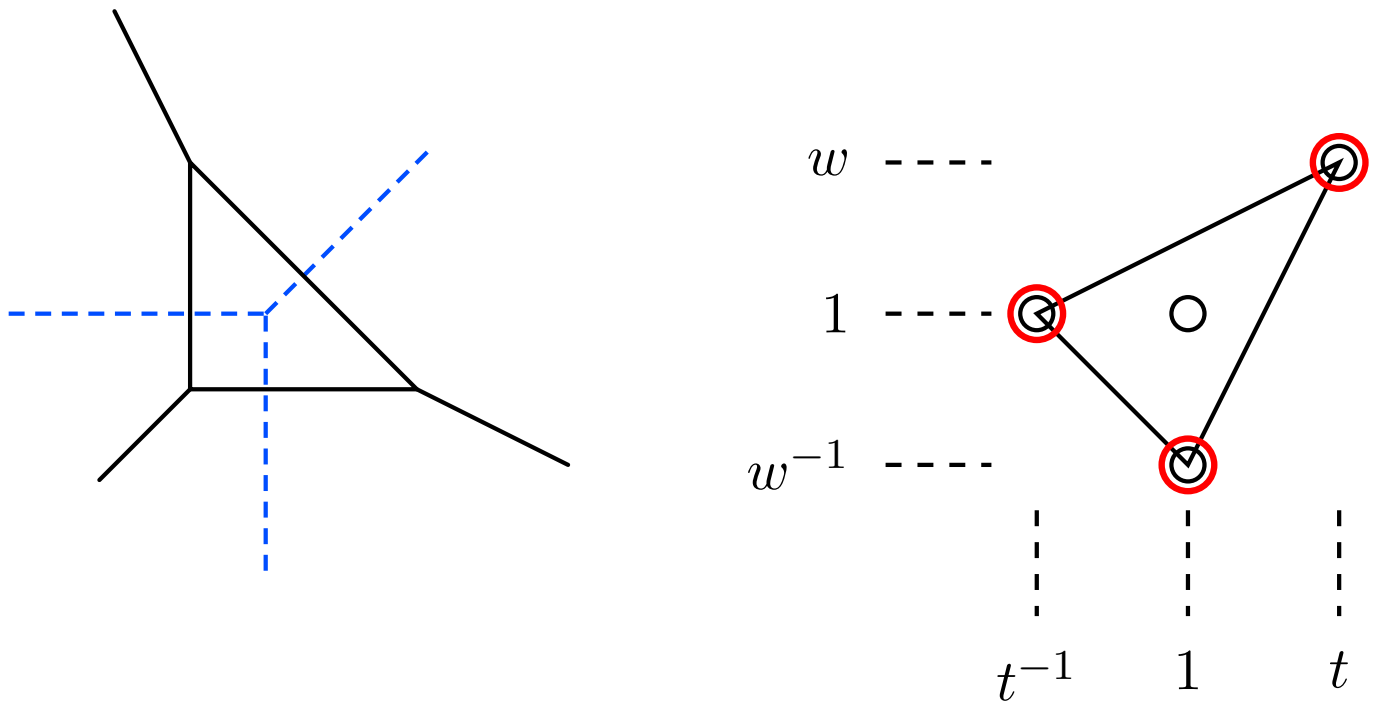
\begin{figure}[htbp]
    \centering
    \begin{tikzpicture}[scale=1]
        \begin{scope}[shift={(-3, -0.5)}]
            \draw[dualweb] (-1.2,0.5) -- (0.5,0.5);
            \draw[dualweb] (0.5,-1.2) -- (0.5,0.5);
            \draw[dualweb] (0.5,0.5) -- (1.6,1.6);

            \draw[web] (0,0) -- (1.5,0) -- (0,1.5) -- cycle;

            \draw[web] (0,0) -- (-0.6,-0.6);
            \draw[web] (0,1.5) -- (-0.5,2.5);
            \draw[web] (1.5,0) -- (2.5,-0.5);
        \end{scope}

        \begin{scope}[shift={(3.6, 0)}]
            \draw[poly] (-1,0) -- (0,-1) -- (1,1) -- cycle;

            \node[pt] at (0,0) {};
            \node[pt] at (-1,0) {}; \node[redpt] at (-1,0) {};
            \node[pt] at (0,-1) {}; \node[redpt] at (0,-1) {};
            \node[pt] at (1,1) {};  \node[redpt] at (1,1) {};

            \draw[labeldash] (-2, 1) -- (-1.3, 1);   \node[left] at (-2.1, 1) {$w$};
            \draw[labeldash] (-2, 0) -- (-1.3, 0);   \node[left] at (-2.1, 0) {$1$};
            \draw[labeldash] (-2,-1) -- (-1.3,-1);   \node[left] at (-2.1,-1) {$w^{-1}$};

            \draw[labeldash] (-1,-1.3) -- (-1,-2);   \node[below] at (-1,-2.1) {$t^{-1}$};
            \draw[labeldash] ( 0,-1.3) -- ( 0,-2);   \node[below] at ( 0,-2.1) {$1$};
            \draw[labeldash] ( 1,-1.3) -- ( 1,-2);   \node[below] at ( 1,-2.1) {$t$};
        \end{scope}
    \end{tikzpicture}
    \caption{5-brane web for local $\mathbb{P}^2$ (left) and its dual graph (right).  Dots of the same color indicate that the corresponding coefficients are identical. In this case, the coefficients are simply 1. The dot in the compact face corresponds to the Coulomb modulus $U$.}
    \label{fig:p2-pair}
\end{figure}

We start with the local $\mathbb{P}^2$ theory. The corresponding 5-brane web diagram and its dual graph are given as in figure \ref{fig:p2-pair}.

In general, each node in the dual graph corresponds to each monomial included in the Seiberg--Witten curve as discussed in \cite{Aharony:1997bh}. From the right diagram in figure~\ref{fig:p2-pair}, we can read off that the Seiberg--Witten curve for the local $\mathbb{P}^2$ theory is given by
\begin{align}\label{eq:SW-localP2-1}
U+\frac1t+\frac1w+tw =0\ ,
\end{align}
where $U$ is the Coulomb branch parameter. The coefficients of the last three terms are fixed to be 1 using the rescaling of $t, w$ and the overall rescaling of the equation.

Or, to simplify subsequent calculations, we change the coordinate $(t,w) \to (1/t,1/w)$ to obtain
\begin{align}\label{eq:SW-localP2}
   F(t,w)= U+t+w+\frac1{tw} =0\ .
\end{align}
This is invariant under the $\Z_3$ transformation \eqref{eq:twz3} without any tuning of the coefficients. Thus, by considering the $\Z_3$ quotient, we can regard \eqref{eq:SW-localP2} also as a curve for the $\Z_3$ S-fold of the local $\mathbb{P}^2$ theory.

With the definition \eqref{eq:Z3-defa}, we rewrite \eqref{eq:SW-localP2} in terms of the coordinates $\{a_i\}$. Although the way we rewrite it is not unique in theory, we should choose an expression that is natural and manifestly invariant under the $\Z_3$ transformation \eqref{eq:z3atrans} as
\begin{align}\label{eq:f0-Z3-P2}
f_0 (\{ a_i \}) := U + a_0+ a_1 + a_2  = 0.
\end{align}
The two equations \eqref{eq:consZ3} and \eqref{eq:f0-Z3-P2} define the Seiberg–Witten curve in $\mathbb{(C^*)}^3/\Z_3$.

It follows from \eqref{eq:xkai} and \eqref{eq:sol3} that
\eqref{eq:f0-Z3-P2} is rewritten in terms of the $\Z_3$ invariant coordinates $\{ T_i \}$ as
\begin{align}\label{eq:SWlocalP2}
h_0(\{T_k\}) := U + T_1 = 0\ . 
\end{align}
Now, the two equations \eqref{eq:constZ3Xi} and \eqref{eq:SWlocalP2} define the Seiberg–Witten curve in $\mathbb{(C^*)}^3$. 
Solving \eqref{eq:SWlocalP2} for $T_1$ and then plugging the result into \eqref{eq:constZ3Xi} give us the Seiberg--Witten curve in $\mathbb{(C^*)}^2$:
\begin{align}\label{eq:SWlocalFin}
H(T_2,T_3):= 
- U^3 T_3  + 3 U T_2 T_3 +T_2^3+T_3^2-27 T_3=0\ .
\end{align}
Following the standard procedure in \cite{Huang:2013yta}, we rewrite \eqref{eq:SWlocalFin} to obtain the Seiberg–Witten curve for the local $\mathbb{P}^2$ theory with $\Z_3$ S-fold in the Weierstrass normal form \eqref{eq:Weierstrass} with
\begin{align}\label{eq:localp2z3}
    g_2^{\mathbb{P}^2/\Z_3}&=\frac{u^4}{12}+18u \ ,\notag\\
    g_3^{\mathbb{P}^2/\Z_3}&=\frac{u^6}{216}-\frac52u^3-27\ .
\end{align}
where  $U=-u$. It follows from \eqref{eq:localp2z3} that the Kodaira fiber structure is $(I_3;I_3^3)$.

\subsubsection*{Comparison with $\E_6$ SCFT}
Our claim is that the Seiberg–Witten curve given in \eqref{eq:localp2z3} can be reproduced from the Seiberg–Witten curve for the rank-1 $E_6$ SCFT by properly tuning the mass parameters. 

In order to guess how exactly we should tune the mass parameters, we consider the Seiberg–Witten curve for the rank-1 $E_6$ SCFT of the form \cite{Bao:2013pwa, Kim:2014nqa}
\begin{align}\label{eq:SWE6-tri}
&w^3-\sum_i \tilde{M}_i w^2 + \sum_i \tilde{L}_i  w^2 t+ \sum_i \tilde{M}^{-1}_i w + U wt\cr
&\qquad \qquad + \sum_i \tilde{L}^{-1}_i wt^2 -1 +\sum_i \tilde{N}_i^{-1} \,t - \sum_i \tilde{N}_i \,t^2 + t^3=0
\end{align}
with 
\begin{align}
\tilde{L}_1 \tilde{L}_2 \tilde{L}_3 = \tilde{M}_1 \tilde{M}_2 \tilde{M}_3 = \tilde{N}_1 \tilde{N}_2 \tilde{N}_3 = 1.
\end{align}
This expression of the Seiberg–Witten curve is based on the 5-brane web diagram that is interpreted as a 5d uplift of the ``${T}_3$ theory'' \cite{Benini:2009gi, Gaiotto:2009we}. 

Here, $\tilde L_i$, $\tilde M_i$, $\tilde N_i$ are the fugacities of $\SU(3)\times \SU(3)\times \SU(3)$ subgroup of $E_6$ and their absolute values correspond to the positions of three $[1,1]$ 7-branes, three $[1,0]$ 7-branes (D7-branes), and three $[0,1]$ 7-branes, respectively. Since all these 7-branes should be at the origin to construct $\Z_3$ S-fold, the tuning of these fugacities should satisfy $|\tilde{L}_i| = |\tilde{M}_i| = |\tilde{N}_i| = 1$ ($i=1,2,3$). Although their phases cannot be determined only from the 5-brane web diagram, it is natural to distribute them at equal distances in the compactified $S^1$, which corresponds to the phases. This indicates that one of the most natural guesses of the tuning would be 
\begin{alignat}{3}\label{eq:tuningT3}
\tilde{L}_1 &= \omega_3^{-1}, &\qquad \tilde{L}_2 &= 1, &\qquad \tilde{L}_3 &= \omega_3 \,, \nn\\
\tilde{M}_1 &= \omega_3^{-1}, &\qquad \tilde{M}_2 &= 1, &\qquad \tilde{M}_3 &= \omega_3 \,, \nn\\
\tilde{N}_1 &= \omega_3^{-1}, &\qquad \tilde{N}_2 &= 1, &\qquad \tilde{N}_3 &= \omega_3 \,,
\end{alignat}
up to Weyl group, which we call \textit{the $E_6$-tuning}. 
Indeed, substituting these tunings into~\eqref{eq:SWE6-tri} and rewriting it in Weierstrass normal form, we reproduce~\eqref{eq:localp2z3}.

For later convenience, we also demonstrate an alternative but equivalent check based on the Seiberg–Witten curve for the rank-1 $\E_6$ SCFT in Weierstrass normal form~\cite{Minahan:1997ch, Eguchi:2002fc, Eguchi:2002nx},
\begin{align}\label{eq:E6-Weierstrass}
     g_2^{\E_6}&=\frac{1}{12}u^4 - \frac{2}{3}\chi_1^{\E_6}u^2
- \bigg(2 \chi_2^{\E_6}-12  \bigg)u
+\frac{4} {3}\chi_1^{\E_6}\chi_1^{\E_6}-4 \chi_3^{\E_6}+4 {\chi_6^{\E_6}}, \notag\\
g_3^{\E_6}&= \frac{1}{216} u^6  -\frac{1}{18}\chi_1^{\E_6}u^4
    - \bigg(\frac{1}{6}\chi_2^{\E_6}+3 \bigg)u^3
    + \bigg(\frac{2}{9}\chi_1^{\E_6}\chi_1^{\E_6}  -\frac{1}{3} \chi_3^{\E_6}
       -\frac{11}{3}\chi_6^{\E_6}\bigg)u^2 \\
       &\quad + \bigg(\frac{2}{3}\chi_1^{\E_6}\chi_2^{\E_6} -4 \chi_5^{\E_6} \bigg)u-\frac{8}{27}\chi_1^{\E_6}\chi_1^{\E_6}\chi_1^{\E_6}+\frac{4}{3}\chi_1^{\E_6}\chi_3^{\E_6}
   -\frac{4}{3}\chi_1^{\E_6}\chi_6^{\E_6}
 +\chi_2^{\E_6}\chi_2^{\E_6}-4\chi_4^{\E_6},\nonumber
\end{align}
where the characters $\chi_1^{\E_6}, \dots, \chi_6^{\E_6}$ correspond to the $27$-, $78$-, $\overline{351}$-, $2925$-, $351$-, and $\overline{27}$-dimensional representations of $E_6$, respectively. By decomposing these $E_6$ characters into the characters of $\SO(10) \times \U(1)$ as in \eqref{eq:E6cha}, we can write these explicitly in terms of the flavor mass parameters $M_1, M_2, \dots, M_5$ and the instanton mass parameter $M_0$, which are the fugacities of the $\SO(10)$ and $\U(1)$, respectively.
The $E_6$-tuning \eqref{eq:tuningT3} is equivalent to
\begin{align}\label{eq:tuningSU2-5F}
    M_1=M_2=1 \ ,\quad M_3=M_4=M_5=\omega^{-1}_3\ ,\quad M_0=-1\ ,
\end{align}
up to the Weyl group, which is explained  in Appendix \ref{sec:rel-fug-1}. With this tuning, each $E_6$ character in \eqref{eq:E6cha} reduces to
\begin{align}\label{eq:E6-tuning}
\chi_1^{\E_6} \to~ 0, \quad
\chi_2^{\E_6} \to~ -3, \quad
\chi_3^{\E_6} \to~ 0, \quad
\chi_4^{\E_6} \to~ 9, \quad
\chi_5^{\E_6} \to~ 0, \quad
\chi_6^{\E_6} \to~ 0.
\end{align}
Substituting \eqref{eq:E6-tuning} into the $E_6$ Seiberg–Witten curve \eqref{eq:E6-Weierstrass}, we again reproduce the Seiberg–Witten curve for local $\mathbb{P}^2$ with the $\Z_3$ S-fold given in \eqref{eq:localp2z3}.

\subsubsection{\texorpdfstring{$\SU(2)+2\F$ with $\Z_3$ S-fold}{SU(2)+2F with Z3 S-fold}}

\begin{figure}[htbp]
    \centering
    \begin{tikzpicture}[scale=1]
        \begin{scope}[shift={(-3.2, -0.4)}]
            \draw[dualweb] (-0.5, 0.4) -- (0.4, 0.4);
            \draw[dualweb] (0.4, -0.5) -- (0.4, 0.4);
            \draw[dualweb] (0.4, 0.4) -- (1.3, 1.3);

            \draw[web] (0,0) -- (1.2,0) -- (0,1.2) -- cycle;

            \draw[web] (0,0) -- (-0.6,-0.6);
            \draw[web] (-0.6,-0.6) -- (-1.1,-0.6);
            \draw[web] (-0.6,-0.6) -- (-0.6,-1.1);
            \draw[labeldash] (-1.3,-0.6) -- (-1.8,-0.6) node[left] {$-\M$};
            \draw[labeldash] (-0.6,-1.3) -- (-0.6,-1.8) node[below] {$-\M$};

            \draw[web] (0,1.2) -- (-0.35,1.9);
            \draw[web] (-0.35,1.9) -- (-0.35,2.35);
            \draw[web] (-0.35,1.9) -- (-0.8,2.35);

            \draw[web] (1.2,0) -- (1.9,-0.35);
            \draw[web] (1.9,-0.35) -- (2.35,-0.35);
            \draw[web] (1.9,-0.35) -- (2.35,-0.8);
        \end{scope}

        \begin{scope}[shift={(3.6, 0)}]
            \draw[poly] (-1,-1) -- (0,-1) -- (1,0) -- (1,1) -- (0,1) -- (-1,0) -- cycle;

            \node[pt] at (0,0) {};
            \node[pt] at (-1,-1) {}; \node[redpt]  at (-1,-1) {};
            \node[pt] at (0,-1)  {}; \node[bluept] at (0,-1)  {};
            \node[pt] at (1,0)   {}; \node[redpt]  at (1,0)   {};
            \node[pt] at (1,1)   {}; \node[bluept] at (1,1)   {};
            \node[pt] at (0,1)   {}; \node[redpt]  at (0,1)   {};
            \node[pt] at (-1,0)  {}; \node[bluept] at (-1,0)  {};

            \draw[labeldash] (-2, 1) -- (-1.3, 1);   \node[left] at (-2.1, 1) {$w$};
            \draw[labeldash] (-2, 0) -- (-1.3, 0);   \node[left] at (-2.1, 0) {$1$};
            \draw[labeldash] (-2,-1) -- (-1.3,-1);   \node[left] at (-2.1,-1) {$w^{-1}$};

            \draw[labeldash] (-1,-1.3) -- (-1,-2);   \node[below] at (-1,-2.1) {$t^{-1}$};
            \draw[labeldash] ( 0,-1.3) -- ( 0,-2);   \node[below] at ( 0,-2.1) {$1$};
            \draw[labeldash] ( 1,-1.3) -- ( 1,-2);   \node[below] at ( 1,-2.1) {$t$};
        \end{scope}
    \end{tikzpicture}
    \caption{5-brane web for $\SU(2)+2\F$ (left) and its dual graph (right). The $\mathbb Z_3$ invariance of the theory requires that there be only one independent mass parameter $-\M$. Dots of the same color indicate that the corresponding coefficients are identical; dots circled in blue correspond to 1 and those circled in red to $\M$. The dot in the compact face corresponds to the Coulomb modulus $U$ of the theory. }
    \label{fig:dp3-pair}
\end{figure}
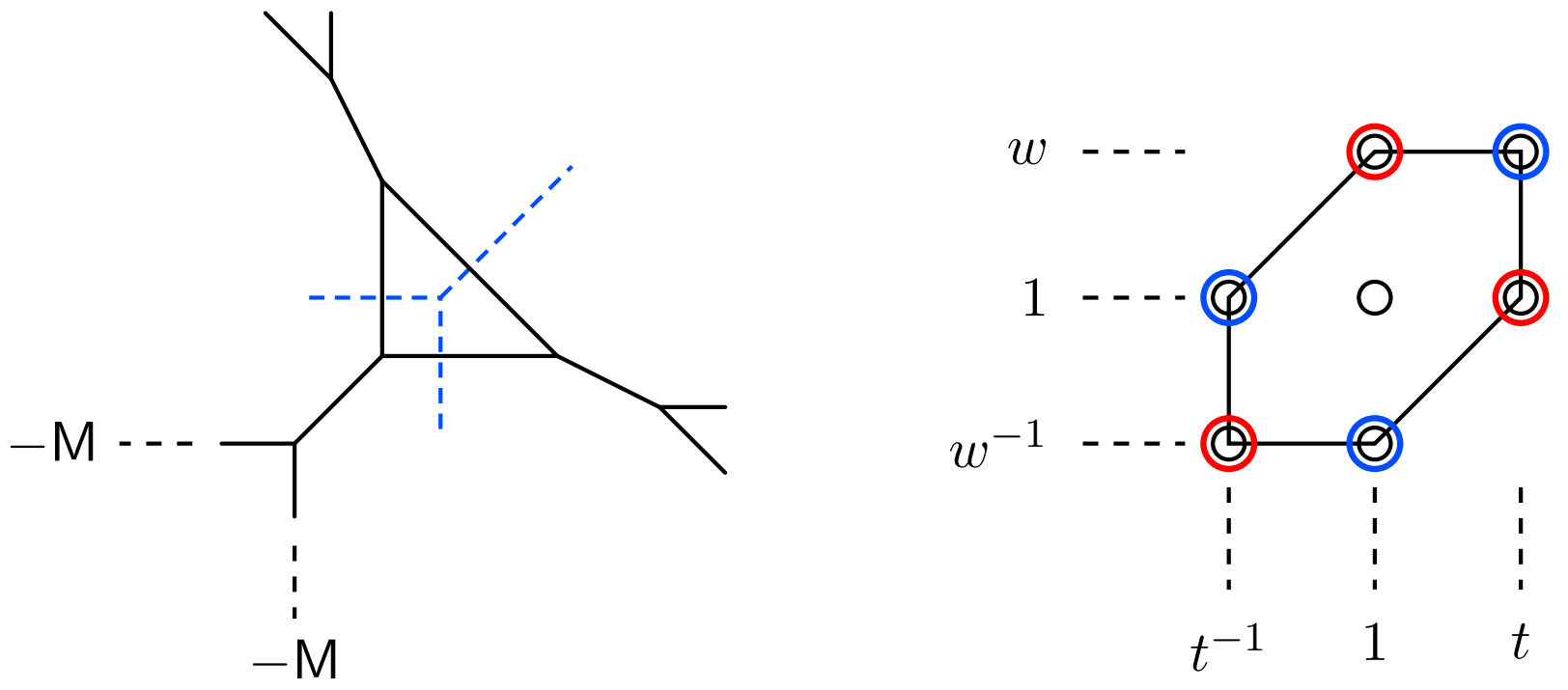

The 5-brane web and its dual graph for the $\Z_3$ S-fold of the $\SU(2)+2\F$ theory are given in figure \ref{fig:dp3-pair}. 
From the dual diagram, the $\Z_3$-invariant Seiberg–Witten curve for SU(2)+2$\F$ under the  transformation \eqref{eq:twz3} is given by

\begin{align}\label{eq:twSU2+2Fz3}
   F(t,w)=U+\M\bigg(t+w+\frac1{tw}\bigg)+\bigg(\frac1t+\frac1w+tw\bigg)\ .
   \end{align}
   Here due to the $\Z_3$ invariance of the theory, the two flavor mass parameters  are equal, which are denoted by $-\M$. As a consequence of the condition \eqref{eq:consZ3}, the Seiberg–Witten curve \eqref{eq:twSU2+2Fz3} takes the form
\begin{align}\label{eq:su22fai}
    f_0(\{a_i\}) := U+\M(a_0+a_1+a_2)+(a_0a_1+a_0a_2+a_1a_2)=0\ .
\end{align}
By substituting \eqref{eq:Z3aixi} into \eqref{eq:su22fai}, followed by the substitution of solution \eqref{eq:sol3} into the resulting equation, we arrive at
\begin{align}\label{eq:su22fXi}
   h_0(\{T_k\}):= U + \M \, T_1+\frac13(T_1^2-T_2)=0\ .
\end{align}
Upon substituting the solution for $T_2$ obtained from \eqref{eq:su22fXi} into \eqref{eq:constZ3Xi}, one obtains the Seiberg–Witten curve for $\SU(2)+2\F$ with $\Z_3$ S-fold:
\begin{align}\label{eq:su22fz3sw}
H(T_1,T_3):&=  27 \M^3 T_1^3 + 81 \M^2 T_1^2 U + 27 \M^2 T_1^4+81 \M T_1 U^2+54 \M T_1^3 U+9 \M T_1^5-9 \M T_3
   T_1^2\notag\\
&+27 T_1^2 U^2+9 T_1^4 U-9 T_3 T_1 U+T_1^6-2 T_3 T_1^3+T_3^2-27 T_3+27U^3\ .
\end{align}

As in the previous case, the application of the method of \cite{Huang:2013yta} to \eqref{eq:su22fz3sw} yields the following Weierstrass normal form:
\begin{align}\label{eq:E7z3curve}
    g_2^{2\F/\Z_3}&=\frac{u^4}{12}+18u^2-18u\,\widetilde\chi_1^{\SU(2)} \ ,\notag\\
    g_3^{2\F/\Z_3}&=\frac{u^6}{216}-\frac{5u^4}{2}+\frac{5u^3}2\widetilde\chi_1^{\SU(2)}-27u^2+54\,\widetilde\chi_1^{\SU(2)}u-27\big(\widetilde\chi_1^{\SU(2)}\big)^2\ ,
\end{align}
where $u=\frac{U}{\sqrt{\M}}$.
Here, we have introduced the character for the fundamental representation of SU(2), given in the usual form:
\begin{align}\label{eq:cha-su2}
\widetilde\chi_1^{\SU(2)}=\M^{3/2}+\frac1{\M^{3/2}}.    
\end{align}
The corresponding Kodaira fiber configuration is given by $(I_2;I_3^3I_1)$. One can see that using the Mordell--Weil lattice formalism~\cite{Caorsi:2018ahl, Xie:2022lcm, Oguiso:1991mw}, 
the corresponding flavor Lie algebra is $\mathfrak{su}(2)$. See Appendix~\ref{app:SW-geometry} for more details. 

\subsubsection*{Comparison with $\E_7$ SCFT}
In the following, we verify that the Seiberg--Witten curve given in \eqref{eq:E7z3curve} can be reproduced from the Seiberg–Witten curve for the rank-1 $E_7$ SCFT by properly tuning the mass parameters. Start from the Seiberg–Witten curve for the SU(2)+6$\F$ theory with the physical mass parametrization \cite{Minahan:1997ch, Eguchi:2002fc, Eguchi:2002nx}, in Weierstrass form, 
\begin{align}\label{eq:E7}
    g_2^{\E_7}&=\frac{1}{12}u^4 - \bigg(\frac{2}{3}\chi_1^{\E_7}-\frac{50}{3}\bigg)u^2
- \bigg(2 \chi_2^{\E_7}-12 \chi_7^{\E_7} \bigg)u
\notag\\
&\quad+\frac{4} {3}\chi_1^{\E_7}\chi_1^{\E_7}-\frac83\chi_1^{\E_7} -4 {\chi_3^{\E_7}}+4 {\chi_6^{\E_7}}+ \frac{28}{3}, \nonumber\\
g_3^{\E_7}&= \frac{1}{216} u^6  -\bigg(\frac{1}{18}\chi_1^{\E_7}+\frac{47}{18}\bigg)u^4
    - \bigg(\frac{1}{6}\chi_2^{\E_7}+3\chi_7^{\E_7} \bigg)u^3\notag\\
    &\quad+ \left(\frac{2}{9}\chi_1^{\E_7}\chi_1^{\E_7} +\frac{20}{9}\chi_1^{\E_7}  -\frac{1}{3} \chi_3^{\E_7}
       -\frac{11}{3}\chi_6^{\E_7}-\frac{196}{9}\right)u^2\notag\\
       &\quad+ \bigg(\frac{2}{3}\chi_1^{\E_7}\chi_2^{\E_7} +\frac{10}{3}\chi_2^{\E_7}-4 \chi_5^{\E_7} - 16\chi_7^{\E_7}\bigg)u
\notag\\
&\quad
 - \frac{8}{27}\chi_1^{\E_7}\chi_1^{\E_7}\chi_1^{\E_7} -\frac{28}{9}\chi_1^{\E_7}\chi_1^{\E_7}
   +\frac{4}{3}\chi_1^{\E_7}\chi_3^{\E_7}
   -\frac{4 }{3}\chi_1^{\E_7}\chi_6^{\E_7}
   +\frac{40}{9}\chi_1^{\E_7}\notag\\
   &\quad+\chi_2^{\E_7}\chi_2^{\E_7}+\frac{20}{3}\chi_3^{\E_7}
   -4\chi_4^{\E_7} -\frac83\chi_6^{\E_7} -\frac{136}{27}\ .
   \end{align}
We now perform the $E_6$-tuning with one flavor with mass $\M$, which takes the form: 
\begin{align}\label{eq:tuning-z3-2f}
    M_1=M_2=1\ ,\quad M_3=M_4=M_5=\omega_3\ ,\quad M_6=\M\ ,\quad M_0=-\M^{1/2}\ .
\end{align}
This is constructed in such a way that this goes back to the configuration of \eqref{eq:tuningSU2-5F} after the decoupling of the flavor, which corresponds to the limit $\M \to \infty$ while $M_0/M_6^{1/2}$ is fixed and identified as the instanton factor of the theory after the decoupling.  

Imposing the $E_6$-tuning on the $\E_7$ characters, explicitly given in \eqref{eq:E7cha}, we find that each $\E_7$ character reduces to: 
\begin{align}
    \chi_1^{\E_7} &\to~ -2, &
    \chi_2^{\E_7} &\to~ 3\widetilde\chi_1^{\SU(2)}, &
    \chi_3^{\E_7} &\to~ 5, \nonumber\\
    \chi_4^{\E_7} &\to~ 9\big(\widetilde\chi_1^{\SU(2)}\big)^2-1, &
    \chi_5^{\E_7} &\to~ -8\widetilde\chi_1^{\SU(2)}, &
    \chi_6^{\E_7} &\to~ 0, \nonumber\\
    \chi_7^{\E_7} &\to~ -\widetilde{\chi}_1^{\SU(2)}\  . \label{eq:E7-cha-z3}
\end{align}
This is consistent with the $\mathfrak{su}(2)$ flavor algebra identified above. Substituting \eqref{eq:E7-cha-z3} into \eqref{eq:E7}, the resulting curve agrees with the Seiberg–Witten curve for SU(2)+2$\F$  with the $\Z_3$ S-fold given in \eqref{eq:E7z3curve}. 
 
\subsubsection{\texorpdfstring{$\SU(2)+5\F$ with $\Z_3$ S-fold}{SU(2)+5F with Z3 S-fold}}

The Seiberg–Witten curve of  $\SU(2)+5\F$, which is rank 1 $\E_6$ SCFT, is given in \eqref{eq:SWE6-tri}, which we write here again as
\begin{align}
\bar{F}(t,w) &:= 
w^{-2} t^{-1} + \sum_i \tilde{M}_i w^{-1} t^{-1} + \sum_i \tilde{L}_iw^{-1} + \sum_i \tilde{M}^{-1}_i  + U \cr
&\quad + \sum_i \tilde{L}^{-1}_i  t + t^{-1} +\sum_i \tilde{N}_i^{-1} \, w + \sum_i \tilde{N}_i \,t w + t^2 w = 0 \ .
\end{align}
Here, we redefined $t \to -t$, $w \to - w^{-1}$, $U \to -U$ and multiplied the equation overall by $-w/t$. 
Imposing the invariance under \eqref{eq:twz3}, we find the constraints 
\begin{align}\label{eq:MLN-rel}
\tilde{L}_i = \tilde{N}_i = \tilde{M}_i^{-1} =: {\sf M_i}
\qquad (i=1,2,3)
\end{align}
up to the exchange of parameters $\tilde{L}_i \leftrightarrow \tilde{L}_j$, $\tilde{M}_i \leftrightarrow \tilde{M}_j$, $\tilde{N}_i \leftrightarrow \tilde{N}_j$. 
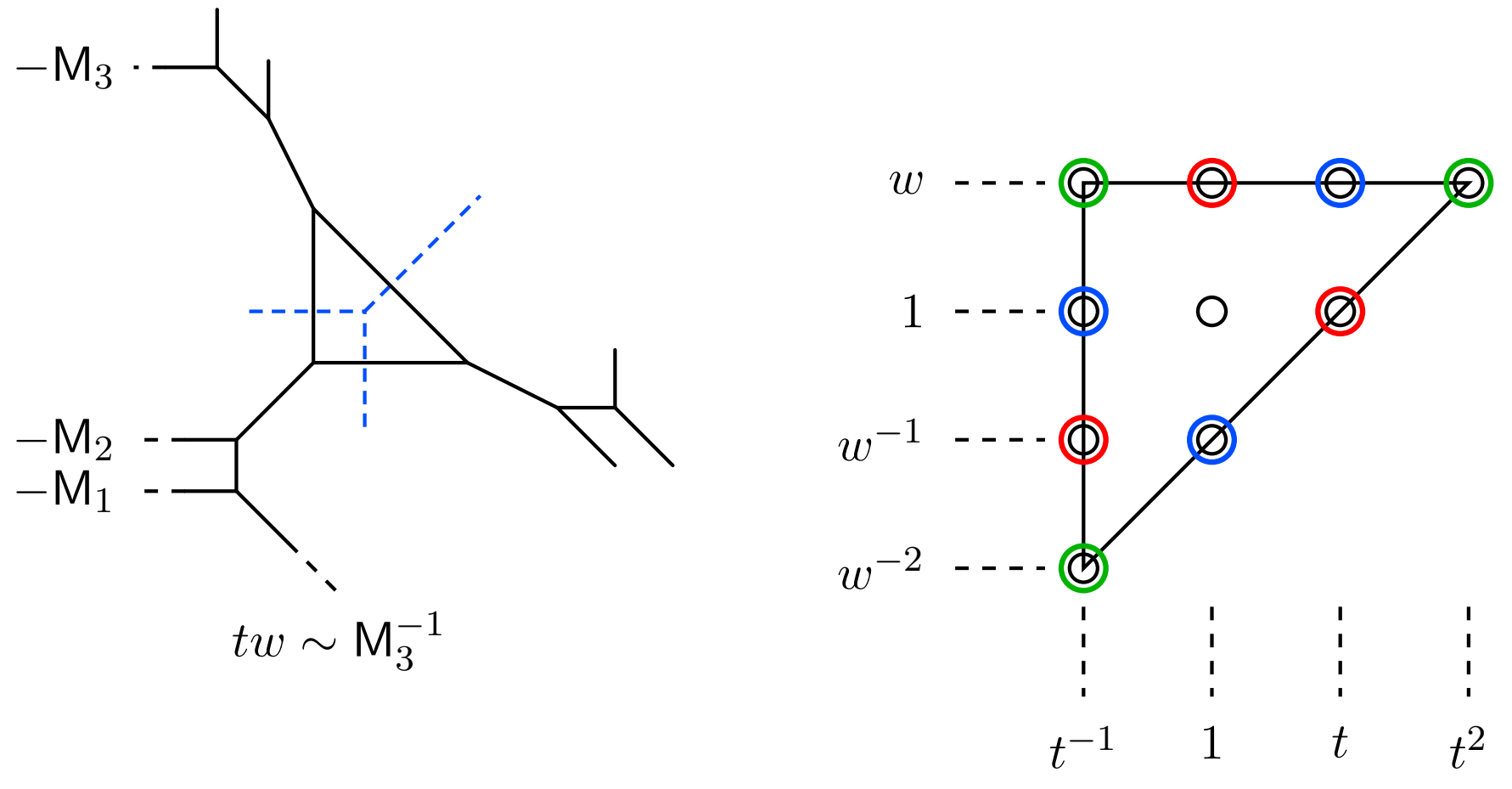
\begin{figure}[htbp]
    \centering
    \begin{tikzpicture}[scale=1]
        \begin{scope}[shift={(-3.4, -0.4)}]
            \draw[dualweb] (-0.5, 0.4) -- (0.4, 0.4);
            \draw[dualweb] (0.4, -0.5) -- (0.4, 0.4);
            \draw[dualweb] (0.4, 0.4) -- (1.3, 1.3);

            \draw[web] (0,0) -- (1.2,0) -- (0,1.2) -- cycle;

            \draw[web] (0,0) -- (-0.6,-0.6);
            \draw[web] (-0.6,-0.6) -- (-1.0,-0.6);
            \draw[labeldash] (-1.0,-0.6) -- (-1.4,-0.6) node[left] {$-\M_2$};
            \draw[web] (-0.6,-0.6) -- (-0.6,-1.0);
            \draw[web] (-0.6,-1.0) -- (-1.0,-1.0);
            \draw[labeldash] (-1.0,-1.0) -- (-1.4,-1.0) node[left] {$-\M_1$};
            \draw[web] (-0.6,-1.0) -- (-0.2,-1.4);
            \draw[labeldash] (-0.2,-1.4) -- (0.2,-1.8) node[below] {$tw \sim \M_3^{-1}$};

            \draw[web] (0,1.2) -- (-0.35,1.9);
            \draw[web] (-0.35,1.9) -- (-0.35,2.35);
            \draw[web] (-0.35,1.9) -- (-0.75,2.3);
            \draw[web] (-0.75,2.3) -- (-0.75,2.75);
            \draw[web] (-0.75,2.3) -- (-1.15,2.3);
            \draw[labeldash] (-1.15,2.3) -- (-1.4,2.3) node[left] {$-\M_3$};

            \draw[web] (1.2,0) -- (1.9,-0.35);
            \draw[web] (1.9,-0.35) -- (2.35,-0.8);
            \draw[web] (1.9,-0.35) -- (2.35,-0.35);
            \draw[web] (2.35,-0.35) -- (2.35,0.1);
            \draw[web] (2.35,-0.35) -- (2.8,-0.8);
        \end{scope}

        \begin{scope}[shift={(3.6, 0)}]
            \draw[poly] (-1,1) -- (2,1) -- (-1,-2) -- cycle;

            \node[pt] at (0,0) {};

            \node[pt] at (-1,1)  {}; \node[greenpt] at (-1,1)  {};
            \node[pt] at (2,1)   {}; \node[greenpt] at (2,1)   {};
            \node[pt] at (-1,-2) {}; \node[greenpt] at (-1,-2) {};

            \node[pt] at (0,1)   {}; \node[redpt]  at (0,1)   {};
            \node[pt] at (1,1)   {}; \node[bluept] at (1,1)   {};
            \node[pt] at (1,0)   {}; \node[redpt]  at (1,0)   {};
            \node[pt] at (0,-1)  {}; \node[bluept] at (0,-1)  {};
            \node[pt] at (-1,-1) {}; \node[redpt]  at (-1,-1) {};
            \node[pt] at (-1,0)  {}; \node[bluept] at (-1,0)  {};

            \draw[labeldash] (-2, 1) -- (-1.3, 1);   \node[left] at (-2.1, 1) {$w$};
            \draw[labeldash] (-2, 0) -- (-1.3, 0);   \node[left] at (-2.1, 0) {$1$};
            \draw[labeldash] (-2,-1) -- (-1.3,-1);   \node[left] at (-2.1,-1) {$w^{-1}$};
            \draw[labeldash] (-2,-2) -- (-1.3,-2);   \node[left] at (-2.1,-2) {$w^{-2}$};

            \draw[labeldash] (-1,-2.3) -- (-1,-3);   \node[below] at (-1,-3.1) {$t^{-1}$};
            \draw[labeldash] ( 0,-2.3) -- ( 0,-3);   \node[below] at ( 0,-3.1) {$1$};
            \draw[labeldash] ( 1,-2.3) -- ( 1,-3);   \node[below] at ( 1,-3.1) {$t$};
            \draw[labeldash] ( 2,-2.3) -- ( 2,-3);   \node[below] at ( 2,-3.1) {$t^2$};
        \end{scope}
    \end{tikzpicture}
    \caption{5-brane web for $\SU(2)+5\F$ (left) and its dual graph (right). The $\mathbb Z_3$ invariance of the theory requires that there should be two independent mass parameters, denoted by $\M_1,\M_2$. Dots sharing a color indicate equal coefficients: dots circled in blue correspond to $\chi_1^{\SU(3)}$, those circled in red to $\chi_2^{\SU(3)}$, and those circled in green to 1. The dot in the compact face corresponds to the Coulomb modulus $U$.}
    \label{fig:twostep-pair}
\end{figure}

The 5-brane web and its dual graph for the $\Z_3$ S-fold of the $\SU(2)+5\F$ theory are given in figure \ref{fig:twostep-pair}. Introducing the SU(3) characters in terms of the mass parameters $\sf M_i$
\begin{align}\label{eq:SU3-chara}
\chi^{\SU(3)}_1 
&= \sum_{i=1}^3 \M_i
= \M_1 + \M_2 + \frac{1}{\M_1 \M_2},
\cr
\chi^{\SU(3)}_2 
&= \sum_{i=1}^3 \M_i^{-1}
= \frac{1}\M_{1} + \frac{1}{\M_2} + \M_1 \M_2\ ,
\end{align}
the $\Z_3$-invariant Seiberg–Witten curve for $\SU(2)+5\F$ under the  transformation \eqref{eq:twz3} is, from the dual diagram in figure \ref{fig:twostep-pair}, given by \cite{Acharya:2021jsp}
\begin{align}\label{eq:su25fz3}
    F(t,w)=U+ {\chi^{\SU(3)}_1 }\Big(tw+\frac1t+\frac1w\Big)+\Big(\frac wt+\frac{1}{tw^2}+t^2w\Big)+{\chi^{\SU(3)}_2 }\Big(t+w+\frac1{tw}\Big)\ .
\end{align}

By replacing the variables $t$ and $w$ in \eqref{eq:su25fz3} by $a_i$, $i=0,1,2$, we have
\begingroup
\begin{align}\label{eq:su25fz3-f0}
f_0(\{ a_i \}) 
&:=U + \chi_1^{\SU(3)} (a_0 a_1 + a_1 a_2 + a_2 a_0 )+ ( a_0^2 a_1 + a_1^2 a_2 + a_2^2 a_0 )+\chi_2^{\SU(3)}  (a_0 + a_1 + a_2 )\ .
\end{align}
\endgroup
Subsequently, substituting \eqref{eq:Z3aixi} and \eqref{eq:sol3} into \eqref{eq:su25fz3-f0}, one finds:
\begin{align}\label{eq:su25swz3}
 h_0(\{T_k\}):= \omega_3 T_3^2 + \omega_3^2T_2^3-3\chi_1^{\SU(3)}\! T_3(T_2- T_1^2)+9T_3 \big(\chi_2^{\SU(3)}\!  T_1+ U\big)\!+T_3 T_1^3=0.
\end{align}
Now, the two equations \eqref{eq:constZ3Xi} and \eqref{eq:su25swz3} define the Seiberg–Witten curve in $\mathbb{(C^*)}^3$. We eliminate $T_3$ from these two equations to obtain a single equation with only $T_1, T_2$ as
\begin{align}\label{eq:SW25fz3f}
&H(T_1,T_2) :=
\cr
   & T_2^3
-3\left[
\left(\chi_1^{\SU(3)}\right)^2
+\chi_1^{\SU(3)}T_1
+T_1^2
\right]T_2^2
+\biggl[
6\left(\chi_1^{\SU(3)}\right)^2T_1^2
+18\chi_1^{\SU(3)}\chi_2^{\SU(3)}T_1
\notag\\
&
+6\chi_1^{\SU(3)}T_1^3+18\chi_1^{\SU(3)}U
-27\chi_1^{\SU(3)}
+9\chi_2^{\SU(3)}T_1^2
+9T_1U
+3T_1^4
-54T_1
\biggr]T_2
\notag\\
&
-3\left(\chi_1^{\SU(3)}\right)^2T_1^4-18\chi_1^{\SU(3)}\chi_2^{\SU(3)}T_1^3
-18\chi_1^{\SU(3)}T_1^2U
-3\chi_1^{\SU(3)}T_1^5
+27\chi_1^{\SU(3)}T_1^2
\notag\\
&
-27\left(\chi_2^{\SU(3)}\right)^2T_1^2-54\chi_2^{\SU(3)}T_1U
-9\chi_2^{\SU(3)}T_1^4
+81\chi_2^{\SU(3)}T_1
-9T_1^3U
-T_1^6\notag\\
&
+27T_1^3
-27U^2
+81U
-243=0\ .
\end{align}
To determine its Weierstrass normal form, the following variable transformation is applied to reduce the degree of \eqref{eq:SW25fz3f} to three:
\begin{align}
T_1\to T_1\ ,  \qquad T_2\to T_2-\frac13C\ , 
\end{align}
where $C$ is the coefficient of the $T_2^2$ term in \eqref{eq:SW25fz3f}. Finally, applying the approach in \cite{Huang:2013yta} to the resulting cubic curve, the Weierstrass normal form is given by
\begin{align}\label{eq:E8z3curve}
g_2^{5\F/\Z_3}&=\frac{u^4}{12}+(18 \chi^{G_2} _1-1584)u^2-(2016 \chi^{G_2} _1-18 \chi^{G_2} _2-118836)u\notag\\
&\quad+56376 \chi^{G_2} _1-972 \chi^{G_2} _2-2506788\ , \notag\\
g_3^{5\F/\Z_3}&=\frac{u^6}{216}-4u^5-\bigg(\frac{5 \chi^{G_2} _1}{2}-868\bigg)u^4+\bigg(496 \chi^{G_2} _1-\frac{5 \chi^{G_2} _2}{2}-85085\bigg)u^3\notag\\
&\quad-\bigg(27 \big(\chi^{G_2} _1\big){}^2+36018 \chi^{G_2} _1-351 \chi^{G_2} _2-4303233\bigg)u^2\notag\\
&\quad+\bigg(3132 \big(\chi^{G_2}_1\big){}^2-54 \chi^{G_2} _2 \chi^{G_2} _1+1123308 \chi^{G_2} _1-15660 \chi^{G_2} _2-109828764\bigg)u\notag\\
&\quad-90828 \big(\chi^{G_2} _1\big){}^2+3132 \chi^{G_2} _2 \chi^{G_2} _1-12471624 \chi^{G_2} _1-27 \big(\chi^{G_2}_2\big){}^2\notag\\
&\quad+215028 \chi^{G_2}_2+1121561208\ , 
\end{align}
where $U=-u+60$ and 
\begin{align}\label{eq:cha-G2}
\chi_1^{G_2}&= \chi_1^{\SU(3)}\chi_2^{\SU(3)} - 2\ ,
\cr
\chi_2^{G_2}&= \Big(\chi_1^{\SU(3)}\Big)^3
+ \Big(\chi_2^{\SU(3)}\Big)^3
- 5 \Big( \chi_1^{\SU(3)}\chi_2^{\SU(3)} - 1 \Big)\,
\end{align}
are the characters of the 7- and 14-dimensional representations of $G_2$, respectively.
The corresponding Kodaira fiber configuration is given by $(I_1;I_1^2I_3^3)$, from which one can identify the flavor algebra to be $\mathfrak{g}_2$. See Appendix~\ref{app:SW-geometry}.

\subsubsection*{Comparison with $\E_8$ SCFT}

Starting from the Seiberg–Witten curve for the SU(2)+7$\F$ theory \cite{Minahan:1997ch, Eguchi:2002fc, Eguchi:2002nx},
\begin{align}\label{eq:E8}
g_2^{\E_8}=&~\frac{1}{12}u^4 - \Big(\frac{2}{3}\chi_1^{\E_8}-\frac{50}{3}\chi_8^{\E_8}+1550\Big)u^2 
\notag\\
&+\big(70 \chi_1^{\E_8}-2 \chi_2^{\E_8}+12 \chi_7^{\E_8} -1840 {\chi_8^{\E_8}}+115010\big)u\nonumber\\
  &+\frac{4} {3}{\chi_1^{\E_8}}{\chi_1^{\E_8}}-\frac{8 }{3}{\chi_1^{\E_8}}
   {\chi_8^{\E_8}}-1824 {\chi_1^{\E_8}}+112
   {\chi_2^{\E_8}}-4 {\chi_3^{\E_8}}+4 {\chi_6^{\E_8}}
    -680 \chi_7^{\E_8}\nonumber\\
    &+\frac{28}{3}\chi_8^{\E_8} \chi_8^{\E_8} +50744 \chi_8^{\E_8}-2399276, \nonumber\\
g_3^{\E_8}=&~
 \frac{1}{216} u^6 - 4u^5 - \Big(\frac{1}{18}\chi_1^{\E_8}+\frac{47}{18}\chi_8^{\E_8}-\frac{5177}{6} \Big)u^4 \nonumber\\
&
- \Big(-\frac{107}{6}\chi_1^{\E_8}+\frac{1}{6}\chi_2^{\E_8}+3\chi_7^{\E_8} -\frac{1580}{3}\chi_8^{\E_8}+\frac{504215}{6}\Big)u^3 \nonumber\\
&
   - \bigg(-\frac{2}{9}\chi_1^{\E_8}\chi_1^{\E_8}-\frac{20}{9}\chi_1^{\E_8} \chi_8^{\E_8}+\frac{5866}{3}\chi_1^{\E_8}-\frac{112}{3}\chi_2^{\E_8}+\frac{1}{3} \chi_3^{\E_8} \nonumber\\
   &\qquad +\frac
   {11}{3}\chi_6^{\E_8}-\frac{1450}{3}\chi_7^{\E_8}+\frac{196}{9}\chi_8^{\E_8}\chi_8^{\E_8}+39296\chi_8^{\E_8}-\frac{12673792}{3} \bigg)u^2\nonumber\\
&
    - \bigg(\frac{94}{3}\big(\chi_1^{\E_8}\big)^2-\frac{2 }{3}\chi_1^{\E_8}\chi_2^{\E_8}+\frac{718}{3}\chi_1^{\E_8}\chi_8^{\E_8}-\frac{270736}{3}\chi_1^{\E_8}-\frac{10 }{3}\chi_2^{\E_8}\chi_8^{\E_8}\notag\\
    &\qquad+2630 \chi_2^{\E_8}-52\chi_3^{\E_8}+4 \chi_5^{\E_8} -416 \chi_6^{\E_8}+16
   \chi_7^{\E_8} \chi_8^{\E_8}+25880
   \chi_7^{\E_8}\notag\\
   &\qquad-\frac{7328
   }{3}\big(\chi_8^{\E_8}\big)^2-\frac{3841382
   }{3}\chi_8^{\E_8}+107263286 \bigg)u \nonumber\\
   &- \frac{8}{27}\big(\chi_1^{\E_8}\big)^3-\frac{28
   }{9}\big(\chi_1^{\E_8}\big)^2 \chi_8^{\E_8}+1065
   \big(\chi_1^{\E_8}\big)^2-\frac{118 }{3}\chi_1^{\E_8}
   \chi_2^{\E_8}+\frac{4 }{3}\chi_1^{\E_8}
   \chi_3^{\E_8}\nonumber\\
   &
   -\frac{4 }{3}\chi_1^{\E_8}\chi_6^{\E_8} +\frac{8 }{3}\chi_1^{\E_8}
   \chi_7^{\E_8}+\frac{40}{9}\chi_1^{\E_8}
   \big(\chi_8^{\E_8}\big)^2+\frac{19264 }{3}\chi_1^{\E_8}
   \chi_8^{\E_8}-\frac{4521802
   }{3}\chi_1^{\E_8}\nonumber\\
   &-\frac{572}{3}\chi_2^{\E_8} \chi_8^{\E_8}+232 \chi_5^{\E_8} +59482
   \chi_2^{\E_8}+\frac{20 }{3}\chi_3^{\E_8}
   \chi_8^{\E_8}-1880 \chi_3^{\E_8}-4
   \chi_4^{\E_8} \nonumber\\
   &-\frac{8
   }{3}\chi_6^{\E_8} \chi_8^{\E_8}-11808
   \chi_6^{\E_8}+\frac{2740 }{3}\chi_7^{\E_8}
   \chi_8^{\E_8}+460388
   \chi_7^{\E_8}-\frac{136}{27}\big(\chi_8^{\E_8}\big)^3\cr
   &-\frac{205492
   }{3}\big(\chi_8^{\E_8}\big)^2 -\frac{45856940}{3}\chi_8^{\E_8}+\big(\chi_2^{\E_8}\big)^2+1091057493 ,
\end{align}
where the dimensions of the representations associated with the characters of $\chi^{\E_8}_i$, $(i=1,2,\dots,8$), are individually, 
\begin{align}\label{eq:cha-dim-e8}
    \{3875, 147250, 6696000, 6899079264, 146325270, 2450240, 30380, 248\}.
\end{align}

The $E_6$-tuning with two additional flavors takes the form:
\begin{align}\label{eq:tuning-e8-z3}
    M_1=\!M_2\!=1,\quad M_3=\!M_4=\!M_5\!=\omega_3,\quad M_6=\!\frac1{\M_1\M_2},\quad M_7=\!\M_1,\quad M_0=\!-\M_2^{1/2}.
\end{align}
Substituting the $E_6$-tuning into the $E_8$ characters in \eqref{eq:chie8}, we find
\begin{align}\label{eq:E8-cha-tuning}
    \chi_1^{\E_8}&\to~-2\chi_1^{G_2}+1,\notag\\
    \chi_2^{\E_8}&\to~6 \chi^{G_2} _1-3 \chi^{G_2} _2-8,\notag\\
    \chi_3^{\E_8}&\to~ 5 \big(\chi^{G_2}_1\big)^2-5 \chi^{G_2} _1-5 \chi^{G_2}_2-5\ ,\notag\\
    \chi_4^{\E_8}&\to~-\big(\chi^{G_2} _1\big)^3+16 (\chi^{G_2} _1)^2-22 \chi^{G_2} _2 \chi^{G_2} _1-54 \chi^{G_2} _1+9 \big(\chi^{G_2} _2\big)^2+49 \chi^{G_2} _2+54 \ ,\notag\\
    \chi_5^{\E_8}&\to~-8 \big(\chi^{G_2} _1\big)^2+8 \chi^{G_2} _2 \chi^{G_2} _1+24 \chi^{G_2} _1-11 \chi^{G_2} _2-16\ ,\notag\\
    \chi_6^{\E_8}&\to~-3 \chi^{G_2} _1+6 \chi^{G_2}_2+8\ , \notag\\
    \chi_7^{\E_8}&\to~-2 \chi^{G_2} _1+\chi^{G_2} _2+5\ ,\notag\\
    \chi_8^{\E_8}&\to~\chi^{G_2} _1-2\ ,
\end{align}
which is consistent with the $\mathfrak{g}_2$ flavor algebra. One can insert \eqref{eq:E8-cha-tuning} into \eqref{eq:E8}, and see that the resulting curve agrees with the Seiberg--Witten curve for the $\Z_3$ S-fold of  SU(2)+5$\F$ given in \eqref{eq:E8z3curve}. 

\subsection{Seiberg--Witten curves for \texorpdfstring{$\Z_4$ S-fold}{Z4 S-fold}}
\label{sec:Z4Sfold}

For the $\Z_4$ S-fold, $\Z_4$ transformation reads
\begin{align}\label{eq:Z4trans}
       t\to w\to\frac1t\to\frac1w\to t\ .
\end{align}
Now, define 
\begin{align}
a_0=t, \qquad
a_1=w, \qquad
a_2=\frac{1}{t}, \qquad
a_3=\frac{1}{w}, 
\end{align}
and then the $\Z_4$ transformation is rephrased as
\begin{align}\label{eq:z4ai}
    a_0\to a_1\to a_2\to a_3\to a_0\ .
\end{align}
In this case, we have two independent constraints 
\begin{align}\label{eq:2constraints}
f_1(\{a_i\}) := a_0a_2-a_1a_3=0,
\quad 
f_2(\{a_i\}) := a_0a_2+a_1a_3-2=0\ .
\end{align}

Let us define
\begin{align}\label{eq:xkz4}
x_0 = a_0 + a_1 + a_2 + a_3,\qquad
x_1 = a_0 + i a_1 - a_2 - i a_3, \notag\\
x_2 = a_0-a_1+a_2-a_3,\qquad
x_3 = a_0-ia_1-a_2+ia_3,
\end{align}
so that $\Z_4$ acts
on these coordinates as
\begin{align}\label{eq:Z4-on-x}
x_0 \to x_0,
\qquad
x_1 \to -i x_1,
\qquad
x_2 \to - x_2,
\qquad
x_3 \to i x_3.
\end{align}
By inverting \eqref{eq:xkz4}, we obtain
\begin{align}\label{eq:z4ais}
    a_0&=\frac{1}{4} \left(x_0+x_1+x_2+x_3\right),\qquad
    a_1=\frac{1}{4} \left(x_0-i x_1-x_2+i x_3\right)\notag\\
    a_2&=\frac{1}{4} \left(x_0-x_1+x_2-x_3\right),\qquad
    a_3=\frac{1}{4} \left(x_0+i x_1-x_2-i x_3\right)\ .
\end{align}
Thus, by means of \eqref{eq:z4ais},  these two constraints \eqref{eq:2constraints} can be expressed in terms of $x_k$:
\begin{align}\label{eq:z4constraint}
\tilde{f}_1(\{x_i\}) &:= x_1^2+x_3^2-2 x_0 x_2=0\ , \notag\\
\tilde{f}_2(\{x_i\}) &:= x_0^2+x_2^2-2 x_1 x_3-16=0\ .
\end{align}
Then, we introduce four functions $T_i$, $i=1,\dots,4$,
\begin{align}\label{eq:4functions}
    T_1=x_0 \ ,\qquad T_2=x_1x_3 \ ,\qquad T_3=x_2x_3^2 \ ,\qquad T_4=x_2^2\ .
\end{align}
 which are expressed  in terms of $x_k,k=0,\dots,3$, and are invariant under \eqref{eq:Z4-on-x}.
It can be found that the system of these four equations \eqref{eq:4functions} has four solutions. We choose one of them as follows
\begin{align}\label{eq:sol4z4}
    x_0= T_1,\qquad x_1= \frac{T_2 \sqrt[4]{T_4}}{\sqrt{T_3}},\qquad x_2= \sqrt{T_4},\qquad x_3=
   \frac{\sqrt{T_3}}{\sqrt[4]{T_4}}
\end{align}
By substituting \eqref{eq:sol4z4} into \eqref{eq:z4constraint},  the resulting constraints read
\begin{align}\label{eq:su2constraint}
h_1(\{T_k\}) &:= T_4 T_2^2+T_3^2-2 T_1 T_3 T_4=0\ ,  \\
h_2(\{T_k\}) &:= T_1^2 +T_4 -2 T_2-16=0\ .\label{eq:swsu2z4c}
\end{align}

\subsubsection{\texorpdfstring{Pure $\SU(2)_0$ with $\Z_4$ S-fold}{Pure SU(2) with Z4 S-fold}}

\begin{figure}[htbp]
    \centering
    \begin{tikzpicture}[scale=1]
        \begin{scope}[shift={(-2.8, -0.6)}]
            \draw[dualweb] (-0.8, 0.6) -- (2.0, 0.6);
            \draw[dualweb] (0.6, -0.8) -- (0.6, 2.0);

            \draw[web] (0,0) rectangle (1.2,1.2);

            \draw[web] (0,1.2) -- (-0.6,1.8);
            \draw[web] (1.2,1.2) -- (1.8,1.8);
            \draw[web] (0,0) -- (-0.6,-0.6);
            \draw[web] (1.2,0) -- (1.8,-0.6);
        \end{scope}

        \begin{scope}[shift={(3.6, 0)}]
            \draw[poly] (-1,0) -- (0,-1) -- (1,0) -- (0,1) -- cycle;

            \node[pt] at (0,0) {};
            \node[pt] at (-1,0) {}; \node[redpt] at (-1,0) {};
            \node[pt] at (0,-1) {}; \node[redpt] at (0,-1) {};
            \node[pt] at (1,0)  {}; \node[redpt] at (1,0)  {};
            \node[pt] at (0,1)  {}; \node[redpt] at (0,1)  {};

            \draw[labeldash] (-2, 1) -- (-1.3, 1);   \node[left] at (-2.1, 1) {$w$};
            \draw[labeldash] (-2, 0) -- (-1.3, 0);   \node[left] at (-2.1, 0) {$1$};
            \draw[labeldash] (-2,-1) -- (-1.3,-1);   \node[left] at (-2.1,-1) {$w^{-1}$};

            \draw[labeldash] (-1,-1.3) -- (-1,-2);   \node[below] at (-1,-2.1) {$t^{-1}$};
            \draw[labeldash] ( 0,-1.3) -- ( 0,-2);   \node[below] at ( 0,-2.1) {$1$};
            \draw[labeldash] ( 1,-1.3) -- ( 1,-2);   \node[below] at ( 1,-2.1) {$t$};
        \end{scope}
    \end{tikzpicture}
    \caption{5-brane web for local $\mathbb{P}^1 \times \mathbb{P}^1$ (left) and its dual graph (right). Dots marked by red circles indicate that the corresponding coefficients are identical, which are equal to 1 in this theory, and the dot in the compact face corresponds to the Coulomb modulus $U$.}
    \label{fig:f0-unblown}
\end{figure}
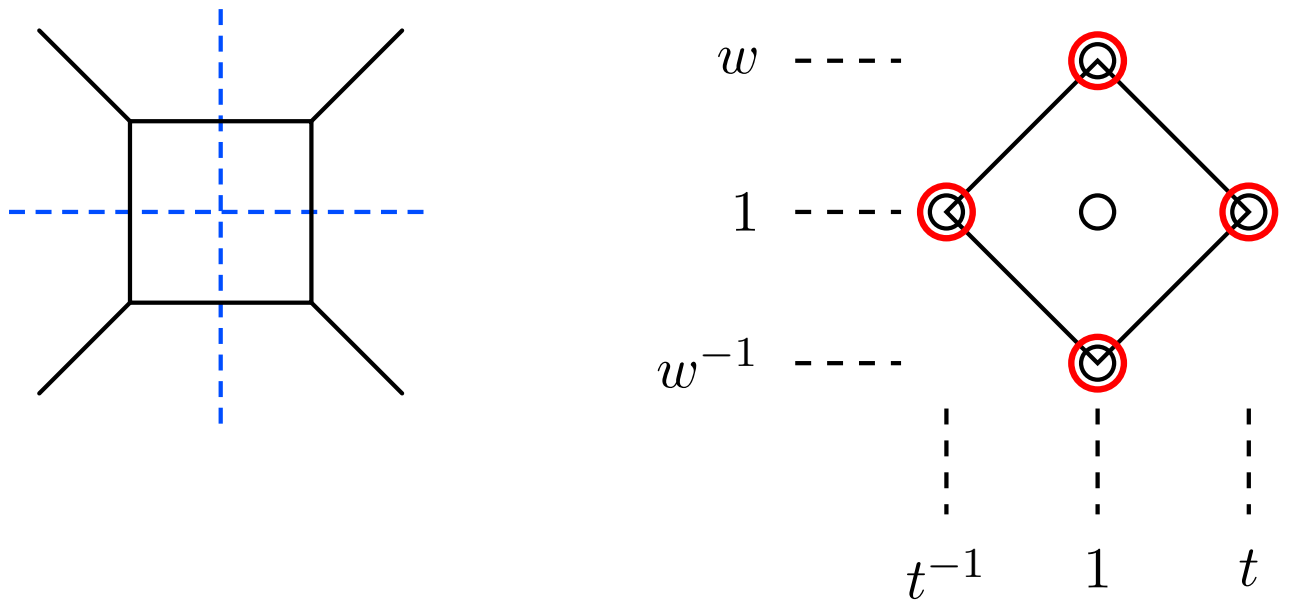
The 5-brane web and its dual graph for the $\Z_4$ S-fold of the $\SU(2)_0$ theory are given in figure \ref{fig:f0-unblown}. From the dual diagram, the $\Z_4$-invariant Seiberg–Witten curve for pure SU(2) under the transformation \eqref{eq:Z4trans} is given by
\begin{align}\label{eq:su2swcurve}
    F(t,w)=U+t+t^{-1}+w+w^{-1}\ .
\end{align}
Replace $t$ and $w$ by $a_i~(i=0,\dots,3$), and then we have
\begin{align}\label{eq:f0su24}
    f_0(\{a_i\}):=U+a_0+a_1+a_2+a_3\ .
\end{align}
According to \eqref{eq:z4ais} and \eqref{eq:sol4z4}, the curve \eqref{eq:f0su24} can be rewritten into
\begin{align}\label{eq:su2z4}
    h_0(\{T_k\}) :=U+T_1=0\ .
\end{align}
Now, the three equations \eqref{eq:su2constraint}, \eqref{eq:swsu2z4c}, and \eqref{eq:su2z4} give the Seiberg–Witten curve embedded in $\mathbb{(C^*)}^4$. 
Solving the system of \eqref{eq:su2constraint} and \eqref{eq:su2z4} for $T_1,T_2$ and substituting the resulting expressions into \eqref{eq:swsu2z4c} yield the desired Seiberg–Witten curve 
\begin{align}
    H(T_3,T_4)\!:=T_4 U^4+2 T_4^2 U^2\!-\!32 T_4 U^2+8 T_3 T_4 U+T_4^3+4 T_3^2\!-32 T_4^2+256 T_4=0\ .
\end{align}
Thus, the Weierstrass normal form is given by
\begin{align}\label{eq:E7z4curve}
     g_2^{\SU(2)_0/\Z_4}&=\frac{u^4}{12}+\frac{896 u^2}{3}+\frac{16384 }{3}\ ,\notag\\
    g_3^{\SU(2)_0/\Z_4}&=\frac{u^6}{216}-\frac{352 u^4}{9}-\frac{90112 u^2}{9}+\frac{2097152}{27}\  ,
\end{align}
where $u=4U$. It follows from \eqref{eq:E7z4curve} that the Kodaira fiber structure is $(I_2;I_4^2I_2)$.

\subsubsection*{Comparison with $\E_7$ SCFT}

Here, we would like to confirm that the Seiberg–Witten curve given in \eqref{eq:E7z4curve} can be reproduced from the Seiberg–Witten curve for the rank-1 $\E_7$ SCFT, which is known to be given by \cite{Kim:2014nqa}
\begin{align}\label{eq:E7SW}
& w^4
+ \left( \chi_{1}^{A_3} (\tilde{L}) t  - \chi_{1}^{A_3} (\tilde{M}) \right) w^3
+ \left( 
\chi_{2}^{A_3} (\tilde{L}) t^2 
+ U t 
+ \chi_{2}^{A_3} (\tilde{M}) 
\right)
w^2
\nonumber \\
&
+ ( t^2 - \chi_1^{A_1}(\tilde{N}) t + 1 )
\left( \chi_3^{A_3}(\tilde{L}) t - \chi_3^{A_3}(\tilde{M}) \right) w
+ \left( t^2 - \chi_1^{A_1}(\tilde{N})t + 1 \right)^2
 = 0,
\end{align}
where we denote the characters
\begin{align}
    \chi_1^{A_3}(\tilde{L}) &= \sum_{k=1}^4 \tilde{L}_k, &
    \chi_2^{A_3}(\tilde{L}) &= \sum_{1\leq k<\ell\leq4} \tilde{L}_k\tilde{L}_{\ell}, &
    \chi_3^{A_3}(\tilde{L}) &= \sum_{k=1}^4 \tilde{L}_k^{-1}, \nonumber\\
    \chi_1^{A_3}(\tilde{M}) &= \sum_{k=1}^4 \tilde{M}_k, &
    \chi_2^{A_3}(\tilde{M}) &= \sum_{1\leq k<\ell\leq4} \tilde{M}_k\tilde{M}_{\ell}, &
    \chi_3^{A_3}(\tilde{M}) &= \sum_{k=1}^4 \tilde{M}_k^{-1}, \nonumber\\
    \chi_1^{A_1}(\tilde{N}) &= \tilde{N}_1+\tilde{N}_2.
\end{align}
with the traceless condition of the SU groups
\begin{align}\label{eq:trace-less-E7}
\tilde{L}_1 \tilde{L}_2 \tilde{L}_3 \tilde{L}_4 = \tilde{M}_1 \tilde{M}_2 \tilde{M}_3 \tilde{M}_4 = \tilde{N}_1 \tilde{N}_2  = 1.
\end{align}
Here, $\tilde{L}_i$, $\tilde{M}_i$, $\tilde{N}_i$ are the fugacities of $\SU(4) \times \SU(4) \times \SU(2)$ subgroup of $E_7$ and their absolute values correspond to the position of four $[1,1]$ 7-branes, four $[1,0]$ 7-branes (D7-branes), and two $[0,1]$ 7-branes, respectively.

A natural guess for the tuning would be given by imposing that the M5-branes originating from these $(p,q)$ 7-branes are located at equal distance in the compactified circle. That is, for example, 
\begin{align}
\tilde{L}_{k+1} = e^{2 \pi i /4} \tilde{L}_{k},
\qquad
\tilde{M}_{k+1} = e^{2 \pi i /4}  \tilde{M}_{k},
\qquad
\tilde{N}_{k+1} = e^{2 \pi i /2}  \tilde{N}_{k}.
\end{align}
The tuning that satisfies these conditions as well as the traceless condition \eqref{eq:trace-less-E7} gives 
\begin{align}\label{eq:E7LMN-tuning}
    \tilde{L}_1 &= e^{-3\pi i/4}, &
    \tilde{L}_2 &= e^{-\pi i/4}, &
    \tilde{L}_3 &= e^{\pi i/4}, &
    \tilde{L}_4 &= e^{3\pi i/4}, \nonumber\\
    \tilde{M}_1 &= e^{-3\pi i/4}, &
    \tilde{M}_2 &= e^{-\pi i/4}, &
    \tilde{M}_3 &= e^{\pi i/4}, &
    \tilde{M}_4 &= e^{3\pi i/4}, \nonumber\\
    \tilde{N}_1 &= e^{-\pi i/2}, &
    \tilde{N}_2 &= e^{\pi i/2}.
\end{align}
up to Weyl group, which we call \textit{the $E_7$-tuning}. With this tuning, we find 
\begin{align}
\chi_1^{A_3}(\tilde{L})
= \chi_2^{A_3}(\tilde{L})
= \chi_3^{A_3}(\tilde{L})
= \chi_1^{A_3}(\tilde{M})
= \chi_2^{A_3}(\tilde{M})
= \chi_3^{A_3}(\tilde{M})
= \chi_1^{A_1}(\tilde{N})
= 0.
\end{align}
Substituting these tunings into \eqref{eq:E7SW} and rewriting it in Weierstrass normal form, we reproduce \eqref{eq:E7z4curve}. 

As done in the $E_6$ case, we discuss an alternative approach. 
Start with the $\E_7$ Seiberg--Witten curve in \eqref{eq:E7}. We tune the mass parameters as
\begin{align}\label{eq:E7-tuning-M}
    M_1=M_2=1\ ,\quad M_3=M_4=-1\ ,\quad M_5=-M_6=M_0=i \ ,
\end{align}
which is equivalent to the $E_7$-tuning \eqref{eq:E7LMN-tuning}, as shown in Appendix \ref{sec:rel-fug-2}.
Imposing them on the $\E_7$ characters in \eqref{eq:E7cha},  we then obtain
\begin{align}\label{eq:E7-Z4-tuning}
    \chi_1^{\E_7}\to-3,\quad\chi_2^{\E_7}\to0,\quad\chi_3^{\E_7}\to5,\quad\chi_4^{\E_7}\to-10,\quad\chi_5^{\E_7}\to0,\quad\chi_6^{\E_7}\to3,\quad\chi_7^{\E_7}\to0\ .
\end{align}
Plug \eqref{eq:E7-Z4-tuning} into \eqref{eq:E7}, and the resulting curve agrees with the Seiberg–Witten curve for the pure $\SU(2)_0$ with the $\Z_4$ S-fold given in \eqref{eq:E7z4curve}.

\subsubsection{\texorpdfstring{ $\SU(2)+4\F$ with $\Z_4$ S-fold}{ SU(2)+4F with Z4 S-fold}}

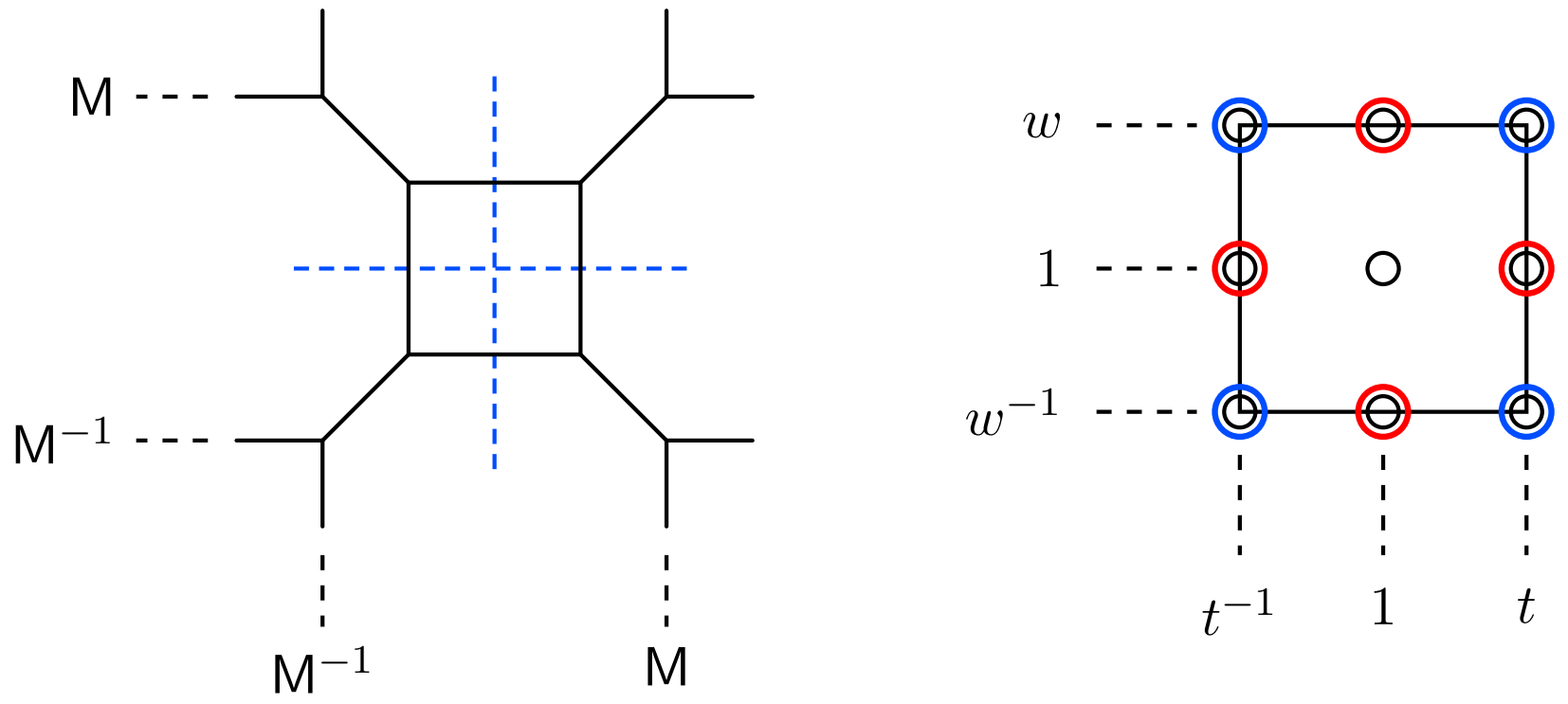
\begin{figure}[htbp]
    \centering
    \begin{tikzpicture}[scale=1]
        \begin{scope}[shift={(-3.2, -0.6)}]
            \draw[dualweb] (-0.8, 0.6) -- (2.0, 0.6);
            \draw[dualweb] (0.6, -0.8) -- (0.6, 2.0);

            \draw[web] (0,0) rectangle (1.2,1.2);

            \draw[web] (0,1.2) -- (-0.6,1.8);
            \draw[web] (-0.6,1.8) -- (-0.6,2.4);
            \draw[web] (-0.6,1.8) -- (-1.2,1.8);
            \draw[labeldash] (-1.4,1.8) -- (-1.9,1.8) node[left] {$\M$};

            \draw[web] (1.2,1.2) -- (1.8,1.8);
            \draw[web] (1.8,1.8) -- (1.8,2.4);
            \draw[web] (1.8,1.8) -- (2.4,1.8);

            \draw[web] (0,0) -- (-0.6,-0.6);
            \draw[web] (-0.6,-0.6) -- (-1.2,-0.6);
            \draw[web] (-0.6,-0.6) -- (-0.6,-1.2);
            \draw[labeldash] (-1.4,-0.6) -- (-1.9,-0.6) node[left] {$\M^{-1}$};
            \draw[labeldash] (-0.6,-1.4) -- (-0.6,-1.9) node[below] {$\M^{-1}$};

            \draw[web] (1.2,0) -- (1.8,-0.6);
            \draw[web] (1.8,-0.6) -- (2.4,-0.6);
            \draw[web] (1.8,-0.6) -- (1.8,-1.2);
            \draw[labeldash] (1.8,-1.4) -- (1.8,-1.9) node[below] {$\M$};
        \end{scope}

        \begin{scope}[shift={(3.6, 0)}]
            \draw[poly] (-1,-1) -- (1,-1) -- (1,1) -- (-1,1) -- cycle;

            \node[pt] at (0,0) {};

            \node[pt] at (-1,0) {}; \node[redpt] at (-1,0) {};
            \node[pt] at (0,-1) {}; \node[redpt] at (0,-1) {};
            \node[pt] at (1,0)  {}; \node[redpt] at (1,0)  {};
            \node[pt] at (0,1)  {}; \node[redpt] at (0,1)  {};

            \node[pt] at (-1,-1) {}; \node[bluept] at (-1,-1) {};
            \node[pt] at (1,-1)  {}; \node[bluept] at (1,-1)  {};
            \node[pt] at (1,1)   {}; \node[bluept] at (1,1)   {};
            \node[pt] at (-1,1)  {}; \node[bluept] at (-1,1)  {};

            \draw[labeldash] (-2, 1) -- (-1.3, 1);   \node[left] at (-2.1, 1) {$w$};
            \draw[labeldash] (-2, 0) -- (-1.3, 0);   \node[left] at (-2.1, 0) {$1$};
            \draw[labeldash] (-2,-1) -- (-1.3,-1);   \node[left] at (-2.1,-1) {$w^{-1}$};

            \draw[labeldash] (-1,-1.3) -- (-1,-2);   \node[below] at (-1,-2.1) {$t^{-1}$};
            \draw[labeldash] ( 0,-1.3) -- ( 0,-2);   \node[below] at ( 0,-2.1) {$1$};
            \draw[labeldash] ( 1,-1.3) -- ( 1,-2);   \node[below] at ( 1,-2.1) {$t$};
        \end{scope}
    \end{tikzpicture}
    \caption{5-brane web for $\SU(2)+4\F$ (left) and its dual graph (right). The $\mathbb Z_4$ invariance of the $\SU(2)+4\bf{F}$ theory requires that it should solely have one independent mass parameter, denoted by $\M$. Dots of the same color have equal coefficients. The red and blue circles correspond to coefficients $-\chi_1^{\SU(2)}$ and 1, respectively; the dot in the compact face corresponds to the Coulomb modulus $U$.}
    \label{fig:square-pair}
\end{figure}

The 5-brane web and its dual graph for the $\Z_4$ S-fold of the $\SU(2)+4\F$ theory are given in figure \ref{fig:square-pair}. From the dual diagram, 
the $\Z_4$-invariant Seiberg–Witten curve for SU(2)+4$\F$  under the transformation \eqref{eq:Z4trans} is given by
\begin{align}\label{eq:su24fz4or}
    F(t,w)= U-\chi_1^{\SU(2)}(t+t^{-1}+w+w^{-1})+tw+tw^{-1}+t^{-1}w+t^{-1}w^{-1}\ .
\end{align}
where $\chi_1^{\SU(2)}$ is the character for the fundamental representation of $\SU(2)$, 
\begin{align}
    \chi_1^{\SU(2)}=\M+\frac1\M\ . 
\end{align}
By replacing variables $t$ and $w$ by $a_i, (i=0,\dots,3)$, we get to
\begin{align}\label{eq:su2-4f-ai}
    f_0( \{ a_i \}):= U-\chi_1^{\SU(2)}\,(a_0 + a_1 + a_2 + a_3)+a_0 a_1 + a_1 a_2 + a_2 a_3 + a_3 a_0\ .
\end{align}

Using \eqref{eq:z4ais} and \eqref{eq:sol4z4}, we can express \eqref{eq:su2-4f-ai} as 
\begin{align}\label{eq:su24swX}
   h_{0}(\{ T_k \}):=4U-4\chi_1^{\SU(2)}T_1+T_1^2-T_4 =0\ .
\end{align}
Solving \eqref{eq:swsu2z4c} and \eqref{eq:su24swX} for $T_2$ and $T_4$ and then plugging the solution into \eqref{eq:su2constraint},  the Seiberg–Witten curve for SU(2)+4$\F$  with the $\Z_4$ S-fold reads
\begin{align}
H(T_1,T_3)&:= \left(-4 \chi_1^{\SU(2)}T_1+T_1^2+4U\right)
\left(-2 \chi_1^{\SU(2)} T_1+T_1^2+2U-8\right)^2 \notag\\
&\quad 
-2T_1T_3\left(-4 \chi_1^{\SU(2)}T_1+T_1^2+4U\right)
+T_3^2=0\ .
\end{align}
The corresponding Weierstrass normal form takes the form:
\begin{align}\label{eq:E8z4curve}
    g_2^{4\F/\Z_4}&=\frac{u^4}{12}
+\frac{56}{3}\left(\chi_1^{\SU(2)}\right)^2 u^2
-\frac{4864}{3}u^2
+\left[
-2048\left(\chi_1^{\SU(2)}\right)^2
+122880
\right]u\notag\\
&\quad
+\frac{64}{3}\left(\chi_1^{\SU(2)}\right)^4+\frac{167936}{3}\left(\chi_1^{\SU(2)}\right)^2
-\frac{7847936}{3}\ , \notag\\
g_3^{4\F/\Z_4}&=\frac{u^6}{216}
-4u^5
+\left[
\frac{7856}{9}
-\frac{22}{9}\left(\chi_1^{\SU(2)}\right)^2
\right]u^4
+\left[
\frac{1408}{3}\left(\chi_1^{\SU(2)}\right)^2
-86016
\right]u^3\notag\\
&\quad
+\left[
-\frac{352}{9}\left(\chi_1^{\SU(2)}\right)^4
-\frac{292352}{9}\left(\chi_1^{\SU(2)}\right)^2
+\frac{39305216}{9}
\right]u^2\notag\\
&\quad
+\left[
\frac{13312}{3}\left(\chi_1^{\SU(2)}\right)^4
+\frac{2818048}{3}\left(\chi_1^{\SU(2)}\right)^2
-111673344
\right]u\notag\\
&\quad+\frac{512}{27}\left(\chi_1^{\SU(2)}\right)^6
-\frac{1134592}{9}\left(\chi_1^{\SU(2)}\right)^4
-\frac{82051072}{9}\left(\chi_1^{\SU(2)}\right)^2\notag\\
&\quad
+\frac{30767316992}{27}\ ,
\end{align}
where $u=60-U$. The corresponding Kodaira fiber configuration is $(I_1;I_4^2I_2I_1)$.

\subsubsection*{Comparison with $\E_8$ SCFT}

Again, start from the Seiberg–Witten curve \eqref{eq:E8} for the SU(2)+7$\mathbf F$ theory, which is given in terms of the $E_8$ characters. We apply the $E_7$-tuning:
\begin{align}\label{eq:tuning-e8-z4}
    M_1=M_2=-M_3=-M_4=1,\quad
    M_5=-M_6=i,\quad M_7=\!-i\M,\quad M_0=e^{\pi i/4}\M^{1/2},
\end{align}
which is consistent with \eqref{eq:E7-tuning-M} after decoupling the flavor $\M$. The $E_8$ characters, whose explicit forms are given in \eqref{eq:chie8}, reduce to \begin{align}\label{eq:cha-e8-tuning}
    &\chi_1^{\E_8}\to-3\left(\chi_1^{\SU(2)}\right)^2+7,\notag\\
    &\chi_2^{\E_8}\to5\left(\chi_1^{\SU(2)}\right)^2-10,\notag\\
    & \chi_3^{\E_8}\to5\left(\chi_1^{\SU(2)}\right)^4-32\left(\chi_1^{\SU(2)}\right)^2+48,\nonumber\\
    &\chi_4^{\E_8}\to-10\left(\chi_1^{\SU(2)}\right)^6+87\left(\chi_1^{\SU(2)}\right)^4-244\left(\chi_1^{\SU(2)}\right)^2+224,\notag\\
    & \chi_5^{\E_8}\to-15\left(\chi_1^{\SU(2)}\right)^4+65\left(\chi_1^{\SU(2)}\right)^2-70,\nonumber\\
    &\chi_6^{\E_8}\to3\left(\chi_1^{\SU(2)}\right)^4-16\left(\chi_1^{\SU(2)}\right)^2+16,\notag\\
    &\chi_7^{\E_8}\to\left(\chi_1^{\SU(2)}\right)^2,
    \notag\\
    &\chi_8^{\E_8}\to\left(\chi_1^{\SU(2)}\right)^2-4 \,.
\end{align}
After substituting \eqref{eq:cha-e8-tuning} into \eqref{eq:E8}, we find that the resulting curve agrees with the Seiberg–Witten curve for SU(2)+4$\F$ with the $\Z_4$ S-fold given in \eqref{eq:E8z4curve}.

\subsection{\texorpdfstring{Seiberg–Witten curves for $\Z_6$ S-fold}{Z6 S-fold}}\label{sec:Z6Sfold}

For the $\Z_6$ S-fold, $\Z_6$ transformation is given by
\begin{align}\label{eq:61trans}
    t\to w\to \frac{w}{t}\to \frac1t \to\frac1w\to \frac{t}{w}\to t\ .
\end{align}
We define 
\begin{align}\label{eq:Z6-defa}
a_0=t, \quad
a_1=w, \quad
a_2=t^{-1} w, \quad
a_3=t^{-1}, \quad
a_4=w^{-1}, \quad
a_5=w^{-1}t, 
\end{align}
so that the $\Z_6$ transformation is rewritten as
\begin{align}\label{eq:z6trans}
    a_0\to a_1\to a_2\to a_3\to a_4\to a_5\to a_0\ .
\end{align}

The embedding \eqref{eq:Z6-defa} of $\mathbb{(C^*)}^2$ parametrized by the coordinates $t,w$ into $\mathbb{(C^*)}^6$ parametrized by the coordinates $a_0,a_1,\dots, a_5$ is equivalently given, for example, by the following four constraints
\begin{align}\label{eq:consZ6}
a_0 a_3-1 = 0, \qquad
a_1 a_4-1 = 0, \qquad
a_2 a_5-1 = 0, \qquad
a_0 a_2-a_1 = 0.
\end{align}
Indeed, if we set $a_0=t$, $a_1=w$, and then solve \eqref{eq:consZ6} for the other coordinates $a_i$ ($i=2,3,4,5$), we reproduce \eqref{eq:Z6-defa}. 

However, in order to have the expression for the constraints to show the $\Z_6$ transformation \eqref{eq:z6trans} manifestly, it is convenient to introduce redundant, a total of nine constraints, including \eqref{eq:consZ6} as 
\begin{align}\label{eq:Z6-redandant-const}
& a_{n} a_{n+3} - 1 = 0, \quad (n=0,1,2), \qquad
& a_{n} a_{n +2} - a_{n+1} = 0,
\,\,\, (n=0,1,2,3,4,5) ,
\end{align}
where we denote $a_6:=a_0, \, a_7:=a_1$. We find that each constraint is transformed to each other so that the set of constraints \eqref{eq:Z6-redandant-const} respects the $\Z_6$ structure.
We can explicitly check that all these constraints are satisfied if we substitute \eqref{eq:Z6-defa}, and thus, give an equivalent embedding as \eqref{eq:consZ6}.

Since we later would like to rewrite these constraints using $\Z_6$ invariant coordinates, it would be useful to further rewrite the set of these constraints \eqref{eq:Z6-redandant-const} into a form where each equation is invariant under the $\Z_6$ transformation \eqref{eq:z6trans} by itself as
\begin{align}\label{eq:consts-Z6}
f_{1} (\{ a_i \}) 
&:=\sum_{n=0}^2 (a_{n} a_{n+3} - 1 ) = 0,
\cr
f_{2} (\{ a_i \}) 
&:=
\sum_{n=0}^2 \omega_6^{2n} (a_{n} a_{n+3} - 1 ) = 0,
\cr
f_{3} (\{ a_i \}) 
&:=
\sum_{n=0}^2 \omega_6^{-2n} (a_{n} a_{n+3} - 1 ) = 0,
\cr
f_{4} (\{ a_i \}) 
&:=
\sum_{n=0}^5 (a_{n} a_{n+2} - a_{n+1}) = 0,
\cr
f_{5} (\{ a_i \}) 
&:=
\sum_{n=0}^5 (-1)^n (a_{n} a_{n+2} - a_{n+1}) = 0.
\end{align}
where $\omega_6=e^{2\pi i/6}$.
If we impose only the first four constraints, we obtain one more solution, in addition to \eqref{eq:Z6-defa}. The last constraint has been added to obtain the unique solution \eqref{eq:Z6-defa}. Thus, the set of equations \eqref{eq:consts-Z6} give constraints equivalent to \eqref{eq:Z6-defa}.

Define
\begin{align}\label{eq:xk06}
    x_k=\sum_{i=0}^5\omega_6^{ki}a_i, \qquad
    (k=0,1,\dots,5)
\end{align}
so that $\Z_6$ acts on these coordinates as
\begin{align}\label{eq:z6-x}
    x_0\to x_0,\quad x_1\to\omega_6^5x_1,\quad x_2\to\omega_6^4 x_2,\quad x_3\to \omega_6^3x_3,\quad x_4\to \omega_6^2x_4,\quad x_5\to\omega_6x_5\ .
\end{align}
Inverting \eqref{eq:xk06}, we obtain
\begin{align}\label{eq:z6ais}
    a_{i}=\frac{1}{6}\sum_{k=0}^5\omega_6^{-ki}x_k,
    \qquad
    (i=0,1,\dots,5).
\end{align}
Plugging \eqref{eq:z6ais} into \eqref{eq:consts-Z6}, we find
\begin{align}\label{eq:const-Z6-x}
\tilde{f}_{1} (\{ x_k \}) 
&:= x_0^2 - 2 x_1 x_5 + 2 x_2 x_4 - x_3^2 -36 = 0\ ,
\cr
\tilde{f}_{2} (\{ x_k \}) 
&:=x_4^2-x_1^2+2 x_0 x_2-2 x_3 x_5=0 \ ,
\cr
\tilde{f}_{3} (\{ x_k \}) 
&:=x_2^2-x_5^2-2 x_1 x_3+2 x_0 x_4=0\ ,
\cr
\tilde{f}_{4} (\{ x_k \}) 
&:=x_0^2-6 x_0+x_3^2-x_2 x_4-x_1 x_5=0\ ,
\cr
\tilde{f}_{5} (\{ x_k \}) 
&:=2 x_0 x_3-x_1 x_2+6 x_3-x_4 x_5=0\ .
\end{align}
Then construct six functions $T_i$, $i=1,2,\dots,6$, 
\begin{align}\label{eq:6functions}
    T_1=x_0 \ ,\quad T_2=x_1x_5 \ ,\quad T_3=x_2x_4 \ ,\quad T_4=x_3^2 \ ,\quad  T_5=x_1x_2x_3\ ,\quad T_6=x_2^3\ , 
\end{align}
which are expressed by $x_k,~(k=0,1,\dots,5)$.
 The system \eqref{eq:6functions} is manifestly invariant under the transformation \eqref{eq:z6-x} and admits six solutions. Without loss of generality, one of these solutions is chosen  as follows:
\begin{align}\label{eq:soluz6}
    x_0=T_1 ,\quad x_1=\frac{T_5}{\sqrt{T_4}T_6^{1/3}},\quad x_2=T_6^{1/3},\quad x_3=\sqrt{T_4} ,\quad x_4=\frac{T_3}{T_6^{1/3}} ,\quad x_5=\frac{T_2\sqrt{T_4}T_6^{1/3}}{T_5} .
\end{align}
Substituting \eqref{eq:soluz6} into \eqref{eq:const-Z6-x}, we find:
\begin{align}\label{eq:h5consts}
h_{1} (\{ T_k \}) 
&:= T_1^2 - 2 T_2 + 2 T_3 - T_4 -36 = 0\ ,
\cr
h_{2} (\{ T_k \}) 
&:=T_3^2 T_4 T_5-T_5^3+2 T_1 T_4 T_6 T_5-2 T_2 T_4^2 T_6=0\ ,
\cr
h_{3} (\{ T_k \}) 
&:=2 T_1 T_3 T_5^2-2 T_5^3+T_6 T_5^2-T_2^2 T_4 T_6=0\ , 
\cr
h_{4} (\{ T_k \}) 
&:=T_1^2-6 T_1-T_2-T_3+T_4=0\ ,
\cr
h_{5} (\{ T_k \}) 
&:=2 T_1 T_4 T_5-T_5^2+6 T_4 T_5-T_2 T_3 T_4=0\ .
\end{align}

\subsubsection{\texorpdfstring{ $\SU(2)+2\F$ with $\Z_6$ S-fold}{ SU(2)+2F with Z6 S-fold}}

\begin{figure}[htbp]
    \centering
    \begin{tikzpicture}
        \begin{scope}[shift={(-2.8, 0)}, scale=0.9]
            \draw[dualweb] (-1.7, 1.7) -- (1.7, -1.7);
            \draw[dualweb] (-1.0, -2.0) -- (1.0, 2.0);
            \draw[dualweb] (-2.0, -1.0) -- (2.0, 1.0);

            \draw[web] (-1,-1) -- (0,-1) -- (1,0) -- (1,1) -- (0,1) -- (-1,0) -- cycle;

            \draw[web] (-1,-1) -- (-1.7,-1.7);
            \draw[web] (0,-1)  -- (0,-2.0);
            \draw[web] (1,0)   -- (2.0,0);
            \draw[web] (1,1)   -- (1.7,1.7);
            \draw[web] (0,1)   -- (0,2.0);
            \draw[web] (-1,0)  -- (-2.0,0);
        \end{scope}

        \begin{scope}[shift={(3.6, 0)}, scale=1.0]
            \draw[poly] (-1,1) -- (0,1) -- (1,0) -- (1,-1) -- (0,-1) -- (-1,0) -- cycle;

            \node[pt] at (0,0) {};

            \node[pt] at (-1,1) {}; \node[redpt] at (-1,1) {};
            \node[pt] at (0,1)  {}; \node[redpt] at (0,1)  {};
            \node[pt] at (1,0)  {}; \node[redpt] at (1,0)  {};
            \node[pt] at (1,-1) {}; \node[redpt] at (1,-1) {};
            \node[pt] at (0,-1) {}; \node[redpt] at (0,-1) {};
            \node[pt] at (-1,0) {}; \node[redpt] at (-1,0) {};

            \draw[labeldash] (-2, 1) -- (-1.3, 1);   \node[left] at (-2.1, 1) {$w$};
            \draw[labeldash] (-2, 0) -- (-1.3, 0);   \node[left] at (-2.1, 0) {$1$};
            \draw[labeldash] (-2,-1) -- (-1.3,-1);   \node[left] at (-2.1,-1) {$w^{-1}$};

            \draw[labeldash] (-1,-1.3) -- (-1,-2);   \node[below] at (-1,-2.1) {$t^{-1}$};
            \draw[labeldash] ( 0,-1.3) -- ( 0,-2);   \node[below] at ( 0,-2.1) {$1$};
            \draw[labeldash] ( 1,-1.3) -- ( 1,-2);   \node[below] at ( 1,-2.1) {$t$};
        \end{scope}
    \end{tikzpicture}
    \caption{5-brane web for SU(2)+2$\mathbf F$ (left) and its dual graph (right). The $\mathbb Z_6$ invariance of the theory requires that both flavors be massless. Dots marked by the red circle indicate that the corresponding coefficients are identical. They are equal to 1. The dot in the compact face corresponds to the Coulomb modulus $U$ of the theory. }
    \label{fig:dp3-hex-pair}
\end{figure}
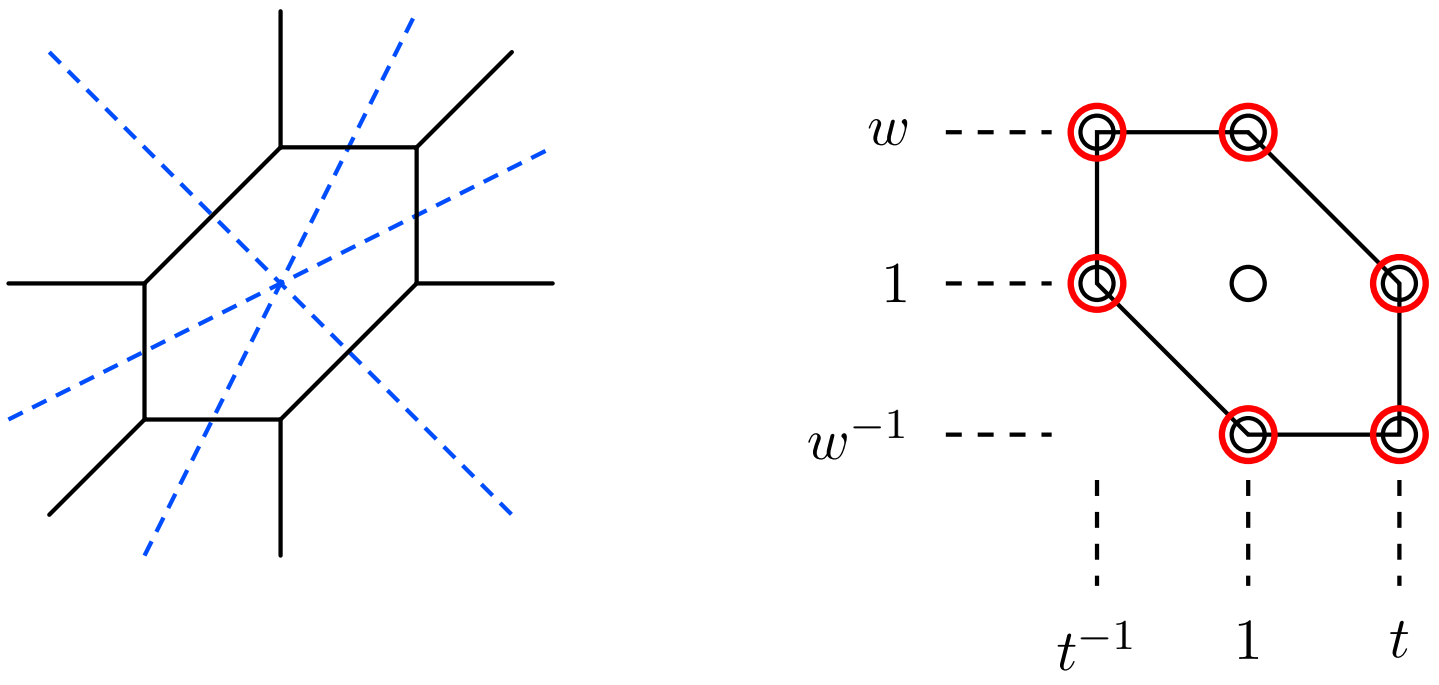

The 5-brane web and its dual graph for the $\Z_6$ S-fold of the $\SU(2)+2\F$ theory are given in figure \ref{fig:dp3-hex-pair}. 
 
From the dual diagram, 
the $\Z_6$-invariance of SU(2)+2$\mathbf F$ theory under the transformation \eqref{eq:61trans} forces the Seiberg–Witten curve to be
\begin{align}\label{eq:su22f-z6-curve}
    F(t,w)=U+\bigg(t+\frac1t+w+\frac1w+\frac{w}{t}+\frac{t}{w}\bigg)\ .
\end{align}
Since $x_0=\sum_{i=0}^5a_i$, it follows from \eqref{eq:soluz6} that the original Seiberg–Witten curve
takes the form
\begin{align}\label{eq:sw26fz6}
    h_0(\{T_k\}) :=U+T_1=0\ .
\end{align}

We solve the system consisting of  $h_1,h_3,h_4$ and $h_0$ for $T_1,T_3,T_4$ and $T_6$. Then we substitute the resulting solution into $h_2$ and $h_5$. We find that the resulting two equations are satisfied simultaneously if the following equation is satisfied
\begin{align}
H(T_2,T_5)&:= 6 T_2 U^4+42 T_2 U^3-6 T_5 U^3-17 T_2^2 U^2-108 T_2 U^2-6 T_5 U^2-60 T_2^2 U\notag\\    
&\quad -648 T_2 U +8 T_2 T_5 U+144 T_5 U+12 T_2^3+252 T_2^2+T_5^2\notag\\    
&\quad 
+1296 T_2-24 T_2 T_5-216 T_5 = 0 \,, 
\end{align}
which is the Seiberg--Witten curve for the $\Z_6$ S-fold.
The corresponding Weierstrass normal form is given by
\begin{align}\label{eq:E8z6curve}
     g_2^{2\F/\Z_6}&=\frac{u^4}{12}- 1602 u^2 + 120816 u -2561220 \ ,
\\
    g_3^{2\F/\Z_6}&=\frac{u^6}{216}-4 u^5+\frac{1741 u^4}{2}-85576 u^3+4338522 u^2-110917728 u+1133518104,\nn
\end{align}
where $u=60-U$. The corresponding Kodaira fiber configuration is $(I_1;I_6I_3I_2)$.

\subsubsection*{Comparison with $\E_8$ SCFT}

Here, we would like to confirm that the Seiberg–Witten curve given in \eqref{eq:E8z6curve} can be reproduced from the Seiberg–Witten curve for the rank-1 $\E_8$ SCFT, which is known to be given by \cite{Kim:2014nqa}, \small
\begin{align}\label{eq:E8SW}
& w^6 
- \left( \chi^{A_5}_1(\tilde{M}) t - 2 \chi^{A_2}_1(\tilde{L}) \right) w^5
 + \left[ \chi_2^{A_5} (\tilde{M}) t^2 
 - \left( \chi_1^{A_5} (\tilde{M}) \chi_1^{A_2} (\tilde{L}) 
 + \chi_5^{A_5} (\tilde{M}) + \chi_1^{A_1} (\tilde{N}) \chi_2^{A_2} (\tilde{L}) \right)t \right.
\nonumber\\
&\left.
 +\left(\chi_1^{A_2}(\tilde{L})\right)^2+2\chi_2^{A_2}(\tilde L) \right] w^4-\left[\chi_3^{A_5}(\tilde{M})t^3-Ut^2+\left(\chi_1^{A_5}(\tilde {M})\chi_2^{A_2}(\tilde{L})+\chi_5^{A_5}(\tilde{M})\chi_1^{A_2}(\tilde{L})\right.\right.\nonumber\\
 &\left.\left.+3\chi_1^{A_1}(\tilde{N})+\chi_1^{A_1}(\tilde{N})\chi_1^{A_2}(\tilde{L})\chi_2^{A_2}(\tilde{L})\right)t-2\left(\chi_1^{A_2}(\tilde{L})\chi_2^{A_2}(\tilde{L})+1\right)\right]w^3+\left(t^2-\chi_1^{A_1}(\tilde{N})t+1\right)\nonumber\\
 &\times\left[\chi_4^{A_5}(\tilde{M})t^2-\left(\chi_1^{A_5}(\tilde{M})+\chi_5^{A_5}(\tilde{M})\chi_2^{A_2}(\tilde{L})+\chi_1^{A_1}(\tilde N)\chi_1^{A_2}(\tilde{L})\right)t+\left(\chi_2^{A_2}(\tilde{L})\right)^2+2\chi_1^{A_2}(\tilde{L})\right]w^2\nonumber\\
 &-\left(t^2-\chi_1^{A_1}(\tilde{N})t+1\right)^2\left(\chi_5^{A_5}(\tilde{M})t-2\chi_2^{A_2}(\tilde{L})\right)w+\left(t^2-\chi_1^{A_1}(\tilde{N})t+1\right)^3= 0,
\end{align}
\normalsize
where we denote the characters
\begin{align}
\chi_n^{A_5}(\tilde{M}) &= \sum_{1 \le i_1 < i_2 < \cdots < i_n \le 6} \tilde{M}_{i_1} \tilde{M}_{i_2} \dots \tilde{M}_{i_n}, \qquad (n=1,2,3,4,5)
\cr
\chi_n^{A_2}(\tilde{L}) &= \sum_{1 \le i_1 < i_2 < \cdots < i_n \le 3} \tilde{L}_{i_1} \tilde{L}_{i_2} \dots \tilde{L}_{i_n}, \qquad (n = 1,2)
\cr
\chi_1^{A_1}(\tilde{N}) &= \tilde{N}_1+\tilde{N}_2,
\end{align}
with the traceless condition of the SU groups
\begin{align}\label{eq:trace-less}
\tilde{M}_1 \tilde{M}_2 \dots  \tilde{M}_6 =\tilde{L}_1 \tilde{L}_2 \tilde{L}_3 =  \tilde{N}_1 \tilde{N}_2  = 1.
\end{align}
Here, $\tilde{M}_i$, $\tilde{L}_i$, $\tilde{N}_i$ are the fugacity of $\SU(6) \times \SU(3) \times \SU(2)$ subgroup of $E_8$ and their absolute values correspond to the position of six $[1,0]$ 7-branes, three $[1,1]$ 7-branes, and two $[0,1]$ 7-branes, respectively. 

Again, a natural guess for the tuning would be given by imposing that the M5-brane originating from these $[p,q]$ 7-branes is located at an equal distance in the compactified circle. That is, for example, 
\begin{align}
\tilde{M}_{k+1} = \omega_6 \tilde{M}_{k},
\qquad
\tilde{L}_{k+1} = \omega_6^2  \tilde{L}_{k},
\qquad
\tilde{N}_{k+1} = \omega_6^3  \tilde{N}_{k}.
\end{align}
The tuning that satisfies these conditions as well as the traceless condition \eqref{eq:trace-less} gives 
\begin{align}\label{eq:mass-tune-Z6}
&\tilde{M}_k = e^{(2k-7) \pi i /6}, \quad \,\, (k=1,2,\dots,6)
\cr
&\tilde{L}_k = e^{(2k-4) \pi i /3}, \quad \,\,\, (k=1,2,3)
\cr
&\tilde{N}_k = e^{(2k-3) \pi i /2},  \quad \, (k=1,2)
\end{align}
up to Weyl group, which we call \textit{the $E_8$-tuning}. With this tuning, we find 
\begin{align}
\chi_i^{A_5}(\tilde{M})
= \chi_j^{A_2}(\tilde{L})
= \chi_1^{A_1}(\tilde{N})
= 0 \qquad (i=1,2,\dots, 5, \quad j =1,2).
\end{align}
Substituting the $E_8$-tuning into \eqref{eq:E8SW} and rewriting it in Weierstrass normal form, we reproduce \eqref{eq:E8z6curve}. 

This could also be checked from different perspective. As discussed in Appendix~\ref{sec:rel-fug-3}, the $E_8$-tuning \eqref{eq:mass-tune-Z6} is equivalent to the following tuning of the flavor masses $M_1,\dots,M_7$ and the instanton mass $M_0$ as
\begin{align}\label{eq:E8-tuning-M}
     M_1=M_2=M_3=-M_4=-M_5=1,\quad M_6=M_7=\omega_6^2,\quad M_0=\omega_6\ .
\end{align}
Imposing \eqref{eq:E8-tuning-M}
on the $E_8$ characters shown in \eqref{eq:chie8} yields 
\begin{align}
    \chi_1^{\E_8} &\to 3,  & \chi_2^{\E_8} &\to -8, &
    \chi_3^{\E_8} &\to 15, & \chi_4^{\E_8} &\to 19, \nonumber\\
    \chi_5^{\E_8} &\to-10, & \chi_6^{\E_8} &\to -1, &
    \chi_7^{\E_8} &\to5, & \chi_8^{\E_8} &\to -3.
    \label{eq:e8z6-cha-tun}
\end{align}
Inserting \eqref{eq:e8z6-cha-tun} into \eqref{eq:E8}, we find that the resulting curve agrees with the Seiberg–Witten curve for the SU(2)+2$\F$ theory with the $\Z_6$ S-fold given in \eqref{eq:E8z6curve}.

\bigskip
\section{\texorpdfstring{$\Z_n$}{Zn} S-folds and \texorpdfstring{$n$}{n}-isogenies}\label{sec:346-isogeny}

We have constructed the quotient Seiberg--Witten curves of $\Z_n$ S-folds, with $n=3,4,6$, by implementing the corresponding discrete symmetry actions on the coordinates of the 5-brane web. 
In this section, we would like to present an observation that the Seiberg--Witten curves for 5d $\Z_n$ S-folds can be obtained from cyclic $n$-isogenies, with $n=2,3,4,6$. In the following, we write down our observation more concretely after a brief review of Mordell--Weil groups and cyclic $n$-isogenies. (See \cite{Closset:2021lhd,Furrer:2024zzu} for a more detailed introduction to isogenies.)

In 5d $\mathcal N=1$ rank-1 SCFTs, the Seiberg--Witten curve $E:y^2=4x^3-g_2x-g_3$ fibered over the compactified complex one-dimensional Coulomb branch $\mathbb P^1=\mathbb{C}\cup\{\infty\}$
defines a rational elliptic surface 
\begin{align}
    E\to\mathcal E\xrightarrow{\pi}\mathbb P^1\ ,
\end{align}
over the field $\mathbb C$.
The rational sections of $\pi$ form the Mordell--Weil group $\mathrm{MW}(\mathcal E)$, where the group law is induced from the elliptic curve group law (see Appendix~\ref{app:SW-geometry}) and the identity element
is the zero section $O$, corresponding to the point at infinity.

Treating the Coulomb modulus $u$ as a variable, the Weierstrass equation gives the generic fiber $E/K$, where $K=\mathbb C(u)$ is the field of rational functions. The set of $K$-rational points is defined as 
\begin{align}
    E(K):=\{(x,y)\in K^2\,|\, y^2=4x^3-g_2(u)x-g_3(u)\}\cup\{O\}.
\end{align}
The set $E(K)$ equipped with the group law is identified with $\mathrm{MW}(\mathcal E)$.

We are particularly interested in cyclic subgroups of order $n$ of $\mathrm{MW}(\mathcal E)$. Such a subgroup is generated by a point $P \in E(K)$ of exact order $n$ and can be written as 
\begin{align}
G=\langle P\rangle=\{O,P,...,(n-1)P\},
\end{align}
where $nP=O$. 
The quotient map $\varphi_n:E\to\widetilde E=E/G$ is a cyclic $n$-isogeny, whose kernel is given by $G$. We have the following observation:

\vspace{0.5cm}

\begin{tcolorbox}[colback=white, colframe=black, boxrule=0.5pt, arc=4pt]
\textbf{Observation:} For each pair $(\mathcal{T}, n)$ of rank 1 parent theory $\mathcal{T}$ and $n \in \{2, 3,4,6\}$ considered in this work and in \cite{Kim:2026yeh}, we find:
\begin{itemize}
\item The Mordell--Weil group $\mathrm{MW}(\mathcal E)$ contains a unique cyclic subgroup $G$ of order~$n$. 
\item The quotient elliptic curve $E/G$ obtained by the cyclic $n$-isogeny $\varphi_n$ is isomorphic to the Seiberg--Witten curve for the $\mathbb Z_n$ S-fold theory $\mathcal T/\mathbb Z_n$.
\end{itemize}
\end{tcolorbox}

More related mathematical background, including Mordell--Weil groups and the quotient construction of isogenies, together with our notation, is collected in Appendix~\ref{app:SW-geometry}.

\subsection{Division polynomials and V\'elu's formula }\label{sec:div-and-velu}
To verify the observation stated above, we first review division polynomials~\cite{Silverman2009arithmetic} and V\'elu's formulas~\cite{Closset:2023pmc,Velu1971}. The division polynomial is used to find the $n$-torsion points explicitly in terms of the coordinates $x,y$ used in the Weierstrass form. For our purpose, we choose only the points that are $K$-rational and find that all such points belong to an identical cyclic subgroup $G$ for all the examples we study. V\'elu's formulas are used to construct the quotient curve $E/G$ in the Weierstrass form by using these torsion points.
Then, we can compare the quotient curve $E/G$
obtained by the $n$-isogenies with that of the corresponding S-fold curve obtained from the 5-brane web quotient.

\subsubsection{Division polynomials}\label{sec:div-poly}

In order to find the $n$-torsion points $Q$ explicitly using the coordinates used in the Weierstrass form $E:y^2=4x^3-g_2x-g_3$,
we use the following division polynomials $\psi_n$ defined by
\begin{align}\label{eq:division-equations}
    \psi_0&=0,\quad
    \psi_1=1,\quad
    \psi_2=y,\quad
    \psi_3=
    3x^4-\frac{3}{2}g_2x^2-3g_3x-\frac{1}{16}g_2^2,
    \notag\\
    \psi_4&=
    \frac{y}{32}
    \left(
        64x^6-80g_2x^4-320g_3x^3
        -20g_2^2x^2-16g_2g_3x
        +g_2^3-32g_3^2
    \right),
\end{align}
together with the recurrence relations
\begin{align}\label{eq:psi-m}
    \psi_{2m+1}
    &=\psi_{m+2}\psi_m^3-\psi_{m-1}\psi_{m+1}^3,
    & m\geq 2,\notag\\
    \psi_{2m}
    &=\frac{\psi_m}{\psi_2}
    \left(
        \psi_{m+2}\psi_{m-1}^2
        -\psi_{m-2}\psi_{m+1}^2
    \right),
    & m\geq 3.
\end{align}
These expressions follow from the standard division polynomials after adapting them to our Weierstrass normalization \cite{Billingsley:2021specialization}. Here, for $n\geq2$, the polynomial $\psi_n^2$, expressed solely in $x$ using the curve equation $E$, has roots precisely at the $x$-coordinates of the nonzero $n$-torsion points~\cite{Silverman2009arithmetic}. The $y$ coordinate of the corresponding point can be obtained by using the elliptic curve in the Weierstrass normal form.

For our purpose, we choose only $K$-rational roots out of all the roots of the division polynomial and then check that the corresponding $y$ coordinates are also $K$-rational. After identifying one point as a generator $P$ of a cyclic group $G$, we check that all the remaining points are generated by $P$ by using the elliptic curve group law, showing the uniqueness of the cyclic subgroup of order $n$.

\subsubsection{V\'elu's formula}
For a given elliptic curve expressed in Weierstrass normal form $y^2=4x^3-g_2x-g_3$ with the cyclic subgroup $G=\{O,P,...,(n-1)P\}$ of the corresponding Mordell--Weil group, V\'elu's formulas provide an explicit construction of the quotient curve $\widetilde{E} = E/G$. 

For a torsion point $Q=(x_0,y_0)\in G\setminus\{O\}$, with $y_0=0$ in 2-torsion cases, we define 
\begin{align}\label{tPwP}
    \alpha(Q)=\begin{cases}
3x_0^2-\frac14g_2,\quad \mathrm{ord}(Q)=2,\\        6x^2_0-\frac12g_2,\quad \mathrm{ord}(Q)>2,
    \end{cases}\quad \beta(Q)=y_0^2+\alpha(Q)x_0\ .
\end{align}
In general,
\begin{align}
    G\setminus\{O\}=\mathcal G_2\sqcup G_+\sqcup G_-\ ,
\end{align}
where $G[2]:=\{Q\in G\,|\, 2Q=O\}=\mathcal G_2\sqcup\{O\}$ is the 2-torsion subgroup and $G_+=-G_-$. Then the new curve $\widetilde E:y^2=4x^3-\widetilde g_2x-\widetilde g_3$ can be determined by
\begin{align}\label{tildeg23}
    \widetilde g_2=g_2+20\alpha(G),\quad \widetilde g_3=g_3+28\beta(G)
\end{align}
where 
\begin{align}\label{eq:alpha-betaG}
    \alpha(G)=\sum_{Q\in \mathcal G_2\sqcup G_+}\alpha(Q),\quad \beta(G)=\sum_{Q\in\mathcal G_2\sqcup G_+}\beta(Q).
\end{align}

For our purpose, we should show
\begin{align}
    \widetilde g_2=g_2^{\mathcal T/\Z_n},\quad \widetilde g_3=g_3^{\mathcal T/\Z_n}\ ,
\end{align}
after fixing the normalization by rescaling $g_2 \to c^4 g_2$, $g_3 \to c^6 g_3$ induced by the coordinate change $x \to c^2 x$, $y \to c^3 y$.

\subsection{Isogeny realization of the \texorpdfstring{$\Z_n$}{Zn} S-fold curves}

In the following, we apply the methods discussed in section~\ref{sec:div-and-velu} to show that the boxed observation proposed above holds for all the $\Z_n$ S-folds ($n=2,3,4,6$) for rank-1 SCFTs.

\subsubsection{2-isogenies and \texorpdfstring{$\Z_2$}{Z2} S-folds}

Here, we construct explicit 2-isogenies from the parent Seiberg–Witten curves and show that the resulting curves reproduce those of the  $\Z_2$ S-folds obtained in \cite{Kim:2026yeh}.

Let $P=(x_0,y_0)\in\mathrm{MW}(\mathcal E)$ be a section of exact order 2, and consider the cyclic subgroup $G=\langle P\rangle=\{O,P\}$. Since $P=-P$, we have $y_0=0$, and the curve equation gives
\begin{align}\label{eq:fx-z2}
f(x_0)=0,\qquad
    f(x)=4x^3-g_2x-g_3.
\end{align}
 Consequently, the $K$-rational root $x_0$ of $f(x)$ corresponds to the nonzero $K$-rational 2-torsion points $(x_0,0)$. 

If $f(x)$ admits a factorization over $K$ of the form
\begin{align}
    f(x)=(ax+b)h(x)\ ,
\end{align}
where $a,b\in K,a\neq0$, the linear factor yields a $K$-rational root $x_0=-b/a$. The remaining two roots are $K$-rational if and only if the discriminant of the quadratic factor $h(x)$ is a square in $K$. In all cases considered below, $f(x)$ has exactly one $K$-rational root, which determines the unique nonzero $K$-rational 2-torsion point $P=(x_0,0)$. 

From \eqref{tPwP} and \eqref{tildeg23} it follows that
\begin{align}\label{eq:g2-z2-gen}
    g_2^{\mathcal T/\Z_2}=-4g_2+60x_0^2,\quad g_3^{\mathcal T/\Z_2}=84x_0^3-7g_2x_0+g_3\ ,
\end{align}
where $g_2,g_3$ are the coefficients of the curve of the parent theory $\mathcal T=\SU(2)+N_f \F$, with $N_f=0,2,4,6$.

\subsubsection*{2-isogeny of the pure \texorpdfstring{$\SU(2)$}{SU2} theory}
As a simple but representative example, let us explicitly compute 2-isogeny of the pure $\SU(2)$ theory.
Denote the $\Z_2$ symmetry of 5-brane web coordinates by
\begin{align}
    t\mapsto\frac1t,\quad w\mapsto\frac1w\ .
\end{align}
The Seiberg–Witten curve obtained from 5-brane web of the pure $\SU(2)$ theory under this action reads 
\begin{align}
    F_{\SU(2)_0}(t,w)=U+\sqrt{q}\Big(w+\frac1w\Big)+t+\frac1t=0,
\end{align}
whose Weierstrass normal form is given by
\begin{align}
    E: y^2=4x^3-g^{\SU(2)_0}_2x-g_3^{\SU(2)_0}\ ,
\end{align}
with
\begin{align}\label{eq:su20-z2}
    g^{\SU(2)_0}_2&=\frac{u^4}{12}-\frac{2}{3}\Big(\sqrt{q}+\frac1{\sqrt{q}}\Big)u^2+\frac{4 q}{3}+\frac{4}{3
   q}-\frac{4}{3}\ ,\notag\\
   g^{\SU(2)_0}_3&=\frac{u^6}{216}-\frac1{18}\Big(\sqrt{q}+\frac1{\sqrt{q}}\Big)u^4+\frac29\Big(q+\frac1q+\frac12\Big)u^2-\frac{8 q^{3/2}}{27}-\frac{8}{27 q^{3/2}}+\frac{4 \sqrt{q}}{9}+\frac{4}{9 \sqrt{q}}\ ,\notag\\
   U&=q^{1/4}u\ .
\end{align}
From \eqref{eq:su20-z2} we find that the $K$-rational root of $f(x)$ is given by
\begin{align}\label{eq:x0-root-z2-0f}
    x_0=-\frac{u^2}{12}+\frac13\Big(\sqrt{q}+\frac1{\sqrt{q}}\Big) \ ,
\end{align}
and therefore its unique $K$-rational 2-torsion is
\begin{align}
    P=\Big(-\frac{u^2}{12}+\frac13\Big(\sqrt{q}+\frac1{\sqrt{q}}\Big),0\Big)\ .
\end{align}
Substituting \eqref{eq:x0-root-z2-0f} into \eqref{eq:g2-z2-gen} yields
\begin{align}
    g_2^{\SU(2)_0/\Z_2}&=\frac{u^4}{12}-\frac23\Big(\sqrt{q}+\frac1{\sqrt{q}}\Big)u^2+\frac{4q}{3}+\frac{4}{3q}+\frac{56}3\ ,\notag\\
    g_3^{\SU(2)_0/\Z_2}&=\frac{u^6}{216}-\frac{1}{18}\Big(\sqrt{q}+\frac1{\sqrt{q}}\Big)u^4+\Big(\frac29\Big(\sqrt{q}+\frac1{\sqrt{q}}\Big)^2-\frac83\Big)u^2\notag\\
    &~+\frac{32}{3} \left(\sqrt{q}+\frac{1}{\sqrt{q}}\right)-\frac{8}{27}
   \left(\sqrt{q}+\frac{1}{\sqrt{q}}\right)^3 \ ,
\end{align}
which is exactly the same as Eq.(3.10) in \cite{Kim:2026yeh}.

\subsubsection*{2-isogeny of \texorpdfstring{$\SU(2)+2\F$}{SU2+2F} theory}

The $\Z_2$-invariant parent Seiberg--Witten curve
of the $\SU(2)+2\F$ theory, obtained from the 5-brane web, reads
\begin{align}
    F_{\SU(2)+2\F}(t,w)=U+\left(t+t^{-1}\right)-\frac1{M_1}\left(tw+t^{-1}w^{-1}\right)+\sqrt{\frac{q}{M_1}}\left(w+w^{-1}\right)=0\ .
\end{align}
Its Weierstrass coefficients are given by
\begin{align}
    g_2^{2\F}&=\frac{u^4}{12}-\frac23\chi_1^{\SU(3)}(y)u^2-4u-4\chi^{\SU(3)}_{2}(y)+\frac43\Big(\chi_1^{\SU(3)}(y)\Big)^2\  ,\notag\\
    g_3^{2\F}&=\frac{u^6}{216}-\frac1{18}\chi_1^{\SU(3)}(y)u^4-\frac13u^3+\Big(\frac29\Big(\chi_1^{\SU(3)}(y)\Big)^2-\frac13\chi_2^{\SU(3)}(y)\Big)u^2\notag\\
    &+\frac43\chi_1^{\SU(3)}(y)u-\frac8{27}\Big(\chi_1^{\SU(3)}(y)\Big)^3+\frac43\chi_1^{\SU(3)}(y)\chi_2^{\SU(3)}(y)
\end{align}
where $U=q^{1/6}M_1^{-1/2}u$, we rescaled
\begin{align}
    g_2^{2\F}\to (M_1^{-2}q^{-2/3})g_2^{2\F},\quad g_3^{2\F}\to (M_1^{-3}q^{-1})g_3^{2\F} \ ,
\end{align}
and $\chi_1^{\SU(3)}(y)$ and $\chi_2^{\SU(3)}(y)$ are the characters of the fundamental and anti-fundamental representations of $\SU(3)$, respectively:
\begin{align}
    \chi_1^{\SU(3)}(y)=y_1+\frac1{y_2}+\frac{y_2}{y_1},\quad\chi_2^{\SU(3)}=\frac1{y_1}+y_2+\frac{y_1}{y_2}\ ,
\end{align}
with
\begin{align}
    y_1=\frac{M_1}{q^{1/3}},\quad y_2=M_1q^{1/3}.
\end{align}
Then substituting the coefficients $g_2^{2\F},g_3^{2\F}$ into \eqref{eq:fx-z2}, we obtain a unique $K$-rational 2-torsion
\begin{align}\label{eq:P-2f-torsion}
    P=\Big(-\frac{u^2}{12}+\frac13\chi_1^{\SU(3)}(y),0\Big)\ .
\end{align}
Substituting~\eqref{eq:P-2f-torsion} into \eqref{eq:g2-z2-gen}, we find that
the resulting curve agrees with Eq.(3.14) in~\cite{Kim:2026yeh}.

\subsubsection*{2-isogeny of \texorpdfstring{$\SU(2)+4\F$}{SU(2)+4F} theory}

The $\Z_2$-invariant parent Seiberg--Witten curve
of the $\SU(2)+4\F$ theory, obtained from the 5-brane web, reads
\begin{align}
    F_{\SU(2)+4\F}(t,w)&=U+\sqrt{\frac{{M_1M_2}}{q}}(q+1)(w+w^{-1})+(tw+t^{-1}w^{-1})+M_1M_2(tw^{-1}+t^{-1}w)\notag\\
     &\quad -(M_1+M_2)(t+t^{-1})=0\ .
\end{align}
Its Weierstrass coefficients are given by
\begin{align}
    g_2^{4\F}&=\frac{u^4}{12}-\frac23(\chi_1^{\SO(7)}+3)u^2-4\chi_2^{\SO(7)}u+\frac43\Big(\Big(\chi_1^{\SO(7)}\Big)^2-3\chi_1^{\SO(7)}-3\chi_3^{\SO(7)}+3\Big)\ ,\notag\\
    g_3^{4\F}&=\frac{u^6}{216}-\frac1{18}\Big(\chi_1^{\SO(7)}+3\Big)u^4-\frac13\chi_2^{\SO(7)}u^3\notag\\
    &+\frac19\Big(2\Big(\chi_1^{\SO(7)}\Big)^2+3\chi_1^{\SO(7)}-3\chi_3^{\SO(7)}-24\Big)u^2\notag\\
    &+\Big(\frac43\chi_1^{\SO(7)}\chi_2^{\SO(7)}-4\chi_2^{\SO(7)}\Big)u-\frac8{27}\Big(\chi_1^{\SO(7)}\Big)^3\notag\\
    &+\frac43\chi_1^{\SO(7)}\chi_3^{\SO(7)}+\frac43\Big(\chi_1^{\SO(7)}\Big)^2-4\chi_3^{\SO(7)}-\frac43\chi_1^{\SO(7)}\ ,
\end{align}
where $U=\sqrt{M_1M_2}u$, we rescaled
\begin{align}
    g_2^{4\F}\to(M_1M_2)^{-2}g_2^{4\F},\quad g_3^{4\F}\to(M_1M_2)^{-3}g_3^{4\F}\ ,
\end{align}
and $\chi_1^{\SO(7)}$, $\chi_2^{\SO(7)}$ and $\chi_3^{\SO(7)}$ are the characters of the 7-, 8-, and $21$-dimensional representations of $\SO(7)$, respectively:
\begin{align}
    \chi_1^{\SO(7)}&=\sum_{i=1}^3\Big(y_i+\frac1{y_i}\Big)+1\ ,\notag\\
    \chi_2^{\SO(7)}&=\prod_{i=1}^3\Big(\sqrt{y_i}+\frac{1}{\sqrt{y_i}}\Big)\ ,\notag\\
    \chi_3^{\SO(7)}&= \sum_{1 \le i < j \le 3} 
   \left( y_i + \frac{1}{y_i} \right)
   \left( y_j + \frac{1}{y_j} \right) 
   + \sum_{i=1}^3 \left( y_i + \frac{1}{y_i} \right)
   + 3 \ ,
\end{align}
with
\begin{align}
    y_1=\frac1{M_1M_2},\qquad y_2=\frac{M_1}{M_2},\qquad y_3=q \ .
\end{align}
Performing a factorization about $f(x)=4x^3-g_2^{4\F}x-g_3^{4\F}$ over $K$ yields a unique $K$-rational 2-torsion point as follows:
\begin{align}\label{eq:P-4f-torsion}
    P=\Big(-\frac{u^2}{12}+\frac13\chi_1^{\SO(7)}-1,0\Big)\ .
\end{align}
Substituting \eqref{eq:P-4f-torsion} into \eqref{eq:g2-z2-gen}, we find that the resulting curve agrees with Eq.(3.20) in~\cite{Kim:2026yeh}.

\subsubsection*{2-isogeny of \texorpdfstring{$\SU(2)+6\F$}{SU(2)+6F} theory}

The $\Z_2$-invariant parent Seiberg--Witten curve
of the $\SU(2)+6\F$ theory, obtained from the
5-brane web, reads
\begin{align}\label{eq:sw-su2-6f-z2}
    &F_{\SU(2)+6\F}(t,w):=\frac{M_2 M_3 }{M_0
   M_1}\left(\frac{w^2}{t}+\frac{t}{w^2}\right)-\left(M_2+M_3+\frac{1}{M_0}+\frac{1}{M_1}\right) \left(t w+\frac{1}{t
   w}\right)\nonumber\\
   &\quad -\left(\frac{M_3 M_2}{M_0}+\frac{M_3 M_2}{M_1}+\frac{M_2}{M_0
   M_1}+\frac{M_3}{M_0 M_1}\right) \left(\frac{t}{w}+\frac{w}{t}\right)+t w^2+\frac{1}{t w^2}+U\nonumber\\
   &\quad +\left(M_3
   M_2+\frac{M_2}{M_0}+\frac{M_2}{M_1}+\frac{M_3}{M_0}+\frac{M_3}{M_1}+\frac{1}{M_0
   M_1}\right) \left(t+\frac{1}{t}\right) -\frac{2 \sqrt{M_2} \sqrt{M_3}}{\sqrt{M_0} \sqrt{M_1}}
   \left(w^2+\frac{1}{w^2}\right)\nonumber\\
   &\quad +\Bigg[\frac{\sqrt{M_3}
   M_2^{3/2}}{\sqrt{M_0} \sqrt{M_1}}+\frac{M_3^{3/2} \sqrt{M_2}}{\sqrt{M_0}
   \sqrt{M_1}}+\frac{\sqrt{M_3} \sqrt{M_2}}{M_0^{3/2}
   \sqrt{M_1}}+\frac{\sqrt{M_3} \sqrt{M_2}}{\sqrt{M_0}
   M_1^{3/2}}+\frac{\sqrt{M_2}}{\sqrt{M_0} \sqrt{M_1}
   \sqrt{M_3}} \nonumber\\
   &\quad+\frac{\sqrt{M_3}}{\sqrt{M_0} \sqrt{M_1} \sqrt{M_2}}+\frac{\sqrt{M_1} \sqrt{M_3} \sqrt{M_2}}{\sqrt{M_0}} +\frac{\sqrt{M_0}
   \sqrt{M_3} \sqrt{M_2}}{\sqrt{M_1}}\Bigg]
   \left(w+\frac{1}{w}\right)=0 \ ,
\end{align}
where $M_0=q$ is the instanton fugacity.
Its Weierstrass coefficients are given by
\begin{align}
    g_2^{6\F}&=\frac{u^4}{12}-\frac23\Big(\chi_1^{F_4}+3\chi_4^{F_4}-22\Big)u^2+(12\chi_4^{F_4}-8\chi_1^{F_4}-4\chi_3^{F_4}+44)u\notag\\
    &+\frac43\Big(\Big(\chi_1^{F_4}\Big)^2+3\Big(\chi_4^{F_4}\Big)^2-3\chi_1^{F_4}\chi_4^{F_4}-3\chi_3^{F_4}-3\chi_2^{F_4}-8\chi_1^{F_4}+18\chi_4^{F_4}+25\Big)\ ,\notag\\
    g_3^{6\F}&=\frac{u^6}{216}-\frac1{18}\Big(\chi_1^{F_4}+3\chi_4^{F_4}+50\Big)u^4+\Big(-\frac13\chi_3^{F_4}-\frac23\chi_1^{F_4}-7\chi_4^{F_4}-\frac{37}3\Big)u^3\notag\\
    &+\frac19\Big(2\Big(\chi_1^{F_4}\Big)^2+3\chi_1^{F_4}\chi_4^{F_4}-24\Big(\chi_4^{F_4}\Big)^2-3\chi_2^{F_4}-75\chi_3^{F_4}-52\chi_1^{F_4}-186\chi_4^{F_4}-283\Big)u^2\notag\\
    &+\frac{4}{3}\Bigl(
\chi_1^{F_4}\chi_3^{F_4}
-3\chi_4^{F_4}\chi_3^{F_4}
+2\bigl(\chi_1^{F_4}\bigr)^2
-3\chi_4^{F_4}\chi_1^{F_4}
-3\bigl(\chi_4^{F_4}\bigr)^2 
-6\chi_2^{F_4}
-16\chi_3^{F_4}
\notag\\
&-7\chi_1^{F_4}
-21\chi_4^{F_4}
-40
\Bigr)u+\frac{4}{27}
\left(\chi_1^{F_4}-3\chi_4^{F_4}-10\right)
\Bigl(
3\chi_4^{F_4}\chi_1^{F_4}
-2\bigl(\chi_1^{F_4}\bigr)^2
+6\chi_4^{F_4}
+4\chi_1^{F_4}\notag\\
&+9\chi_3^{F_4}
+9\chi_2^{F_4}
+25
\Bigr)
\end{align}
where $U=\frac{M_2M_3}{M_0M_1}u$, we rescaled
\begin{align}
    g_2^{6\F}\to (M_0M_1M_2M_3)^{-4}g_2^{6\F},\quad g_3^{6\F}\to (M_0M_1M_2M_3)^{-6}g_3^{6\F},
\end{align}
and
$\chi_i^{F_4}, i=1,2,3,4$, are the characters of 52-, 1274-, 273-, and $26$-dimensional representations, with fugacities:
\begin{align}
    \Big\{M_0M_1,\frac{M_0}{M_1},M_2M_3,\frac{M_2}{M_3}
    \Big\}
\end{align}
Performing a factorization about $f(x)=4x^3-g_2^{6\F}x-g_3^{6\F}$ over $K$ yields
\begin{align}\label{eq:P-6f-torsion}
    P=\Big(x_0,0\Big)\ ,
\end{align}
where
\begin{align}
x_0=-\frac{u^2}{12}-2u+\frac13(\chi_1^{F_4}-3\chi_4^{F_4}-10)\ .
\end{align}
Substituting \eqref{eq:P-6f-torsion} into \eqref{eq:g2-z2-gen}, we find that the resulting curve agrees with Eq. (3.28) in~\cite{Kim:2026yeh}, in which the fugacities in the characters of the $F_4$ symmetry are precisely $\{q,M_1,M_2,M_3\}$.

\subsubsection{3-isogenies and \texorpdfstring{$\Z_3$}{Z3} S-folds}

Let $P\in\mathrm{MW}(\mathcal E)$ be a section of exact order $3$, and consider the cyclic subgroup $G=\langle P\rangle=\{O,P,-P\}$, where both $P$ and $-P$ have order 3.  That is, we find
\begin{align}
    \mathcal G_2=\varnothing,\qquad G_+=\{P\},\qquad G_-=\{-P\} \ ,
\end{align}
with $G_+=-G_-$. Consequently, only the point $P$ contributes to the sum in \eqref{eq:alpha-betaG}. 

The roots of the division polynomial $\psi_3(x)$ in \eqref{eq:division-equations}  give the $x$-coordinates of the points of exact order 3.
To identify a $K$-rational 3-torsion point $P=(x_0,y_0)$, we factor $\psi_3(x)$ over $K$ and solve the linear factor. Plugging the solution $x_0$ into the curve equation yields
\begin{align}
    y_0^2=4x_0^3-g_2x_0-g_3\ .
\end{align}
If the right-hand side is a square in $K$, we obtain $y_0\in K$ and hence a $K$-rational point $P=(x_0,y_0)$ of exact order $3$.

\subsubsection*{3-isogeny of the local \texorpdfstring{$\mathbb{P}^2$}{P2} theory}

The $\Z_3$-invariant parent Seiberg--Witten curve of the local $\mathbb{P}^2$ theory, obtained from the 5-brane web, is given by \eqref{eq:SW-localP2}. Its Weierstrass coefficients read
\begin{align}\label{eq:wei-p2}
    y^2=4x^3-g^{\mathbb{P}^2}_2x-g^{\mathbb{P}^2}_3
\end{align}
where $U$ appearing in \eqref{eq:SW-localP2} is identified as $u$ and
\begin{align}
    g^{\mathbb{P}^2}_2&=\frac {u^4}{12}+2u\ , \qquad g^{\mathbb{P}^2}_3=\frac{u^6}{216}+\frac{u^3}6+1\ .
\end{align}
Using \eqref{eq:division-equations} for this case and performing a factorization on $\psi_3$, we find that the $K$-rational root is only
\begin{align}
    x_0=-\frac{u^2}{12}\ .
\end{align}
Substituting $x_0$ into \eqref{eq:wei-p2}, we obtain
\begin{align}
    y_0^2=-1
\end{align}
Choosing $P=(x_0,i)$ as the generator of $G$, the other point is identified as $-P=(x_0,-i)$. Obviously, all the $K$-rational $3$-torsion points belong to an identical cyclic subgroup $G=\{O,P,-P\}$, indicating the uniqueness of the cyclic subgroup of order 3 of the Mordell--Weil group. From \eqref{tPwP} we find
\begin{align}\label{eq:alpha-beta-p2}
    \alpha(P)=6x^2_0-\frac12g_2^{\mathbb{P}^2}=-u,\qquad \beta(P)=y_0^2+\alpha(P)x_0=\frac{u^3}{12}-1\ .
\end{align}
One can readily see that the 3-isogeny of the Seiberg--Witten curve for local $\mathbb{P}^2$ agrees with the $\Z_3$ S-fold Seiberg--Witten curve \eqref{eq:localp2z3}, by inserting \eqref{eq:alpha-beta-p2} and \eqref{eq:wei-p2} into \eqref{tildeg23} and also by redefining $u$ as $-u$.

\subsubsection*{3-isogeny of \texorpdfstring{$\SU(2)+2\F$}{SU2+2\F} theory}

The $\Z_3$-invariant parent Seiberg--Witten curve of the  $\SU(2)+2\F$ theory, obtained from the 5-brane web, is given by \eqref{eq:twSU2+2Fz3}. Its Weierstrass coefficients read
\begin{align}\label{eq:g2g3o-2f-z3-fd}
    g_2^{2\F}&=\frac{u^4}{12}-2u^2+2\widetilde{\chi}^{\SU(2)}_1u   \ , \notag\\
    g_3^{2\F}&=\frac{u^6}{216}-\frac{1}{6}u^4+\frac16\widetilde{\chi}^{\SU(2)}_1u^3+u^2-2\widetilde{\chi}_1^{\SU(2)}u+(\widetilde{\chi}^{\SU(2)}_1)^2 \ ,
\end{align}
where in the middle step, we identified $U=\sqrt{\M}u$, and rescaled
\begin{align}
    g_2^{2\bf{F}}\to (\sqrt{\M})^{-4}g^{2\bf{F}}_2,\quad g^{2\bf F}_3\to 
    (\sqrt{\M})^{-6}g_3^{2\bf F} \ ,
\end{align}
induced by the coordinate transformation $x\to (\sqrt{\M})^{-2}x$ and $y\to (\sqrt{\M})^{-3}y$. 
The generator $P$ for this case is determined by $ (x_0,y_0)$ in terms of factoring $\psi_3$, with
\begin{align}
x_0=-\frac{u^2}{12},\qquad y_0^2=-u^2+2\widetilde{\chi}_1^{\SU(2)}u-(\widetilde{\chi}_1^{\SU(2)})^2\ .
\end{align}
We confirm that there are no other roots of $\psi_3$ that are rational functions of $u$.
Then from \eqref{tPwP} we find
\begin{align}\label{eq:al-beta-2f-z3}
    \alpha(P)&=u^2-\widetilde\chi_1^{\SU(2)}u, \nn\\ 
    \beta(P)&=-\frac{u^4}{12}+\frac1{12}\widetilde{\chi}_1^{\SU(2)}u^3-u^2+2\widetilde{\chi}_1^{\SU(2)}u-(\widetilde{\chi}_1^{\SU(2)})^2.
\end{align}
Inserting \eqref{eq:al-beta-2f-z3} and \eqref{eq:g2g3o-2f-z3-fd} into \eqref{tildeg23} gives rise to agreement with \eqref{eq:E7z3curve}.

\subsubsection*{3-isogeny of \texorpdfstring{$\SU(2)+5\F$}{SU(2)+5\F} theory}

The $\Z_3$-invariant parent Seiberg--Witten curve of $\SU(2)+5\F$ theory, obtained from the 5-brane web, is given by \eqref{eq:su25fz3}. Its Weierstrass coefficients read
\begin{align}\label{eq:g2g3o-5f-z3-fd}
    g^{5\F}_2&=\frac{u^4}{12}-2(\chi_1^{G_2}+2)u^2
    +2(\chi_2^{G_2}+8\chi_1^{G_2}+2)u-12(2\chi_1^{G_2}+\chi_2^{G_2}-1),\notag\\
    g_3^{5\F}&=\frac{u^6}{216}-\frac{1}{6}(2+\chi_1^{G_2})u^4
    +\frac{8\chi_1^{G_2}+\chi_2^{G_2}+26}{6}u^3
    +((\chi_1^{G_2})^2-10\chi_1^{G_2}-\chi_2^{G_2}-19)u^2\notag\\
    &-2(-14 - 18\chi_1^{G_2} + 2(\chi_1^{G_2})^2 - 4\chi_2^{G_2} + \chi_1^{G_2}\chi_2^{G_2})u
    +(-4 + 2\chi_1^{G_2} + \chi_2^{G_2})^2\notag\\
    &-24 - 24\chi_1^{G_2} - 12\chi_2^{G_2}\ ,
\end{align}
where $U$ appearing in \eqref{eq:su25fz3} is identified with $u$, and $\chi_1^{G_2}$ and $\chi_2^{G_2}$ are given by \eqref{eq:cha-G2}. The generator $P$ for this case is given by $ (x_0,y_0)$ in terms of factoring $\psi_3$, with
\begin{align}
    x_0&=-\frac{u^2}{12}+u-3\ ,\notag\\
    y_0^2&=-\Bigl[u^2-(\chi_1^{G_2}+5)u+8 + 2\chi_1^{G_2} + \chi_2^{G_2}\Bigr]^2 \ .
\end{align}
We again confirm that there are no other roots of $\psi_3$ that are rational functions of $u$.

Then from \eqref{tPwP} we find
\begin{align}\label{eq:al-beta-5f-z3}
    \alpha(P)&=-(u-6)\Bigl[u^2-(\chi_1^{G_2}+5)u+8 + 2\chi_1^{G_2} + \chi_2^{G_2}\Bigr]\ ,\notag\\
    \beta(P)&=\frac{1}{12}
\Bigl[u^2-(\chi_1^{G_2}+5)u+8 + 2\chi_1^{G_2} + \chi_2^{G_2}\Bigr]\notag\\
&\times
\Bigl[(u-6)^3
-12\bigl(u^2-(\chi_1^{G_2}+5)u+8 + 2\chi_1^{G_2} + \chi_2^{G_2}\bigr)\Bigr]\,.
\end{align}
It follows from \eqref{eq:al-beta-5f-z3}, \eqref{eq:g2g3o-5f-z3-fd}, and shifting $u$ to $-u+60$ that the resulting quotient curve from the 3-isogeny agrees with the $\Z_3$ S-fold curve~\eqref{eq:E8z3curve}.

\subsubsection{4-isogenies and \texorpdfstring{$\Z_4$}{Z4} S-folds}

Let $P\in\mathrm{MW}(\mathcal E)$ be a section of exact order $4$, and consider the cyclic subgroup $G=\langle P\rangle=\{O,P,2P,3P\}$. Since $4P=O$,  we find $G=\{O,P,2P,-P\}$, where the sections $P,2P$ have order $4,2$, respectively. As a consequence,
\begin{align}
    \mathcal G_2=\{2P\}, \qquad G_+=\{P\}, \qquad G_-=\{-P\}\ ,
\end{align}
with $G_+=-G_-$, and
\begin{align}
    G\setminus\{O\}=\mathcal G_2\sqcup G_+\sqcup G_-=\{2P\}\sqcup\{P\}\sqcup\{-P\}\ .
\end{align}
Then, to completely determine $\alpha(G)$ and $\beta(G)$ listed in \eqref{eq:alpha-betaG}, we 
therefore need the coordinates
\begin{align}
    P=(x_1,y_1)\,,  \qquad 2P=(x_0,0)\ .
\end{align}

To this end, we consider $\psi^2_4(x)$, expressed solely in $x$ using the curve equation. The division polynomial $\psi_4$ in \eqref{eq:division-equations} takes the form
\begin{align}
    \psi_4=\frac{y}{32}p_4(x)\ ,
\end{align}
with
\begin{align}
    p_4(x)=64x^6-80g_2x^4-320g_3x^3
        -20g_2^2x^2-16g_2g_3x
        +g_2^3-32g_3^2\ .
\end{align}
Using the curve equation, we obtain
\begin{align}
    \psi_4(x)^2=\frac{1}{1024}\bigl(4x^3-g_2x-g_3\bigr)p_4(x)^2.
\end{align}
The equation
$\psi_4(x)^2=0$ thus separates into two branches: $p_4(x)^2=0$ and $4x^3-g_2x-g_3=0$. On a smooth fiber, the former gives the $x$-coordinates of points of exact order $4$, while the latter gives those of the nonzero 2-torsion point.

We choose a root $x_1\in K$ of $p_4(x)^2=0$ such that
\begin{align}
    y_1^2=4x_1^3-g_2x_1-g_3
\end{align}
admits a solution $y_1\in K$. The resulting point $P=(x_1,y_1)$ has exact order $4$.  To determine the nonzero 2-torsion points, we factor $y^2(x)=4x^3-g_2x-g_3$ and solve its linear factor, obtaining $x_0\in K$. We then verify that $(x_0,0)=2P$ to ensure that the selected 4-torsion and 2-torsion points belong to the same cyclic subgroup $G$ using the elliptic curve group law in \eqref{eq:add-law-x}-\eqref{eq:add-law-y}.

\subsubsection*{4-isogeny of the pure $\SU(2)$ theory}

The $\Z_4$-invariant parent Seiberg--Witten curve of the $\SU(2)$ theory, obtained from the 5-brane web, is given by \eqref{eq:su2swcurve}. Its Weierstrass coefficients read
\begin{align}\label{eq:g2g3or-0f-z4}
    g_2^{\SU(2)_0}&=\frac{u^4}{12}-\frac{4 u^2}{3}+\frac{4}{3}\ ,\notag\\
    g_3^{\SU(2)_0}&=\frac{u^6}{216}-\frac{u^4}{9}+\frac{5 u^2}{9}+\frac{8}{27}\ ,
\end{align}
where $U$ appearing in \eqref{eq:su2swcurve} is identified with $u$.
Then  the rational root of $p_4(x)$ and the Weierstrass normal equation $y(x)^2$ give us the 4-torsion point $P$
\begin{align}
    x_1&=-\frac{u^2}{12}-\frac{1}{3}\ ,\quad  y_1^2=-u^2\ ,
\end{align}
while the rational root of $y(x)^2$ gives us the 2-torsion point $2P$
\begin{align}
    x_0&=\frac{2}{3}-\frac{u^2}{12}\ ,\qquad y_0=0 \ .
\end{align}
Using the elliptic curve group law, we find $2P=(x_0,0)$.

From \eqref{tPwP} and \eqref{eq:alpha-betaG} we obtain
\begin{align}\label{eq:al-be-0f-z4}
    \alpha(G)&=\alpha(2P)+\alpha(P)=1+u^2\ ,\quad
    \beta(G)=\beta(2P)+\beta(P)=-\frac{u^4}{12}-\frac{17}{12}u^2+\frac23\ .
\end{align}
Plugging \eqref{eq:al-be-0f-z4} and \eqref{eq:g2g3or-0f-z4} into \eqref{tildeg23}, we find that the resulting quotient curve agrees with \eqref{eq:E7z4curve} by redefining $u$ as $\frac14u$, and rescaling
\begin{align}
    g_2^{\SU(2)_0}\to 256g^{\SU(2)_0}_2,\quad g^{\SU(2)_0}_3\to 4096 g^{\SU(2)_0}_3
\end{align}
by implementing the coordinate transformation $x\to16x$ and $y\to 64 y$.

\subsubsection*{4-isogeny of $\SU(2)+4\F$ theory}

The $\Z_4$-invariant parent Seiberg--Witten curve of the $\SU(2)+4\bf{F}$ theory, obtained from the 5-brane web, is given by \eqref{eq:su24fz4or}. Its Weierstrass coefficients read
\begin{align}\label{eq:g2g3-4f-z4fol}
    g_2^{4\F}&=\frac{u^4}{12}-\frac43(s+1)u^2
    +8su+\frac43(s^2-16s+16)\ ,\notag\\
    g_3^{4\F}&=\frac{u^6}{216}-\frac{s+1}{9}u^4
    +\frac23su^3+\frac{5s^2-8s-16}{9}u^2\notag\\
    &\quad-\frac{16}{3}s(s-2)u
    +\frac8{27}(s-2)(s^2+32s-32)\ ,
\end{align}
where 
\begin{align}
    s=\Big(\chi_1^{\SU(2)}\Big)^2\equiv\Big(\M+\frac1\M\Big)^2\ ,
\end{align}
and in the middle step, we identified $U$ appearing in \eqref{eq:su24fz4or} with $u$, and  rescaled 
\begin{align}
g_2^{4\bf{F}}\to \M^{-4}g^{4\bf{F}}_2,\quad g_3^{4\bf F}\to \M^{-6}g^{4\bf F}_3  \ ,   
\end{align}
by implementing the coordinate transformation $x\to \M^{-2}x$ and $y\to \M^{-3}y$.
Then the rational root of $p_4(x)$ and the Weierstrass normal equation $y(x)^2$ give us the 4-torsion point $P$
\begin{align}
    x_1&=-\frac{u^2}{12}+u-\frac{s+4}{3},\qquad y_1^2=-(u-4)^2(u-s)^2,
\end{align}
while the rational root of $y(x)^2$ gives us the 2-torsion point $2P$
\begin{align}
    x_0&=\frac{2 \M^2}{3}+\frac{2}{3 \M^2}-\frac{u^2}{12}\ ,\qquad y_0=0\ .
\end{align}
Using the elliptic curve group law, we find $2P=(x_0,0)$.

From \eqref{tPwP} and \eqref{eq:alpha-betaG} we obtain
\begin{align}\label{eq:al-be-4f-z4}
    \alpha(G)&=-u^3 + (s + 9) u^2 - 2 (5 s + 8) u + s (s + 16)\ ,\notag\\
    \beta(G)&=-\frac{1}{12} \left(17 s^2+344 s+528\right) u^2+\frac{4}{3} \left(7 s^2+50 s+16\right)
   u\notag\\
   &\quad +\frac{2}{3} s \left(s^2-34 s-32\right)-\frac{1}{12} (s+33) u^4+\frac{1}{6} (25
   s+112) u^3+\frac{u^5}{12}\ .
\end{align}
Substituting \eqref{eq:al-be-4f-z4} and \eqref{eq:g2g3-4f-z4fol} into \eqref{tildeg23}, we find agreement with \eqref{eq:E8z4curve} after shifting $u\to-u+60$.

\subsubsection{6-isogenies and \texorpdfstring{$\Z_6$}{Z6} S-folds}

Let $P\in\mathrm{MW}(\mathcal E)$ be a section of exact order $6$, and consider the cyclic subgroup $G=\langle P\rangle=\{O,P,\dots,5P\}$. Since $6P=O$,  we find $G=\{O,P,2P,3P,-P,-2P\}$, where the sections $P,2P,3P$ have order $6,3,2$, respectively. Therefore, we have
\begin{align}
    \mathcal G_2=\{3P\}, \qquad G_+=\{P,2P\}, \qquad G_-=\{-P,-2P\}\ ,
\end{align}
with $G_+=-G_-$, and
\begin{align}
    G\setminus\{O\}=\mathcal G_2\sqcup G_+\sqcup G_-=\{3P\}\sqcup\{P,2P\}\sqcup\{-P,-2P\}\ .
\end{align}
Then, to completely determine $\alpha(G)$ and $\beta(G)$ listed in \eqref{eq:alpha-betaG}, we 
accordingly need the coordinates
\begin{align}
    P=(x_2,y_2)\ , \qquad 2P=(x_1,y_1) \ ,  \qquad 3P=(x_0,0)\ .
\end{align}

For this purpose, we consider $\psi_6^2$, expressed only in $x$ using the curve equation. The division polynomial $\psi_6$ obtained from \eqref{eq:division-equations}-\eqref{eq:psi-m} takes the form
\begin{align}
    \psi_6=\frac{\psi_3}{\psi_2}(\psi_5\psi_2^2-\psi_4^2)=\frac1{2^{16}}p_6(x)p_3(x)y\ ,
\end{align}
with
\begin{align}
    p_6(x)&=-4096 x^{12}+22528 g_2 x^{10}+225280 g_3 x^9+42240 g_2^2 x^8+135168 g_2 g_3 x^7-5888 g_2^3 x^6\notag\\
    &\quad+454656
   g_3^2 x^6+16896 g_2^2 g_3 x^5+2960 g_2^4 x^4-61440 g_2 g_3^2 x^4-20480 g_3^3 x^3\notag\\
   &\quad+1280
   g_2^3 g_3 x^3-360 g_2^5 x^2+9984 g_2^2 g_3^2 x^2+14336 g_2 g_3^3 x-528 g_2^4 g_3 x+3
   g_2^6\notag\\
   &\quad+8192 g_3^4-384 g_2^3 g_3^2 \ ,\notag\\
   p_3(x)&=-48 x^4+24 g_2 x^2+48 g_3 x+g_2^2\ .
\end{align}
Using the curve equation, $\psi_6(x)^2=0$ hence separates into three branches: $p_6(x)^2=0,$ $p_3(x)^2=0$ and $4x^3-g_2x-g_3=0$. On a smooth fiber, these branches give the $x$-coordinates of points of exact order 6, nonzero points of order 3, and nonzero points of order 2, respectively.

We choose a root $x_2\in K$ of $p_6(x)=0$ such that
\begin{align}
    y_2^2=4x^3_2-g_2x_2-g_3
\end{align}
 admits a solution $y_2\in K$. The resulting point $P=(x_2,y_2)$ has exact order 6.
 To determine the remaining points, we select $K$-rational roots $x_1$ of $p_3(x)$ and $x_0$ of $4x^3-g_2x-g_3$, and choose $y_1\in K$ satisfying
 \begin{align}
     y_1^2=4x_1^3-g_2x_1-g_3\ .
 \end{align}
 We then verify that
 \begin{align}
     (x_1,y_1)=2P,\qquad (x_0,0)=3P\ ,
 \end{align}
using the elliptic curve group law in \eqref{eq:add-law-x}-\eqref{eq:add-law-y}. This ensures that the selected points belong to the same cyclic subgroup $G$.

The $\Z_6$-invariant parent Seiberg--Witten curve of  $\SU(2)+2\F$ theory, obtained from the 5-brane web, is given by \eqref{eq:su22f-z6-curve}. Its Weierstrass coefficients read
\begin{align}\label{eq:weier-2f-z6}
g_2^{2\F}&=\frac{u^4}{12}-2 u^2+4 u\ ,\notag\\
g_3^{2\F}&=\frac{u^6}{216}-\frac{u^4}{6}+\frac{u^3}{3}+u^2-4 u+4\ ,
\end{align}
where $U$ appearing in \eqref{eq:su22f-z6-curve} is identified with $u$.
Thus, using a $K$-rational root of $p_6(x)=0$ and the Weierstrass equation, we obtain the $6$-torsion point $P$:
\begin{align}
    x_2=-\frac{u^2}{12}+u-2,\qquad
    y_2^2=-(-3 + u)^2 (-2 + u)^2\ .
\end{align}
Similarly, a $K$-rational root of $p_3(x)=0$, together with the Weierstrass equation, yields the coordinate of the $3$-torsion point $2P$:
\begin{align}
    x_1&=-\frac{u^2}{12},\qquad
    y_1^2=-(u-2)^2\ .
\end{align}
Finally, a $K$-rational root of $4x^3-g_2x-g_3=0$  determines the nonzero 2-torsion point $3P$ with coordinates:
\begin{align}
    x_0&=1 - \frac{u^2}{12}\ ,\qquad y_0=0\ .
\end{align}
Using the elliptic curve group law, we find that they indeed belong to the same cyclic subgroup $G$. 

From \eqref{tPwP} and \eqref{eq:alpha-betaG} we obtain
\begin{align}\label{eq:al-be-z6}
    \alpha(G)
    &=-u^3+10 u^2-29 u+27\ ,\notag\\
    \beta(G)
    &=\frac{u^5}{12}-\frac{17 u^4}{6}+\frac{281 u^3}{12}-\frac{337 u^2}{4}+139 u-85\ .
\end{align}
Substituting \eqref{eq:al-be-z6} and \eqref{eq:weier-2f-z6} into \eqref{tildeg23}, we find the resulting curve from the 6-isogeny agrees with \eqref{eq:E8z6curve} after shifting $u$ to $-u+60$.

\section{Other quotient curve: \texorpdfstring{$\Z_5$}{Z5} quotient of \texorpdfstring{$E_4$}{E4} SCFT}\label{sec:Otherquotient}
As discussed in the preceding section, $n$-isogenies are a powerful tool for finding the Seiberg--Witten curves for the $\Z_n$ S-folds. Recall that $\Z_n$ ($n=2,3,4,6$) is a subgroup of $\mathrm{SL}(2,\Z)$, but $\Z_5$ is not. So, the $\Z_5$ S-fold is not possible. On the other hand, it was suggested in~\cite{Furrer:2024zzu} that $\Z_5$ folding of the $E_4$ theory could be possible from the perspective of base change. In this section, we consider the 5-isogeny of the $E_4$ theory and construct the resulting $\Z_5$-quotient curve. As $\Z_5$ is not a subgroup of $\mathrm{SL}(2,\Z)$, the resulting curve is not the Seiberg--Witten curve for the $\Z_5$ S-fold. For this reason, we call it the $\Z_5$-quotient curve. We also provide an interpretation from the 5-brane web coordinate ($t,w$). 

The Seiberg--Witten curve for the $\SU(2)+3\F$ theory takes the form \cite{Kim:2014nqa}:
\begin{equation}\label{eq:SU2+3F-WS}
E:\quad y^2=4x^3-g^{3\F}_2(u)\,x-g^{3\F}_3(u)\ , 
\end{equation}
where in the massless limit,  
\begin{align}
g^{3\F}_2(u)&=\frac{u^4}{12}-\frac{10}{3}u^2-10u-\frac{20}{3},\\
g^{3\F}_3(u)&=\frac{u^6}{216}-\frac{5}{18}u^4-\frac56u^3
+\frac{20}{9}u^2+\frac{38}{3}u+\frac{395}{27}.
\end{align}

As done in section \ref{sec:346-isogeny}, we repeat the same procedure for 5-isogeny. For $P\in\mathrm{MW}(\mathcal E)$ a section of exact order $5$, we consider the cyclic group $G=\langle P\rangle=\{O,\pm P,\pm 2P\}$, and solve 
\begin{align}\label{eq:psi5=0}
    \psi_5&=\frac1{4096}\Big(20480 x^{12}-63488 g_2 x^{10}-389120 g_3 x^9-26880 g_2^2 x^8+61440 g_2 g_3 x^7+19200 g_2^3 x^6\notag\\
    &-61440
   g_3^2 x^6+44544 g_2^2 g_3 x^5-2000 g_2^4 x^4+122880 g_2 g_3^2 x^4+102400 g_3^3
   x^3-1280 g_2^3 g_3 x^3\notag\\
   &+200 g_2^5 x^2-\!3840 g_2^2 g_3^2 x^2-\!10240 g_2 g_3^3 x+400 g_2^4
   g_3 x+g_2^6-\!4096 g_3^4+128 g_2^3 g_3^2\Big)=0. 
\end{align}
Two rational solutions for \eqref{eq:psi5=0} are given by
\begin{align}\label{eq:Z5-P}
    x_1=-\frac{u^2}{12}-u-\frac73,\quad y_1=\pm i(3+u)^2\ . 
\end{align}
and
\begin{align}
    x_2=-\frac{u^2}{12}+\frac23,\quad y_2=\pm i(3+u)\ .
\end{align}
These are the torsion sections discussed in~\cite{Closset:2021lhd}. It follows from the elliptic curve group law that two solutions $(x_1,y_1)$ and $(x_2,y_2)$ correspond to $P$ and $2P$ points, respectively.
Finally, 
\begin{align}
 \alpha(G) & =u^3+11u^2+38u+42,\cr   
 \beta(G) & = -\frac{u^5}{12}-\frac{35 u^4}{12}-\frac{55 u^3}{2}-\frac{685 u^2}{6}-\frac{671 u}{3}-170 \ .
\end{align}
One then finds the $\Z_5$ quotient curve for $\SU(2)+3\F$ is given by
\begin{equation}\label{eq:Z5curve}
E_{\Z_5}:\quad y^2=4x^3-g^{3\F/\Z_5}_2(u)\, x-g^{3\F/\Z_5}_3(u),
\end{equation}
with
\begin{align}\label{eq:Z5-quot-curve}
g^{3\F/\Z_5}_2(u)&=\frac{u^4}{12}
-\frac{4750}{3}u^2+118750u-\frac{7512500}{3}\ ,\\
g^{3\F/\Z_5}_3(u)&=\frac{u^6}{216}-4u^5
+\frac{15625}{18}u^4
-\frac{510625}{6}u^3 \notag\\
&~~~
+\frac{38750000}{9}u^2
-\frac{329843750}{3}u
+\frac{30341796875}{27}\ , 
\end{align}
where we shifted $u$ by $u-60$. 
The corresponding Kodaira fiber configuration is $(I_1; I_5^2I_1)$, with $\mathrm{MW}_{\text{tor}}=\mathbb Z_5$.

Consider the prepotential relation of the $\Z_n$ S-folds. The prepotential \eqref{eq:IMSprep} for rank-1 theory can be simplified in a suitable massless limit,  
\begin{align}
    \mathcal{F}\big(\SU(2)+N_f\F\big) =\frac{8-N_f}{6}\phi^3\ . 
\end{align}
Then, it follows from \eqref{eq:1/nF} that 
\begin{align}\label{eq:Z5-prep}
    \mathcal{F}\big(\SU(2)+3\F/\Z_{5}\big) = \frac1{5}\,\mathcal{F}\big(\SU(2)+3\F\big) = \frac16\phi^3 \quad \Longleftrightarrow\quad  \mathcal{F}\big(\SU(2)+7\F\big) \ .
\end{align}
This relation suggests that one may find the $\Z_5$ quotient curve from the Seiberg--Witten curve for the $\SU(2)+7\F$ theory by tuning the $\E_8$ mass parameters, in a special way that is different from the $E_8$-tuning discussed in the preceding sections. 

The following specialization of the $\E_8$ mass fugacities
\begin{align}\label{eq:Z5-tuning}
    M_1=M_2=M_3=1,\quad M_4=M_5=M_6=M_7=M_0^2=e^{4\pi i/5}. 
\end{align}
yields
 \begin{align}
    \chi_1^{\E_8} &\to 0,  & \chi_2^{\E_8} &\to 0, &
    \chi_3^{\E_8} &\to 0, & \chi_4^{\E_8} &\to 14, \nonumber\\
    \chi_5^{\E_8} &\to20, & \chi_6^{\E_8} &\to -10, &
    \chi_7^{\E_8} &\to5, & \chi_8^{\E_8} &\to -2, 
    \label{eq:e8z5-cha-tun}
 \end{align}
and accordingly the $E_8$ Seiberg--Witten curve \eqref{eq:E8} with this specialization then reduces to the $\Z_5$ quotient curve \eqref{eq:Z5curve}.

We now give an interpretation of the $\Z_5$ quotient curve from the viewpoint of the 5-brane web diagram. The Seiberg--Witten curve for the parent massless $\SU(2)+3\F$ theory is given by
\begin{align}\label{eq:parent-Z5-tw}
(w+1) w^{-1} t + ( w - U  + 2 w^{-1})  + (w+1)^2 w^{-1} t^{-1} = 0.
\end{align}
This curve turns out to be invariant under the following $\Z_5$ transformation
\begin{align}\label{eq:Z5-tw}
\sigma: t \to w \to \frac{1+w}{t} \to \frac{1+t+w}{tw}
\to \frac{1+t}{w} \to t,
\end{align}
which is known as the Lyness 5-cycle \cite{Lyness1942} and is also used in an analogous setup in \cite{Hone:2021, Ducat:2021}. 
We check that the Seiberg--Witten 1-form $\lambda_{\mathrm{SW}} = \log t d \log w$ is invariant under this action up to total derivative terms as
\begin{align}
\lambda_{\mathrm{SW}} \to \lambda_{\mathrm{SW}} 
+ d \left( \log w \log (1+w) + \mathrm{Li}_2 (-w) - \log t \log w \right),
\end{align}
which is consistent with the expectation that the quotient by the $\mathbb Z_5$ action \eqref{eq:Z5-tw} gives the cyclic 5-isogeny.

To check it more explicitly, we see that the Seiberg--Witten curve \eqref{eq:parent-Z5-tw} agrees with the elliptic curve in the Weierstrass form \eqref{eq:SU2+3F-WS} under the following coordinate change
\begin{align}
x &= (U+3)w^{-1}+ \frac{1}{12}(8-U^2),
\cr
y &= -i (U+3) w^{-1} \left( 2(1+w^{-1})t + w - U + 2w^{-1} \right) \ ,
\end{align}
with $u=U$. Under this relation, we can check that the $\Z_5$ action \eqref{eq:Z5-tw} acts on the $(x,y)$ coordinates as
\begin{align}\label{eq:P-generated}
&x \to \lambda^2 - x_1 -x =: x',
\qquad
y \to - 2 \lambda x' - 2 \nu,
\cr
&\lambda := \frac{y-y_1}{2(x-x_1)},
\qquad
\nu := \frac{y_1 x- x_1 y}{2(x-x_1)}\ ,
\end{align}
where $(x_1, y_1)$ is the coordinates of $P$ given in \eqref{eq:Z5-P} with the minus sign chosen for $y_1$.
Thus, \eqref{eq:P-generated} is exactly the action generated by the torsion point $P$. 
This implies that the quotient by the $\mathbb Z_5$ action \eqref{eq:Z5-tw} gives the cyclic 5-isogeny. 

We present the $\mathbb Z_5$-quotient curve in terms of the 5-brane web coordinates:
\begin{align}\label{eq:tuning-E8-z5}
     -t^3Q_{\text{tuned}}(w)^2+t^2Q_{\text{tuned}}(w)S_{\text{tuned}}(w)+tR_{\text{tuned}}(w)+P_{\text{tuned}}(w)=0\ ,
\end{align}
where
\begin{align}
    Q_{\mathrm{tuned}}(w)
    &=(w-\zeta)^3,\notag\\
    S_{\mathrm{tuned}}(w)
    &=3w^3-\sigma_1w^2+\sigma_2\,w-3,\notag\\
    R_{\mathrm{tuned}}(w)
    &=-3w^6+2\sigma_1 w^5-\sigma_3w^4-Uw^3
    -\sigma_4\,w^2+2\sigma_2\,w-3,\notag\\
    P_{\mathrm{tuned}}(w)
    &=(w-1)^4(w-\zeta)^2,
\end{align}
with $\zeta:=e^{4\pi i/5}$ and
\begin{align}
    \sigma_1&=4+5\zeta,\quad \sigma_2= (\sigma_1)^*,\quad
    \sigma_3
      =5\left(1+3\zeta+\zeta^2-\zeta^3\right), \quad \sigma_4=(\sigma_3)^*. \nonumber
\end{align}
This is obtained by applying the $E_8$ mass specialization \eqref{eq:Z5-tuning} to the $E_8$ Seiberg--Witten curve,  \eqref{eq:sw-e8-mass}, given in the 5-brane web coordinates. As expected, with the shift $U = u-60$, the Weierstrass normal form for \eqref{eq:tuning-E8-z5} agrees with the $\Z_5$ quotient curve \eqref{eq:Z5curve}.


\acknowledgments{We thank Hee-Cheol Kim and Kimyeong Lee for useful discussions. SK is supported by National Natural Science Foundation of China grant No. 12250610188 and also by the Brain Pool Program funded by the Ministry of Science and ICT through the National Research Foundation of Korea (RS-2025-25457100). SL is also supported by National Natural Science Foundation of China grant No. 12250610188. MS is supported in part by the National Natural Science Foundation of China (Grant No. 12475105). FY is supported by start-up research grant RK21121 from Huzhou Normal University. Assistance on many details in section~\ref{sec:Otherquotient} was provided by ChatGPT-6.}


\appendix
\section{Explicit maps for
\texorpdfstring{$\Z_3$ and $\Z_4$}{Z3 and Z4}
S-folds}
\label{app:n-to-1-map}

In this Appendix, we aim to demonstrate the same procedure for computing the Seiberg–Witten curves for the $\Z_3$ and $\Z_4$ S-folds by directly employing 3-to-1 and 4-to-1 maps, analogous to the approach used for the $\Z_2$ S-fold \cite{Kim:2026yeh}. 

\subsection{\texorpdfstring{3-to-1 map in $\Z_3$ S-fold}{3-to-1 map in Z3 S-fold}}

For the $\Z_3$ S-fold, $\Z_3$ transformation is given by \eqref{eq:twz3}. 
Under this transformation, we find a pair of $\Z_3$ invariant functions whose defining equations have  three solutions:
\begin{align}\label{eq:3to1map}
    X&=\frac{w-1}{tw-w}+\frac{tw-w}{1-tw}+\frac{1-tw}{w-1}\ ,\notag\\
    Y&=\frac{tw+1}{1-tw}+\frac{t+1}{t-1}+\frac{w+1}{w-1}\ .
\end{align}
Although this system satisfies the desired properties (i.e., being 3-to-1 and $\Z_3$ invariant), the three solutions are too complicated to be used  directly to determine the Seiberg–Witten curves for the $\Z_3$ S-fold. 

However, our objective can still be realized through the application of the following coordinate transformation:
\begin{align}\label{eq:coortrans}
    T=\frac{1-tw}{w-1},\qquad W=\frac{1+t}{1-t}\ .
\end{align}
After substituting the coordinate transformation \eqref{eq:coortrans} into the 3-to-1 map \eqref{eq:3to1map}, the system of equations simplifies to:
\begin{align}\label{eq:XandY}
        X=\frac{-1-3T+T^3}{T+T^2},\qquad Y=\frac{-1+W}T+W+T(1+W)\ .
\end{align}
It is not difficult to find that \eqref{eq:XandY} is indeed $\Z_3$ invariant under these new variables $(T,W)$:
\begin{align}
    (T,W)\to\bigg(-\frac{T+1}{T},-W-T(1+W)\bigg)\to\bigg(-\frac1{1+T}\ ,\frac{1-(T+1)W}{T}\bigg)\to(T,W)\ ,
\end{align}
which is related to \eqref{eq:twz3} by \eqref{eq:coortrans}.

The selection of one of the three solutions to
the system \eqref{eq:XandY}, followed by its substitution into the $\Z_3$-invariant Seiberg–Witten curves, reproduces the Seiberg–Witten curves shown in section \ref{sec:Z3Sfold}.

\subsection{\texorpdfstring{4-to-1 map in $\Z_4$ S-fold}{4-to-1 map in Z4 S-fold}}
For the $\Z_4$ S-fold, under this transformation given by \eqref{eq:Z4trans}, we find a pair of $\Z_4$ invariant functions whose defining equations have four solutions:
\begin{align}\label{eq:Z4XY}
    X&=\frac{(t-w)^2+(-1+t w)^2}{(-1+t+w-t w)^2}  \ , \notag\\
    Y&=\frac{t^2 w^2-t^2-w^2+1}{(t-w) (t w-1)}\ .
\end{align}
Similar to the $\Z_3$ S-fold,
the direct solution of this system results in the generation of four highly complicated solutions, which prevents their utilization in the determination of the Seiberg–Witten curve for the $\Z_4$ S-fold.

For the $\Z_4$ case, four computationally tractable solutions can still be obtained through the following coordinate transformation:
\begin{align}\label{eq:TWtrans}
        T=\frac{t}{1-t}\ ,\qquad W=\frac w{1-w}\ .
\end{align}
Under this coordinate transformation, the system of two equations in \eqref{eq:Z4XY} simplifies to
\begin{align}\label{eq:XYnew}
    X&=1+2 T+2 T^2+2 W+2 W^2,\notag\\
    Y&=-\frac{(1+2 T) (1+2 W)}{(T-W) (1+T+W)}\ .
\end{align}
It can be confirmed that \eqref{eq:XYnew} admits precisely four solutions and exhibits $\Z_4$ invariance under
\begin{align}
        T\to W\to-(T+1)\to-(W+1)\to T\ ,
\end{align}
which is derived from \eqref{eq:Z4trans} by applying \eqref{eq:TWtrans}. 

We here explicitly present one of the four solutions to \eqref{eq:XYnew}:
\begin{align}\label{eq:solZ4}
    T=-\frac{1}{2} \sqrt{X-\frac{2 X}{\sqrt{Y^2+4}}}-\frac{1}{2},\qquad W=-\frac{X Y}{2
   \sqrt{Y^2+4} \sqrt{X-\frac{2 X}{\sqrt{Y^2+4}}}}-\frac{1}{2}\ .
\end{align}
Without loss of generality, substitution of this solution \eqref{eq:solZ4} into the $\Z_4$-invariant Seiberg–Witten curves yields the Seiberg–Witten curves shown in section \ref{sec:Z4Sfold}.

\section{SW geometry, Kodaira fiber configuration and Mordell--Weil group}\label{app:SW-geometry}

In this appendix, we briefly review some necessary mathematical background: rational elliptic surfaces (Seiberg--Witten geometry)~\cite{Furrer:2024zzu,Closset:2021lhd,Closset:2023pmc} and Kodaira fiber classification listed in~\cite{Furrer:2024zzu}, Mordell--Weil group~\cite{Closset:2021lhd,Oguiso:1991mw,schuttshioda} and cyclic isogeny~\cite{Rosen:2016extensions}. This background serves two purposes:
\begin{itemize}
    \item To determine the flavor symmetry of $\mathbb Z_{n}$ S-fold curve in section~\ref{sec:SW-for-zn}, we first need to analyze the Kodaira fiber configuration after expressing the Seiberg--Witten curve in the Weierstrass normal form, and then determine the corresponding Mordell--Weil (MW) group and lattice, by using the main theorem in \cite{Oguiso:1991mw}, which encode the flavor root system of the theory \cite{Caorsi:2018ahl, Xie:2022lcm}; 
    \item To investigate  the relation between the quotient Seiberg--Witten curves of $\mathbb Z_n$ S-folds and their parent curves in section~\ref{sec:346-isogeny},  we use cyclic isogenies. 
\end{itemize}

\subsection{Seiberg--Witten geometry}

Note that throughout this work, we focus on 5d $\mathcal N=1$ rank-1 SCFTs. After a suitable change of variables, the Seiberg--Witten curve can thus be expressed in the Weierstrass normal form
\begin{align}
   E:\quad y^2=4x^3-g_2(u,M_i)x-g_3(u,M_i)\ .
\end{align}
Here $M_i$ are the exponentiated mass parameters of the theory. At fixed $M_i$, the Coulomb branch $\mathcal B$ is the complex $u$-plane, $\mathcal B\simeq\mathbb C$, with the Coulomb modulus $u$ as its global coordinate. The Weierstrass coefficients $g_2$ and $g_3$ are functions of Coulomb modulus $u$ as well as $M_i$. 
When the mass parameters are fixed, the Seiberg--Witten curve $E$ fibered over the complex one-dimensional Coulomb branch $\mathcal{B}$ defines a rational elliptic surface (RES) over the field $\mathbb C$:
\begin{align}\label{eq:def-SW-geo}
    E\to\mathcal E\xrightarrow{\pi}\mathcal B,
\end{align}
where $\mathcal E$ is the total space of Seiberg--Witten curves over the Coulomb branch, including the singularities~\cite{Furrer:2024zzu,schuttshioda}, and $\pi:\mathcal E\to\mathcal B$ is a genus-one fibration, with $F_u:=\pi^{-1}(u)$. For simplicity, we  call $\mathcal E$ the rational elliptic surface. \eqref{eq:def-SW-geo} is also referred to as the Seiberg--Witten geometry. 
Adding the point $u=\infty$ to the complex $u$-plane $\mathcal B$ gives the compactified base $\mathbb{P}^1_{\mathbb C}$. For simplicity, we denote it by $\mathbb P^1$ without ambiguity. 

We denote by $P$ a section of fibration $\pi:\mathcal E\to\mathbb P^1$. In this work, we consider non-isotrivial RESs. That is, each fibration $\pi$ has at least two non-isomorphic smooth fibers. By the Mordell--Weil theorem for function fields (the Lang–N\'eron theorem~\cite{Conrad:2006LangNeron}), the sections of each such fibration form a finitely generated Abelian group, denoted by $\mathrm{MW}(\mathcal E)$. The zero section denoted by $O$ is the identity of the group. As a result, the MW group admits the decomposition
\begin{align}
    \mathrm{MW}(\mathcal E)=\mathbb Z^r\oplus \mathrm{MW}(\mathcal E)_{\mathrm{tor}},
\end{align}
where $r$ is the rank of $\mathrm{MW}$ group and $\mathrm{MW}(\mathcal E)_{\mathrm{tor}}$ is the torsion subgroup of the MW group~\cite{Oguiso:1991mw,Closset:2021lhd}. The quotient $\mathrm{MW}(\mathcal E)/\mathrm{MW}(\mathcal E)_{\mathrm{tor}}$, equipped with the canonical height pairing, is called the Mordell--Weil lattice~\cite{schuttshioda}. In particular, a section $P\in\mathrm{MW}(\mathcal E)$ is an $n$-torsion section if $nP=P+P\cdots+P=O$.

For a given Seiberg--Witten curve $E$, the discriminant of the Weierstrass form is defined~via
\begin{align}
    \Delta(u):=g_2(u)^3-27g_3(u)^2\ .
\end{align}
At finite values of $u$ where the Weierstrass coefficients $g_2,g_3$ are regular, 
the fiber $F_u\cong E_u$ is smooth if $\Delta(u)\neq0$, while the fiber $F_u$ is singular if $\Delta(u)=0$. Moreover, the fiber at $u=\infty$ is denoted by $F_\infty=\pi^{-1}(\infty)$. For all the 5d theories, the elliptic fibration $\pi:\mathcal{E}\to\mathbb P^1$ is singular at $u=\infty$ \cite{Closset:2021lhd}.  The possible singularities of the Seiberg--Witten geometry are characterized by the Kodaira classification of singular fibers, in terms of the vanishing orders of $g_2,g_3$, and $\Delta(u)$. The different types of fibers are listed in \cite{Furrer:2024zzu}.

Using the Kodaira fiber classification, we find that, for generic values of the mass parameters, the $\mathbb Z_{3}$ S-fold curves in section~\ref{sec:Z3Sfold} have Kodaira fiber configurations $(I_3;I_3^3),(I_2;I_3^3I_1)$ and $(I_1;I_3^3I_1^2)$, respectively; the two $\mathbb Z_4$ S-fold curves in section~\ref{sec:Z4Sfold} have Kodaira fiber configurations $(I_2;I_4^2I_2)$ and $(I_1;I_4^2I_2I_1)$, respectively; the $\mathbb Z_6$ S-fold curve in section~\ref{sec:Z6Sfold} has Kodaira fiber configuration $(I_1;I_6I_3I_2)$. 
Then, we use the Oguiso-Shioda classification~\cite{Oguiso:1991mw} (i.e., the main theorem) to determine the Mordell--Weil groups and lattices of the associated rational elliptic surfaces. Finally, extracting the flavor root systems using the prescription of~\cite{Caorsi:2018ahl,Xie:2022lcm}, we find: 
\begin{itemize}
\item for the local $\mathbb P^2/\mathbb Z_3$ theory with $(I_3;I_3^3)$, $G_F=\varnothing$ and  $\mathrm{MW}(\mathcal E)_{\mathrm{tor}}=\mathbb Z_3\oplus\mathbb{Z}_3$;
    \item for $\SU(2)+2\bf{F}/\mathbb{Z}_3$ theory with  $(I_2;I_3^3I_1)$, $G_F=\SU(2)$ and  $\mathrm{MW}(\mathcal E)_{\mathrm{tor}}=\mathbb Z_3$;
    \item for $\SU(2)+5\bf{F}/\mathbb{Z}_3$ theory with  $(I_1;I_3^3I_1^2)$, $G_F=G_2$ and $\mathrm{MW}(\mathcal E)_{\mathrm{\mathrm{tor}}}=\mathbb Z_3$;
    \item for $\SU(2)_0/\mathbb Z_4$ theory with $(I_2;I_4^2I_2)$,  $G_F=\varnothing$ and $\mathrm{MW}_{\mathrm{tor}}(\mathcal E)=\mathbb Z_4\oplus\mathbb Z_2$;
    \item for $\SU(2)+4\bf{F}/\mathbb{Z}_4$ theory with $(I_1;I_4^2I_2I_1)$, $G_F=\SU(2)$ and $\mathrm{MW}_{\mathrm{tor}}(\mathcal E)=\mathbb Z_4$;
    \item for $\SU(2)+2\bf{F}/\mathbb{Z}_6$ theory with $(I_1;I_6I_3I_2)$, $G_F=\varnothing$ and $\mathrm{MW}_{\mathrm{tor}}(\mathcal E)=\mathbb Z_6$;
\end{itemize}
where $G_F$ is the flavor symmetry  ($G_F=\varnothing$ means no flavor symmetry) and $\mathrm{MW}(\mathcal E)_{\mathrm{tor}}$ denotes the torsion subgroup of its corresponding elliptic surface, which is related to the 1-form symmetry of the theory~\cite{Xie:2022lcm,Closset:2023pmc,Furrer:2024zzu}.

\subsection{Mordell-Weil group action}

Regarding $u$ as a variable, the same Weierstrass equation defines the generic fiber $E/K:y^2=4x^3-g_2x-g_3$, where $K=\mathbb C(u)$ is the field of rational functions, and $g_2,g_3\in K$.
Define its set of $K$-rational points by
\begin{align}
    E(K):=\{(x,y)\in K^2\,|\, y^2=4x^3-g_2x-g_3\}\cup\{O\}\ ,
\end{align}
where $O$ is the point at infinity.
The elliptic curve group law \cite{schuttshioda, Schuett:2009pej, Closset:2021lhd, Silverman2009arithmetic} makes $E(K)$ an Abelian group with identity $O$ and inverse $-(x,y)=(x,-y)$. For $P_1=(x_1,y_1)$ and $P_2=(x_2,y_2)$ with $P_2\neq-P_1$, the $x$-coordinate of their sum $P_3=P_1+P_2$ is given by
\begin{align}\label{eq:add-law-x}
    x_3&=\lambda^2-x_1-x_2\ ,
\end{align}
where
\begin{align}
    \lambda=\begin{cases}
       \displaystyle\frac{y_2-y_1}{2(x_2-x_1)},\quad x_1\neq x_2\ ,\\[8pt]
         \displaystyle\frac{12x_1^2-g_2}{4y_1},\quad P_1=P_2,\, y_1\neq0 \ ,
    \end{cases}
\end{align}
while the $y$-coordinate of the sum is given by
\begin{align}
    y_3=-2\lambda x_3-2\nu\ ,
\end{align}
where
\begin{align}\label{eq:add-law-y}
    \nu=\begin{cases}
       \displaystyle \frac{y_1x_2-y_2x_1}{2(x_2-x_1)},\quad x_1\neq x_2\ ,\\[8pt]
       \displaystyle\frac{-4x_1^3-g_2x_1-2g_3}{4y_1},\quad P_1=P_2,\, y_1\neq0\ .
    \end{cases}
\end{align}
The group of rational points $E(K)$ is traditionally called the Mordell--Weil group of $E/K$, denoted by $E(K)$. The correspondence between sections of $\pi:\mathcal E\to\mathbb P^1$ and $K$-rational points of the generic fiber gives a natural group isomorphism~\cite{schuttshioda}:
\begin{align}
    \mathrm{MW}(\mathcal E)\cong E(K).
\end{align} 
 Thus, we identify these two groups, and use the same symbol $P$ for a $K$-rational point and its corresponding section.

 \subsection{Cyclic \texorpdfstring{$n$}{n}-isogeny}

Let $P\in\mathrm{MW}(\mathcal E)$ be a torsion section of exact order $n\in\mathbb N$ of the elliptic fibration $\pi:\mathcal E\to\mathbb P^1$. It generates a cyclic subgroup $G:=\langle P\rangle$ of the torsion subgroup as follows
\begin{align}
    \langle P\rangle=\{O,P,\dots,(n-1)P\}\leq\mathrm{MW}(\mathcal E)_{\mathrm{tor}}\leq\mathrm{MW}(\mathcal E)\ .
\end{align}

 Each section $Q\in\mathrm{MW}(\mathcal E)$ defines a translation automorphism $t_Q$ of $\mathcal E$, acting on each smooth fiber as
 \begin{align}
     t_Q:R\mapsto R+Q.
 \end{align}
 The assignment $Q\mapsto t_Q$ gives a group embedding
 \begin{equation}
     t:\begin{cases}
         \mathrm{MW}(\mathcal E)&\hookrightarrow\mathrm{Aut}(\mathcal E),\\
         Q&\mapsto t_Q
     \end{cases}
 \end{equation}
 \cite{Furrer:2024zzu}.  Since $t_{kP}=t_P^k$, this embedding restricts to a group isomorphism $G=\langle P\rangle\cong\langle t_P\rangle\leq\text{Aut}(\mathcal E)$. 

 On the generic fiber, taking the quotient by this translation action gives the elliptic curve
\begin{align}
    \widetilde E=E/G=E/\langle t_P\rangle\ .
\end{align}
The quotient map 
$\varphi_n:E\longrightarrow\widetilde E$ 
is an $n$-to-1 homomorphism, mapping the identity point $O_E$ to $O_{\widetilde E}$. Since $G$ is cyclic of order $n$, this map is a cyclic $n$-isogeny.

\section{5-brane web realization of \texorpdfstring{$\mathbb {Z}_{3,4,6}$}{Z346} S-fold curves via mass tuning of \texorpdfstring{$\E_{6,7,8}$ SCFTs}{E678}}\label{app:Z236-curve-web}

In this appendix, we present the Seiberg–Witten curves for the $\mathbb{Z}_3$, $\mathbb{Z}_4$, and $\mathbb{Z}_6$ S-folds of rank-1 SCFTs via 5-brane webs, extending the framework developed for the $\mathbb{Z}_2$ S-fold~\cite{Kim:2026yeh}. This complements the Weierstrass normal form analysis in the main text. While the main text starts from the Weierstrass normal form of the $\E_{6,7,8}$ SCFTs and imposes the special tuning that realizes the $\Z_{3,4,6}$ S-fold, here we start from the Seiberg–Witten curves obtained directly from the corresponding 5-brane webs and impose the same special tuning. Although the two approaches are both based on the special tuning of the $\E_{6,7,8}$ SCFTs, they are different in the language used to describe the curves.

In tuning the mass parameters to special values, it is essential to make clear the convention and the physical meaning of the parameters.  Since the special tuning is described in terms of the physical masses, one needs to know how the coefficients of the Seiberg–Witten curves are expressed in terms of the physical mass parameters, as we did in~\cite{Kim:2026yeh}, in order to determine the result of the tuning.

For the $\E_{6,7,8}$ Seiberg–Witten curves, the polynomial structure has been known from previous studies \cite{Bao:2011rc, Bao:2013pwa, Kim:2014nqa}. In these studies, however, the curves are often expressed in terms of parameters in which a different subgroup of the $\E_{6,7,8}$ global symmetry is manifest, and the relation to the physical mass parameters, which correspond to the $\SO(2N_f)$ flavor symmetry, is not entirely straightforward%
\footnote{See, however, the discussion on the physical parameters in Appendix A of \cite{Bao:2011rc}.}.  
In this appendix, we provide the explicit form of the coefficients in terms of the physical mass parameters in the 5-brane web parametrization, together with the result of substituting the special tuning. 

\subsection{\texorpdfstring{$\Z_3$ S-fold}{Z3 S-fold}}

\underline{$\E_6$ to local $\mathbb{P}^2/\Z_3$:}
For the $\E_6$ SCFT,  under the physical mass parametrization, the $\E_6$ Seiberg–Witten curve reads
\begin{align}\label{eq:sw-e6-mass}
    F_{\E_6}(t,w)&=(w-M_1)(w-M_2)(w-\mathcal{M})+t^2\big(w-M_3\big)(w-M_4)(w-M_5)\notag\\
    &\quad+\big(M_1M_2M_3M_4M_5+\mathcal{M}\big)t
    +Utw+tw^2\Big(\sum_{i=1}^5M_i+\mathcal{M}\Big)-2tw^3=0\ .
\end{align}
where
\begin{align}\label{eq:M6}
    \mathcal M=M_0\sqrt{M_1M_2M_3M_4M_5} \ .
\end{align}
Now we impose the special tuning \eqref{eq:tuningSU2-5F} on \eqref{eq:sw-e6-mass} and \eqref{eq:M6} and use this square-root branch $\sqrt{\omega_3}=e^{\pi i/3}$ so that
\begin{align}
    \omega_3^2+\omega_3+1=0,\qquad (\sqrt{\omega_3})^{3}=-1 \ .
\end{align}
Then we obtain $\mathcal M=1$ and
\begin{align}\label{eq:sw-e6-tuned-explicit-z3}
    F_{\E_6}^{\text{tuned}}(t,w)
    &=(w-1)^3+t^2(w-\omega_3)^3+t(2+Uw-3\omega_3^2w^2-2w^3)=0\ .
\end{align}
This is the Seiberg–Witten curve for the local $\mathbb{P}^2/\Z_3$ theory obtained from the 5-brane web of $\E_6$ SCFT. 
To bring this curve into the Weierstrass normal form, from \eqref{eq:sw-e6-tuned-explicit-z3}
we find
\begin{align}
    t^2=(-4U-15\omega_3)w^4+(-6\omega_3^2U-36)w^3+(U^2-36\omega_3^2)w^2+4(U-3\omega_3)w \ ,
\end{align}
where, reusing $t$ for the new
coordinate, we made the replacement
\begin{align}
t
&\to
\frac{t
-\left(2+Uw-3\omega_3^2w^2-2w^3\right)
}{
2(w-\omega_3)^3
}\, .
\end{align}
Then 
its Weierstrass normal form, as expected, agrees with \eqref{eq:localp2z3}, 
using the coordinate transformation
\begin{align}
    x=-\frac{U-3\omega_3}w-\frac{U^2-36\omega^2_3}{12},\qquad y=\frac{i(U-3\omega_3)t}{w^2}\ .
\end{align}

\noindent\underline{$\E_7$ to $\SU(2)+2\F/\Z_3$:}
For the $\E_7$ SCFT,  under the physical mass parametrization, the $\E_7$ Seiberg–Witten curve reads
\begin{align}\label{eq:sw-e7-mass-z3}
    F_{\E_7}(t,w)&=(w-M_1)(w-M_2)(w-M_5)(w-\mathcal{M})+t^2(w-M_3)(w-M_4)(w-M_6)(w-M_7)\notag\\
    &\quad-2tw^4+tw^3\Big(\sum_{i=1}^7M_i+\mathcal{M}\Big)+Utw^2+c_{11}tw-2M_7\mathcal{M}t\ ,
\end{align}
where
\begin{align}\label{eq:c11-e7-z3}
&\mathcal MM_7=M_1M_2M_3M_4M_5M_6\ ,\notag\\
&\mathcal M=M_0\sqrt{M_1M_2M_3M_4M_5M_6}\ ,\notag\\
    &c_{11}=\frac{\mathcal M}{M_3M_4M_6}(M_7 M_3 M_4+M_7 M_6 M_4+M_3 M_6 M_4+M_7 M_3 M_6)\notag\\
    &\quad\quad+\frac{M_7}{M_1M_2M_5}(M_1 M_2 M_5+M_1 \mathcal M M_5+M_2 \mathcal M M_5+M_1 M_2\mathcal M)\ .
\end{align}
Imposing the tuning condition \eqref{eq:tuning-z3-2f} on \eqref{eq:sw-e7-mass-z3} and \eqref{eq:c11-e7-z3}, we obtain 
\begin{align}
    \mathcal M=\M,\qquad M_7=1,\qquad c_{11}^{\text{tuned}}=2-3\omega_3\M \,,
\end{align}
and further
\begin{align}
    F_{\E_7}^{\text{tuned}}&=(w-1)^2(w-\omega_3)(w-\M)+t^2(w-\omega_3)^2(w-\M)(w-1)-2tw^4\notag\\
    &\quad +tw^3(3\omega_3+3+2\M)+Utw^2+(2-3\omega_3\M)tw-2\M t=0 \,.
\end{align}
Obviously, bringing this curve into the Weierstrass normal form, it agrees with \eqref{eq:E7z3curve} as expected.

\vspace{0.3cm}
\noindent\underline{$\E_8$ to $\SU(2)+5\F/\Z_3$:}
For the $\E_8$ SCFT,  under the physical mass parametrization, the $\E_8$ Seiberg–Witten curve reads
\begin{align}\label{eq:sw-e8-mass}
    F_{\E_8}(t,w)=-t^{3}Q(w)^2+t^2Q(w)S(w)+tR(w)+P(w)=0\ ,
\end{align}
where
\begin{align}\label{eq:e8-coeffts}
    Q(w)&=\prod_{i=5}^7(w-M_i)\ ,\notag\\
    S(w)&=3w^3-\sigma_1w^2+\sigma_2w-3M_8\ ,\notag\\
    R(w)&=-3w^6+2\sigma_1 w^5-\sigma_3 w^4-U w^3-\sigma_4 w^2+2\sigma_5 w-3M_8^2\ ,\notag\\
    P(w)&=(w-\mathcal{M})(w-M_8)\prod_{i=1}^4(w-M_i)\ ,\notag\\
    \mathcal{M}&=M_0^2\ ,\notag\\ M_8&=M_0\sqrt{M_1M_2M_3M_4M_5M_6M_7} \ ,
\end{align}
with
\begin{align}\label{eq:e8-sigma}
\sigma_1&=\sum_{i=1}^8M_i+\mathcal{M}\ ,\notag\\
    \sigma_2&=1+M_8\Big(\frac{1}{\mathcal M}+\sum_{i=1}^7\frac1{M_i}\Big)\ ,\notag\\
    \sigma_3&=1+\sum_{1\leq i<j\leq 7}M_iM_j
+\mathcal{M}\sum_{i=1}^{7}M_i
+M_8\bigg[
\mathcal{M}+\frac{1}{\mathcal{M}}
+\sum_{i=1}^{7}\Big(M_i+\frac{1}{M_i}\Big)
\bigg]\ ,\notag\\
    \sigma_4&=\bigg(\prod_{i=1}^{7}M_i\bigg)
\bigg[
\mathcal{M}+\sum_{i=1}^{7}\frac{1}{M_i}
+\mathcal{M}\sum_{1\leq i<j\leq 7}\frac{1}{M_iM_j}
\bigg]
+M_8\bigg[
\mathcal{M}+\frac{1}{\mathcal{M}}
+\sum_{i=1}^{7}\Big(M_i+\frac{1}{M_i}\Big)
\bigg]\ ,\notag\\
    \sigma_5&=M_8+\bigg(\prod_{i=1}^{7}M_i\bigg)
\bigg(1+\mathcal{M}\sum_{i=1}^{7}\frac{1}{M_i}\bigg)\ .
\end{align}
Imposing the tuning condition \eqref{eq:tuning-e8-z3} on \eqref{eq:sw-e8-mass}, we find $\mathcal M=\M_2$, $M_8=1$, and
\begin{align}\label{eq:Ftuned-e8-z3}
    F_{\text{tuned}}^{\E_8}=-t^3Q_{\text{tuned}}(w)^2+t^2Q_{\text{tuned}}(w)S_{\text{tuned}}(w)+tR_{\text{tuned}}(w)+P_{\text{tuned}}(w)=0,
\end{align}
where
\begin{align}
    Q_{\text{tuned}}(w)&=(w-\omega_3)\bigg(w-\frac{1}{\M_1\M_2}\bigg)(w-\M_1)\notag\\
    S_{\text{tuned}}(w)&=3w^3-\sigma_1^{\text{tuned}}w^2+\sigma_2^{\text{tuned}}w-3\ ,\notag\\
    R_{\text{tuned}}(w)&=-3w^6+2\sigma^{\text{tuned}}_1 w^5-\sigma_3^{\text{tuned}} w^4-U w^3-\sigma_4^{\text{tuned}} w^2+2\sigma_5^{\text{tuned}} w-3\ ,\notag\\
    P_{\text{tuned}}(w)&=(w-\M_2)(w-1)^3(w-\omega_3)^2\ ,
\end{align}
with
\begin{align}
\sigma_1^{\text{tuned}}&=-3\omega_3^2+\chi_1^{\SU(3)}(\M)\ ,\notag\\
\sigma_2^{\text{tuned}}&=-3\omega_3+\chi_2^{\SU(3)}(\M)=\sigma_5^{\text{tuned}}\ ,\notag\\
\sigma_3^{\text{tuned}}&=3\omega_3-3\omega_3^2\chi_1^{\SU(3)}(\M)+2\chi_2^{\SU(3)}(\M)\ ,\notag\\
\sigma_4^{\text{tuned}}&=3\omega_3^2+2\chi_1^{\SU(3)}(\M)-3\omega_3\chi_2^{\SU(3)}(\M)\ .
\end{align}
Bringing \eqref{eq:Ftuned-e8-z3} into the Weierstrass normal form by using the following coordinate transformation
\begin{align}
    x&=\frac{\xi_1\xi_5w(w-1)}{H(t,w)}-\frac{\xi_2^2-4\xi_1\xi_3}{12} \ , \notag\\
    y&=i\xi_1\xi_5\Bigg(\xi_4+\frac{w(2\xi_1+\xi_2(w-1))}{H(t,w)}\Bigg)\ ,
\end{align}
we recover \eqref{eq:E8z3curve},
where
\begin{align}
    \xi_1&=(u-60)-3\omega_3^2\chi_1^{\SU(3)}(\M)-3\omega_3\chi_2^{\SU(3)}(\M)-3 \ ,\notag\\
    \xi_2&=6 \chi_1^{\SU(3)}(\M) \omega _3+6 \chi_2^{\SU(3)}(\M)-18 \omega _3-12+(u-60)\ ,\notag\\
    \xi_3&=3(\omega_3-1)(\chi_1^{\SU(3)}(\M)-3)\ ,\notag\\
    \xi_4&=\chi_1^{\SU(3)}(\M)-\chi_2^{\SU(3)}(\M)+3+6\omega_3\ ,\notag\\
    \xi_5&=(u-60)-3\omega_3\chi_1^{\SU(3)}(\M)-3\omega_3^2\chi_2^{\SU(3)}(\M)-3\ ,
\end{align}
and
\begin{align}
    H(t,w):=t Q_{\text{tuned}}(w)-(w-1)^3\ ,\quad U=u-60 \ .
\end{align}

\subsection{\texorpdfstring{$\Z_4$}{Z4} S-fold}

\noindent{\underline{$\E_7$ to $\SU(2)_0/\Z_4$}:} Imposing the tuning condition \eqref{eq:E7-tuning-M} on the $\E_7$ SCFT curve \eqref{eq:sw-e7-mass-z3}, we obtain
\begin{align}
    \mathcal M=\omega_4,\qquad M_7=-\omega_4,\qquad c_{11}^{\text{tuned}}=0\ ,
\end{align}
and
\begin{align}
    F_{\E_7}^{\text{tuned}}=(w-1)^2(w-\omega_4)^2+t^2(w+1)^2(w+\omega_4)^2+t(-2w^4+Uw^2-2)=0\ .
\end{align}
Obviously, bringing this curve into the Weierstrass normal form, it agrees with \eqref{eq:E7z4curve} as expected.

\vspace{0.3cm}
\noindent{\underline{$\E_8$ to $\SU(2)+4\F/\Z_4$}:} Imposing the tuning condition \eqref{eq:tuning-e8-z4} on the $\E_8$ SCFT curve \eqref{eq:sw-e8-mass}, we obtain 
\begin{align}
    \mathcal M=\omega_4\M,\qquad M_8=\M\ ,
\end{align}
and further
\begin{align}\label{eq:Fe8-z4}
     F_{\text{tuned}}^{\E_8}=-t^3Q_{\text{tuned}}(w)^2+t^2Q_{\text{tuned}}(w)S_{\text{tuned}}(w)+tR_{\text{tuned}}(w)+P_{\text{tuned}}(w)=0,
\end{align}
where
\begin{align}
    Q_{\text{tuned}}(w)&=(w^2+1)(w+\omega_4\M)\ ,\notag\\
    S_{\text{tuned}}(w)&=3w^3-\M w^2+w-3\M\ ,\notag\\
    R_{\text{tuned}}(w)&=-3w^6+2\M w^5-\M^2w^4-Uw^3-w^2+2\M w-3\M^2\ ,\notag\\
    P_{\text{tuned}}(w)&=(w-\omega_4\M)(w-\M)(w^2-1)^2\ .
\end{align}
By introducing the following coordinate transformation
\begin{align}
    x&=\frac{\zeta_1\zeta_2(w-\M)}{H(t,w)\M^2}-\frac{\zeta_3}{12\M^2}\ ,\notag\\
    y&=\frac{i\zeta_1\zeta_2\big(-2(\M^2-1)H^2(t,w)+\zeta_4 H(t,w)+2\zeta_2(\M-1)(w-1)(w-\M)\big)}{H^2(t,w)\M^3}\ ,
\end{align}
where
\begin{align}
    \zeta_1&=U+4\M^2-4\M+4\ ,\notag\\
    \zeta_2&=U-4\M^2-4\M-4\ ,\notag\\
    \zeta_3&=U^2-72\M U-80\M^4-144\M^2-80\ ,\notag\\
    \zeta_4&=(U+4\M^3+8\M^2-8)w+\M U-8\M^3+8\M+4\ ,\notag\\
    H(t,w)&=\frac{tQ_{\text{tuned}}-(w-1)^2(w-\M)}w\ ,\notag\\
    U&=(u-60)\M\ ,
\end{align}
\eqref{eq:Fe8-z4} can be exactly transformed into \eqref{eq:E8z4curve}, as expected.

\subsection{\texorpdfstring{$\Z_6$}{Z6} S-fold}

\noindent{\underline{$\E_8$ to $\SU(2)+2\F/\Z_6$}:} 
Imposing the tuning condition \eqref{eq:E8-tuning-M} on the $\E_8$ SCFT curve \eqref{eq:sw-e8-mass}, we choose
\begin{align}
    \sqrt{M_1M_2M_3M_4M_5M_6M_7}=-\omega_6^2\ .
\end{align}
Then we find
\begin{align}
    M_8=1,\qquad \mathcal M=\omega_6^2\ ,
\end{align}
and further the tuned $\E_8$ curve as
\begin{align}\label{eq:tuning-E8-z6}
     F_{\text{tuned}}^{\E_8}=-t^3Q_{\text{tuned}}(w)^2+t^2Q_{\text{tuned}}(w)S_{\text{tuned}}(w)+tR_{\text{tuned}}(w)+P_{\text{tuned}}(w)=0,
\end{align}
where
\begin{align}
    Q_{\text{tuned}}(w)&=(w+1)(w-\omega_6^2)^2\ ,\notag\\
    S_{\text{tuned}}(w)&=3w^3-(2+3\omega_6^2)w^2+(2+3\omega_6^4)w-3\ ,\notag\\
    R_{\text{tuned}}(w)&=-3w^6+2(2+3\omega_6^2)w^5+5w^4-Uw^3+5w^2+2(2+3\omega_6^4)w-3\ ,\notag\\
    P_{\text{tuned}}(w)&=(w-\omega_6^2)(w-1)^4(w+1)\ .
\end{align}
By introducing the following coordinate transformation
\begin{align}
    x&=\frac{(u-66)^2w(w-1)}{H(t,w)}-\frac{u^2-204u+8676}{12}\ ,\notag\\
    y&=i(u-66)^2\Bigg(3(\omega_6^2-\omega_6^4)+\frac{w(2(u-66)+(u-78-24\omega_6^2)(w-1))}{H(t,w)}\Bigg)\ ,
\end{align}
where
\begin{align}
    H(t,w)=tQ_{\text{tuned}}(w)-(w-1)^3\ ,\quad U=u-60\ ,
\end{align}
as expected,
\eqref{eq:tuning-E8-z6} can be exactly transformed into \eqref{eq:E8z6curve}.

\section{\texorpdfstring{$E_N$}{EN} embeddings and characters}\label{app:rel-fug}

In this Appendix, we discuss the relation between the fugacities corresponding to different subgroups of the $\E_6$, $\E_7$, and $\E_8$ groups. Then, we show that the tuning of masses \eqref{eq:tuningT3}, \eqref{eq:E7LMN-tuning}, and \eqref{eq:mass-tune-Z6}  are equivalent to  \eqref{eq:tuningSU2-5F}, \eqref{eq:E7-tuning-M}, and \eqref{eq:E8-tuning-M}, respectively. Finally, in section \ref{app:characters}, we write the characters of $\E_6,\E_7$ and $\E_8$ explicitly by decomposing them into one of the subgroups, respectively.

\subsection{\texorpdfstring{$\E_6$}{E6} group}\label{sec:rel-fug-1}
\begin{figure}
    \centering
\begin{tikzpicture}[
  x=10mm,
  y=10mm,
  dynkin edge/.style={
    draw,
    line width=1.5pt,
    line cap=round,
    line join=round
  },
  dynkin node/.style={
    circle,
    draw,
    fill=white,
    minimum size=3.5mm,
    inner sep=0pt,
    outer sep=0pt,
    line width=1.5pt
  },
  root label/.style={
    font=\scriptsize,
    inner sep=1pt,
    label distance=1.5mm
  },
  diagram title/.style={
    font=\normalsize,
    inner sep=1pt
  }
]

\coordinate (a6) at (-2.4,0);
\coordinate (a5) at (-1.2,0);
\coordinate (a4) at (0,0);
\coordinate (a3) at (1.2,0);
\coordinate (a1) at (2.4,0);
\coordinate (a2) at (0,1.05);
\coordinate (a0) at (0,2.10);

\draw[dynkin edge] (a6) -- (a5) -- (a4) -- (a3) -- (a1);
\draw[dynkin edge] (a4) -- (a2) -- (a0);

\node[dynkin node, label={[root label]below:$\alpha_1$}] at (a6) {};
\node[dynkin node, label={[root label]below:$\alpha_3$}] at (a5) {};
\node[dynkin node, label={[root label]below:$\alpha_4$}] at (a4) {};
\node[dynkin node, label={[root label]below:$\alpha_5$}] at (a3) {};
\node[dynkin node, label={[root label]below:$\alpha_6$}] at (a1) {};
\node[dynkin node, label={[root label]right:$\alpha_2$}] at (a2) {};
\node[dynkin node, label={[root label]right:$\alpha_0$}] at (a0) {};

\node[diagram title] at (0,2.65) {$\hat E_6$};

\end{tikzpicture}
\caption{Affine Dynkin diagram for $E_6$.}
\label{fig:E6-Dynkin}
\end{figure}
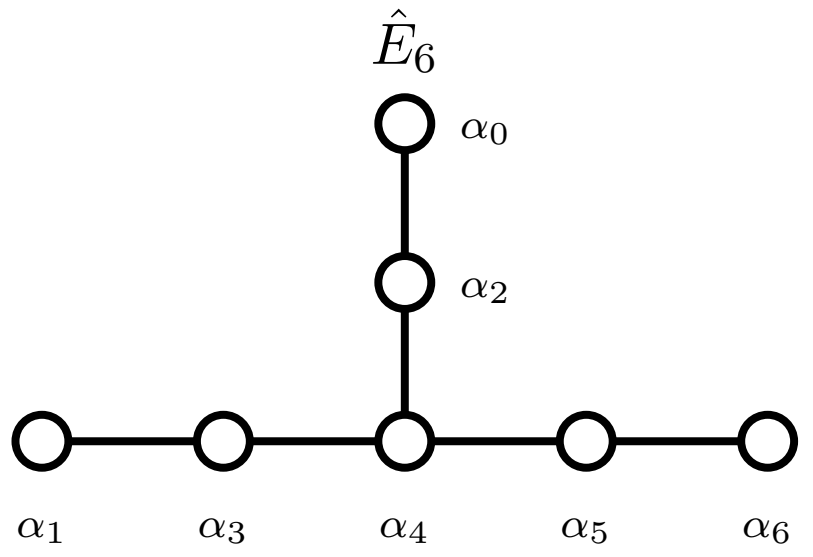

We start with affine Dynkin diagram for $\E_6$ as depicted in figure \ref{fig:E6-Dynkin}. The simple roots $\alpha_0, \alpha_1, \dots, \alpha_6$ are assigned to each node and the inner products of them give Cartan matrix $C_{ij}$
\begin{align}\label{eq:inner-prod}
(\alpha_i, \alpha_j) = C_{ij}
:= \left\{ 
\begin{array}{rl}
2 & \qquad \text{if } \quad i = j \\
-1 & \qquad \text{if } \quad i\text{-th node and } j\text{-th node are connected} \\
0 & \qquad \text{otherwise} \\
\end{array}
\right.
\end{align}

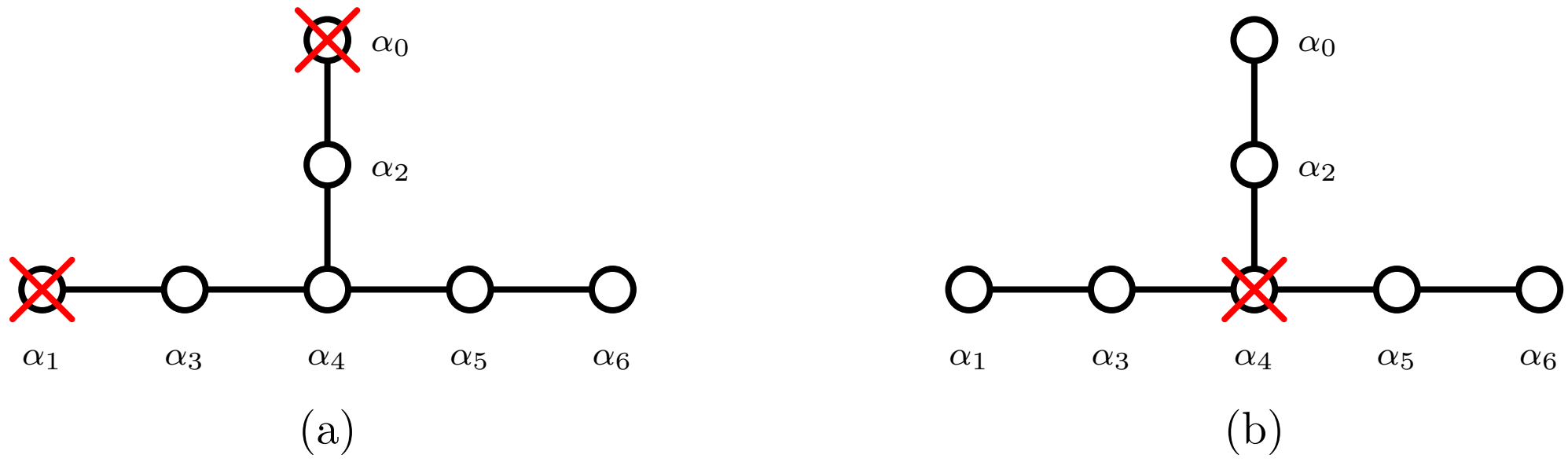
\begin{figure}
    \centering 
\begin{tikzpicture}[
  x=10mm,
  y=10mm,
  dynkin edge/.style={
    draw,
    line width=1.5pt,
    line cap=round,
    line join=round
  },
  dynkin node/.style={
    circle,
    draw,
    fill=white,
    minimum size=3.5mm,
    inner sep=0pt,
    outer sep=0pt,
    line width=1.5pt
  },
  root label/.style={
    font=\scriptsize,
    inner sep=1pt,
    label distance=1.5mm
  },
  diagram title/.style={
    font=\normalsize,
    inner sep=1pt
  }
]

\coordinate (L6) at (-2.4,0);
\coordinate (L5) at (-1.2,0);
\coordinate (L4) at (0,0);
\coordinate (L3) at (1.2,0);
\coordinate (L1) at (2.4,0);
\coordinate (L2) at (0,1.05);
\coordinate (L0) at (0,2.10);

\draw[dynkin edge] (L6) -- (L5) -- (L4) -- (L3) -- (L1);
\draw[dynkin edge] (L4) -- (L2) -- (L0);

\node[dynkin node, label={[root label]below:$\alpha_1$}] at (L6) {};
\node[dynkin node, label={[root label]below:$\alpha_3$}] at (L5) {};
\node[dynkin node, label={[root label]below:$\alpha_4$}] at (L4) {};
\node[dynkin node, label={[root label]below:$\alpha_5$}] at (L3) {};
\node[dynkin node, label={[root label]below:$\alpha_6$}] at (L1) {};
\node[dynkin node, label={[root label]right:$\alpha_2$}] at (L2) {};
\node[dynkin node, label={[root label]right:$\alpha_0$}] at (L0) {};

\draw[dynkin edge,red] ([shift={(-2.5mm,-2.5mm)}]L4) -- ([shift={(2.5mm,2.5mm)}]L4);
\draw[dynkin edge,red] ([shift={(-2.5mm,2.5mm)}]L4) -- ([shift={(2.5mm,-2.5mm)}]L4);

\node at(0,-1.2){(b)};

 \begin{scope}[xshift=-7.8cm]
  \coordinate (R6) at (-2.4,0);
  \coordinate (R5) at (-1.2,0);
  \coordinate (R4) at (0,0);
  \coordinate (R3) at (1.2,0);
  \coordinate (R1) at (2.4,0);
  \coordinate (R2) at (0,1.05);
  \coordinate (R0) at (0,2.10);

  \draw[dynkin edge] (R6) -- (R5) -- (R4) -- (R3) -- (R1);
  \draw[dynkin edge] (R4) -- (R2) -- (R0);

  \node[dynkin node, label={[root label]below:$\alpha_1$}] at (R6) {};
  \node[dynkin node, label={[root label]below:$\alpha_3$}] at (R5) {};
  \node[dynkin node, label={[root label]below:$\alpha_4$}] at (R4) {};
  \node[dynkin node, label={[root label]below:$\alpha_5$}] at (R3) {};
  \node[dynkin node, label={[root label]below:$\alpha_6$}] at (R1) {};
  \node[dynkin node, label={[root label]right:$\alpha_2$}] at (R2) {};
  \node[dynkin node, label={[root label]right:$\alpha_0$}] at (R0) {};

  \draw[dynkin edge,red] ([shift={(-2.5mm,-2.5mm)}]R6) -- ([shift={(2.5mm,2.5mm)}]R6);
  \draw[dynkin edge,red] ([shift={(-2.5mm,2.5mm)}]R6) -- ([shift={(2.5mm,-2.5mm)}]R6);

  \draw[dynkin edge,red] ([shift={(-2.5mm,-2.5mm)}]R0) -- ([shift={(2.5mm,2.5mm)}]R0);
  \draw[dynkin edge,red] ([shift={(-2.5mm,2.5mm)}]R0) -- ([shift={(2.5mm,-2.5mm)}]R0);

  \node at(0,-1.2){(a)};
\end{scope}

\end{tikzpicture}
\caption{Part (a) of this figure shows the first embedding $\E_6\supset\SO(10)\times\U(1)$, while part (b) depicts the second one $\E_6\supset\SU(3)^3$.}\label{fig:e6dyn-brok}
\end{figure}

These simple roots can be constructed explicitly in a way to satisfy \eqref{eq:inner-prod} using the orthonormal basis $\{ e_i \}$ ($i=0,1, \dots, 5$) of the Euclidean vector space containing the root system as 
\begin{align}\label{eq:S-root-SO10-U1}
\alpha_0 
&= - \frac12 ( \sqrt{3} e_0 + e_1 + e_2 + e_3 + e_4 + e_5),
\quad
\alpha_1 
= \frac12 (\sqrt{3} e_0 + e_1 - e_2 - e_3 - e_4 - e_5 )
\cr
\alpha_2
&= e_1 + e_2,
\quad
\alpha_3
= - e_1 + e_2, 
\quad
\alpha_4
= - e_2 + e_3, 
\quad
\alpha_5
= - e_3 + e_4,
\quad
\alpha_6
= - e_4 + e_5.
\end{align}
This is the expression that makes the structure of the $\SO(10)$ subgroup of $\E_6$ manifest. Indeed, if we remove the nodes corresponding to $\alpha_0$ and $\alpha_1$ as in (a) of figure \ref{fig:e6dyn-brok}, we find the Dynkin diagram for SO(10) with its standard expression for the simple roots. Based on the expressions for $\alpha_2, \dots, \alpha_6$, the expressions for $\alpha_0$ and $\alpha_1$ in \eqref{eq:S-root-SO10-U1} have been determined in such a way to satisfy \eqref{eq:inner-prod}. The root $\alpha_0$ corresponding to the extended node can be determined also by the relation
\begin{align}\label{eq:E6-marks-id}
\alpha_0 + \alpha_1 + 2 \alpha_2 + 2 \alpha_3 + 3 \alpha_4 + 2\alpha_5 + \alpha_6  = 0,
\end{align}
where the coefficients of the terms are given by the marks (Kac labels) associated with the corresponding nodes.

Fixing an orthonormal basis for the Euclidean space containing  the root system naturally determines a dual basis for the Cartan sub-algebra. Suppose we denote an element of the Cartan sub-algebra dual to $e_i$ as $H_{e_i}$, an arbitrary element of Cartan sub-algebra $\mathfrak{h}$ can be expressed as a linear combination of them as 
\begin{align}\label{eq:cartan-E6-LC}
\frac{2}{\sqrt{3}} m_0 H_{e_0} + \sum_{i=1}^5 m_i H_{e_i} \in \mathfrak{h}.
\end{align}
Here, the coefficients of this linear combination are identified as the instanton mass $m_0$ and the flavor masses $m_i$ ($i=1,\dots, 5$), respectively. The factor $2/\sqrt{3}$ in \eqref{eq:cartan-E6-LC} appears due to the normalization of the instanton mass $m_0$.
These masses are related to the fugacities $M_i$ of $\SO(10) \times \U(1)$, which is used, for example in \eqref{eq:tuningSU2-5F}, as 
\begin{align}\label{eq:M=expm}
M_i = e^{-m_i}.
\end{align}

In general, a root $\alpha$ can be seen as a linear map from the Cartan sub-algebra to their eigenvalues. That is, 
\begin{align}
[H, X_{\alpha}] = \alpha(H) X_{\alpha},
\end{align}
where $H$ is an element of Cartan subalgebra, $\alpha(H)$ and $X_{\alpha}$ are the eigenvalue and the eigenvector determined by the choice of the root $\alpha$, respectively. 
When the simple roots are expressed as a linear combination 
as in \eqref{eq:S-root-SO10-U1}, the eigenvalue of an element of the Cartan sub-algebra \eqref{eq:cartan-E6-LC} is given as
\begin{align}
\alpha_i = \sum_{j=0}^5 n_i^j e_j \quad \Rightarrow \quad
\alpha_i \left(  \frac{2}{\sqrt{3}} m_0 H_{e_0} + \sum_{j=1}^5 m_j H_{e_j} \right) =  \frac{2}{\sqrt{3}} n_i^0 m_0 + \sum_{j=1}^5 n_i^j m_j .
\end{align}
In other words, we can obtain this eigenvalue by replacing $e_i$ in \eqref{eq:S-root-SO10-U1} by $m_i$:
\begin{align}\label{eq:replace-em}
e_0 \to \frac{2}{\sqrt{3}} m_0, \qquad e_i \to m_i \quad (i=1,\dots 5).
\end{align}
This observation plays a key role to obtain a relation between different fugacities. 

The simple roots identical to \eqref{eq:S-root-SO10-U1} can be expressed differently by using another orthonormal basis $\{ f_i, g_i, h_i \}$, with $i=1,2,3$, satisfying
\begin{align}
(f_i,f_j) = (g_i,g_j) = (h_i,h_j) = \delta_{i,j},
\qquad
(f_i,g_j) = (g_i,h_j) = (h_i,f_j) = 0
\end{align}
as 
\begin{align}\label{eq:S-roots-SU33}
\alpha_0 &= - h_2 + h_3, \quad
\alpha_1 = - g_2 + g_3, \quad
\alpha_2 = - h_1 + h_2, \quad
\alpha_3 = - g_1 + g_2,
\cr
\alpha_4
&= \frac13 (2f_1 - f_2 - f_3 + 2g_1 - g_2 - g_3 + 2h_1 - h_2 - h_3 )\ ,
\cr
\alpha_5
&= -f_1 + f_2,
\quad
\alpha_6
= - f_2 + f_3,
\end{align}
In this expression, the root system is embedded in the 9-dimensional Euclidean vector space but spans only a six-dimensional subspace of it.

This expression makes the structure of the $\SU(3)^3$ subgroup of $E_6$ manifest. 
If we remove the node of $\alpha_4$ as in (b) of figure \ref{fig:e6dyn-brok}, we obtain the Dynkin diagram for $\SU(3)^3$ and their standard expression for their simple roots. Again, the expression of $\alpha_4$ in \eqref{eq:S-roots-SU33} can be determined from the others by using \eqref{eq:E6-marks-id}. 

Analogous to the discussion at \eqref{eq:cartan-E6-LC}, we can parametrize elements of Cartan sub-algebra as 
\begin{align}\label{eq:cartan-E6-lmn}
\sum_{i=1}^3\tilde{m}_i  H_{f_i} 
+ \sum_{i=1}^3 \tilde{\ell}_i H_{g_i}
+ \sum_{i=1}^3 \tilde{n}_i H_{h_i}
\in \mathfrak{h},
\end{align}
where $H_{f_i}, H_{g_i}, H_{h_i}$ are the duals of the orthonormal basis $f_i,g_i,h_i$, respectively. Here, the coefficients $\tilde{\ell}_i, \tilde{m}_i, \tilde{n}_i$ satisfy the traceless condition
\begin{align}\label{eq:tracelesslmn}
\sum_{i=1}^3 \tilde{\ell}_i 
= 
\sum_{i=1}^3 \tilde{m}_i 
= 
\sum_{i=1}^3 \tilde{n}_i 
= 0\ ,
\end{align}
and are related to the fugacities of $\SU(3)^3$ sub-algebra, which appear in \eqref{eq:SWE6-tri}, as 
\begin{align}\label{eq:LMNlmn}
\tilde{L_i} = e^{-\tilde{\ell}_i},
\qquad
\tilde{M_i} =e^{-\tilde{m}_i},
\qquad 
\tilde{N_i} =e^{-\tilde{n}_i}.
\end{align}
Again, analogous to the discussion at \eqref{eq:replace-em}, the eigenvalues of the element of the Cartan sub-algebra \eqref{eq:cartan-E6-lmn} corresponding to the simple roots \eqref{eq:S-roots-SU33} are obtained by the replacement
\begin{align}\label{eq:replace-fgh-lmn}
f_i \to \tilde{m}_i, 
\qquad
g_i \to \tilde{\ell}_i, 
\qquad
h_i \to \tilde{n}_i.
\end{align}

The relation between the fugacities of $\SO(10) \times \U(1)$ and the fugacities of $\SU(3)^3$ is obtained by equating the eigenvalues corresponding to the identical simple roots expressed in different bases \eqref{eq:S-root-SO10-U1} and \eqref{eq:S-roots-SU33} by replacing \eqref{eq:replace-em} and \eqref{eq:replace-fgh-lmn}, respectively. Solving these relations for $m_i$ in terms of $\tilde{\ell}_i, \tilde{m}_i, \tilde{n}_i$, we find
\begin{align}\label{eq:SO10U1-vs-SU33}
    m_0 &= \frac{3}{4}(\tilde{\ell}_3-\tilde{n}_3), &
    m_1 &= \frac{1}{2}(\tilde{\ell}_1-\tilde{\ell}_2-\tilde{n}_1+\tilde{n}_2), &
    m_2 &= \frac{1}{2}(-\tilde{\ell}_1+\tilde{\ell}_2-\tilde{n}_1+\tilde{n}_2), \nonumber\\
    m_3 &= \tilde{m}_1-\frac{1}{2}(\tilde{\ell}_3+\tilde{n}_3), &
    m_4 &= \tilde{m}_2-\frac{1}{2}(\tilde{\ell}_3+\tilde{n}_3), &
    m_5 &= \tilde{m}_3-\frac{1}{2}(\tilde{\ell}_3+\tilde{n}_3).
\end{align}
where we used the traceless condition \eqref{eq:tracelesslmn}.
Note that this relation should be understood up to Weyl group. 

Now we focus on the tuning
\begin{align}
    \tilde{\ell}_1 &= -\frac{2\pi i}{3}, &
    \tilde{\ell}_2 &= 0, &
    \tilde{\ell}_3 &= \frac{2\pi i}{3}, \nonumber\\
    \tilde{m}_1 &= -\frac{2\pi i}{3}, &
    \tilde{m}_2 &= 0, &
    \tilde{m}_3 &= \frac{2\pi i}{3}, \nonumber\\
    \tilde{n}_1 &= -\frac{2\pi i}{3}, &
    \tilde{n}_2 &= 0, &
    \tilde{n}_3 &= \frac{2\pi i}{3}.
\end{align}
corresponding to \eqref{eq:tuningT3}. Using the relation \eqref{eq:SO10U1-vs-SU33}, we find that the corresponding tuning in terms of $m_i$ is given by
\begin{align}\label{eq:E6-tuning-mi}
m_0 = 0, \quad 
m_1 = 0, \quad 
m_2 = \frac{2 \pi i}{3}, \quad 
m_3 = - \frac{4 \pi i}{3}, \quad 
m_4 = - \frac{2 \pi i}{3}, \quad 
m_5 = 0.
\end{align}

Various equivalent tunings can be obtained by applying the $\E_6$ Weyl transformations to \eqref{eq:E6-tuning-mi}. 
Translating the Weyl reflection of each orthonormal basis vector $e_j$ along the direction of simple root $\alpha_i$
\begin{align}\label{eq:Weyl-refl}
e_j \to e_j - \frac{2(\alpha_i, e_j)}{(\alpha_i, \alpha_i)} \alpha_i
\end{align}
into the corresponding Weyl group action $W_{\alpha_i},i=0,1,\dots,6$, on $m_j$ with the replacement \eqref{eq:replace-em}, 
we find 
\begin{align}
&W_{\alpha_0}: 
\left\{
\begin{array}{ll}
m_0 \to m_0 - \frac38 (2m_0+m_1+m_2+m_3+m_4+m_5) \\
m_j \to m_j - \frac14 (2m_0+m_1+m_2+m_3+m_4+m_5) & (j = 1,2, \dots, 5)
\end{array}
\right.
\cr
&W_{\alpha_1}: 
\left\{
\begin{array}{ll}
m_0 \to m_0 - \frac38 (2m_0+m_1-m_2-m_3-m_4-m_5) & \\
m_1 \to m_1 - \frac14 (2m_0+m_1-m_2-m_3-m_4-m_5) & \\
m_j \to m_j + \frac14 (2m_0+m_1-m_2-m_3-m_4-m_5) & (j = 2, \dots, 5)
\end{array}
\right.
\cr
&W_{\alpha_2}:
\left\{
\begin{array}{ll}
m_1 \to - m_2 & \\
m_2 \to - m_1 & \\
m_j \to m_j & (j = 0, 3 , 4, 5)
\end{array}
\right.
\qquad
W_{\alpha_3}:
\left\{
\begin{array}{ll}
m_1 \to m_2 & \\
m_2 \to m_1 & \\
m_j \to m_j & (j = 0, 3 , 4, 5)
\end{array}
\right.
\cr
&W_{\alpha_4}:
\left\{
\begin{array}{ll}
m_2 \to m_3 & \\
m_3 \to m_2 & \\
m_j \to m_j & (j = 0, 1 , 4, 5)
\end{array}
\right.
\qquad \quad
W_{\alpha_5}:
\left\{
\begin{array}{ll}
m_3 \to m_4 & \\
m_4 \to m_3 & \\
m_j \to m_j & (j = 0, 1, 2 , 5)
\end{array}
\right.
\cr
&W_{\alpha_6}:
\left\{
\begin{array}{ll}
m_4 \to m_5 & \\
m_5 \to m_4 & \\
m_j \to m_j & (j = 0,1,2,3)
\end{array}
\right.
\end{align}
These are the generators of the $\E_6$ Weyl group. Note that $W_{\alpha_1}$ can be reproduced from the others and can be omitted.

Combining these actions, we can derive the following $\E_6$ Weyl group actions on $m_k$
\begin{itemize}
\item $
\sigma_{i,j}: \left\{
\begin{array}{ll}
m_i \to m_j  \\
m_j \to m_i  \\
m_k \to m_k & (k \neq i,j) 
\end{array}
\right.
$ $\quad $ for $i,j=1,\dots, 5$ with $i \neq j$
\item $
\lambda_{i,j}: \left\{
\begin{array}{ll}
m_i \to - m_j  \\
m_j \to - m_i  \\
m_k \to m_k & (k \neq i,j) 
\end{array}
\right.
$ $\quad$ for $i,j=1,\dots, 5$ with $i \neq j$
\item 
$
\rho: \left\{
\begin{array}{ll}
m_0 \to m_0 - \frac38 \left( 2m_0 + \sum_{j=1}^5 m_j  \right) \\
m_k \to m_k - \frac14 \left( 2m_0 + \sum_{j=1}^5 m_j  \right) & (k=1,\dots 5) \\
\end{array}
\right.
$
\end{itemize}
The actions $\sigma_{i,j}$ and $\lambda_{i,j}$ originate in $W_{\alpha_2}, \dots,W_{\alpha_6}$ and generate the SO(10) Weyl group, corresponding to the SO(10) subgroup of $\E_6$ in figure \ref{fig:e6dyn-brok} (a). Combined with $W_{\alpha_0}$, which we rename as $\rho$, the whole $\E_6$ Weyl group is generated. Applying these Weyl group actions $\sigma_{1,3} \circ \lambda_{1,5} \circ\rho \circ \lambda_{3,4}$ to \eqref{eq:E6-tuning-mi}, we find
\begin{align}
m_0 = - \pi i, \quad
m_1 = m_2 = 0, \quad 
m_3 = m_4 = m_5 = \frac{2 \pi i}{3}, \quad 
\end{align}
which reproduces \eqref{eq:tuningSU2-5F}. 

\subsection{\texorpdfstring{$\E_7$}{E7} group}
\label{sec:rel-fug-2}

\begin{figure}
    \centering
\begin{tikzpicture}[
  x=11mm,
  y=11mm,
  dynkin edge/.style={
    draw,
    line width=1.5pt,
    line cap=round,
    line join=round
  },
  dynkin node/.style={
    circle,
    draw,
    fill=white,
    minimum size=3.5mm,
    inner sep=0pt,
    outer sep=0pt,
    line width=1.5pt
  },
  root label/.style={
    font=\scriptsize,
    inner sep=1pt,
    label distance=1.5mm
  },
  diagram title/.style={
    font=\normalsize,
    inner sep=1pt
  }
]
\coordinate (a0) at (-3,0);
\coordinate (a1) at (-2,0);
\coordinate (a2) at (-1,0);
\coordinate (a3) at (0,0);
\coordinate (a4) at (1,0);
\coordinate (a5) at (2,0);
\coordinate (a6) at (3,0);
\coordinate (a7) at (0,1.05);

\draw[dynkin edge] (a0) -- (a1) -- (a2) -- (a3) -- (a4) -- (a5) -- (a6);
\draw[dynkin edge] (a3) -- (a7);

\node[dynkin node, label={[root label]below:$\alpha_0$}] at (a0) {};
\node[dynkin node, label={[root label]below:$\alpha_1$}] at (a1) {};
\node[dynkin node, label={[root label]below:$\alpha_3$}] at (a2) {};
\node[dynkin node, label={[root label]below:$\alpha_4$}] at (a3) {};
\node[dynkin node, label={[root label]below:$\alpha_5$}] at (a4) {};
\node[dynkin node, label={[root label]below:$\alpha_6$}] at (a5) {};
\node[dynkin node, label={[root label]below:$\alpha_7$}] at (a6) {};
\node[dynkin node, label={[root label]right:$\alpha_2$}] at (a7) {};

\node[diagram title] at (0,1.70) {$\hat E_7$};
\end{tikzpicture}
    \caption{Affine Dynkin diagram for $E_7$.}
\label{fig:E7-Dynkin}
\end{figure}
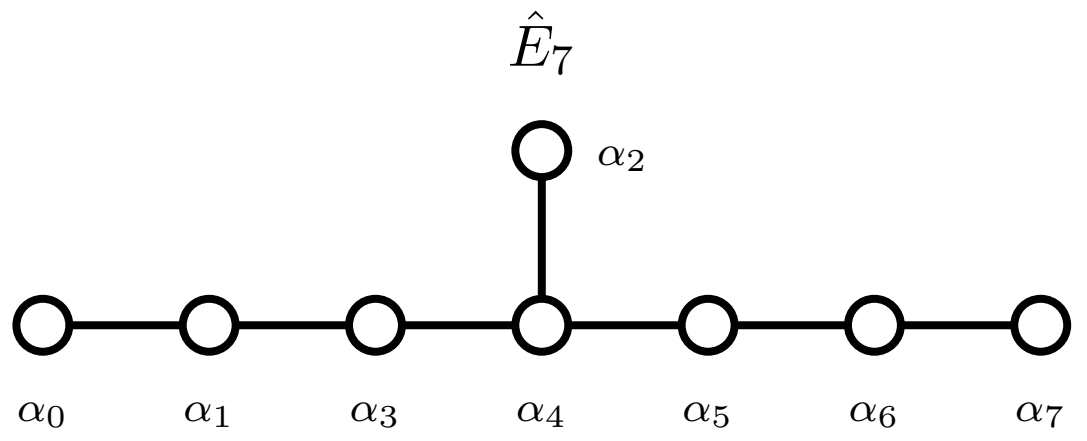

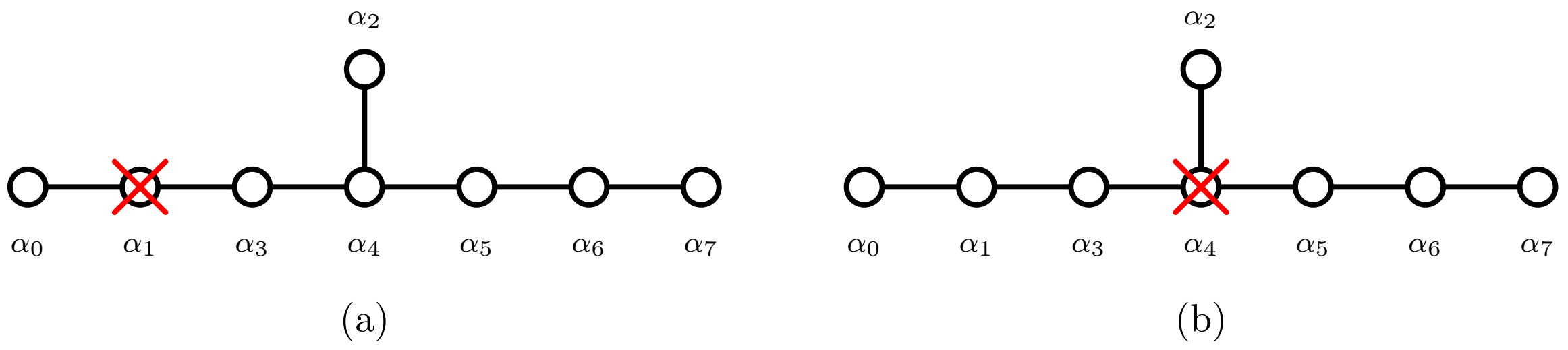
\begin{figure}
    \centering
\begin{tikzpicture}[
  x=11mm,
  y=11mm,
  dynkin edge/.style={
    draw,
    line width=1.5pt,
    line cap=round,
    line join=round
  },
  dynkin node/.style={
    circle,
    draw,
    fill=white,
    minimum size=3.5mm,
    inner sep=0pt,
    outer sep=0pt,
    line width=1.5pt
  },
  root label/.style={
    font=\scriptsize,
    inner sep=1pt,
    label distance=1.5mm
  },
  diagram title/.style={
    font=\normalsize,
    inner sep=1pt
  }
]

\coordinate (L0) at (-3,0);
\coordinate (L1) at (-2,0);
\coordinate (L2) at (-1,0);
\coordinate (L3) at (0,0);
\coordinate (L4) at (1,0);
\coordinate (L5) at (2,0);
\coordinate (L6) at (3,0);
\coordinate (L7) at (0,1.05);

\draw[dynkin edge] (L0) -- (L1) -- (L2) -- (L3) -- (L4) -- (L5) -- (L6);
\draw[dynkin edge] (L3) -- (L7);

\node[dynkin node, label={[root label]below:$\alpha_0$}] at (L0) {};
\node[dynkin node, label={[root label]below:$\alpha_1$}] at (L1) {};
\node[dynkin node, label={[root label]below:$\alpha_3$}] at (L2) {};
\node[dynkin node, label={[root label]below:$\alpha_4$}] at (L3) {};
\node[dynkin node, label={[root label]below:$\alpha_5$}] at (L4) {};
\node[dynkin node, label={[root label]below:$\alpha_6$}] at (L5) {};
\node[dynkin node, label={[root label]below:$\alpha_7$}] at (L6) {};
\node[dynkin node, label={[root label]above:$\alpha_2$}] at (L7) {};

\draw[dynkin edge,red] ([shift={(-2.5mm,-2.5mm)}]L3) -- ([shift={(2.5mm,2.5mm)}]L3);
\draw[dynkin edge,red] ([shift={(-2.5mm,2.5mm)}]L3) -- ([shift={(2.5mm,-2.5mm)}]L3);

\node at (0,-1.2){(b)};

\begin{scope}[xshift=-8.2cm]
  \coordinate (R0) at (-3,0);
  \coordinate (R1) at (-2,0);
  \coordinate (R2) at (-1,0);
  \coordinate (R3) at (0,0);
  \coordinate (R4) at (1,0);
  \coordinate (R5) at (2,0);
  \coordinate (R6) at (3,0);
  \coordinate (R7) at (0,1.05);

  \draw[dynkin edge] (R0) -- (R1) -- (R2) -- (R3) -- (R4) -- (R5) -- (R6);
  \draw[dynkin edge] (R3) -- (R7);

  \node[dynkin node, label={[root label]below:$\alpha_0$}] at (R0) {};
  \node[dynkin node, label={[root label]below:$\alpha_1$}] at (R1) {};
  \node[dynkin node, label={[root label]below:$\alpha_3$}] at (R2) {};
  \node[dynkin node, label={[root label]below:$\alpha_4$}] at (R3) {};
  \node[dynkin node, label={[root label]below:$\alpha_5$}] at (R4) {};
  \node[dynkin node, label={[root label]below:$\alpha_6$}] at (R5) {};
  \node[dynkin node, label={[root label]below:$\alpha_7$}] at (R6) {};
  \node[dynkin node, label={[root label]above:$\alpha_2$}] at (R7) {};

  \draw[dynkin edge,red] ([shift={(-2.5mm,-2.5mm)}]R1) -- ([shift={(2.5mm,2.5mm)}]R1);
  \draw[dynkin edge,red] ([shift={(-2.5mm,2.5mm)}]R1) -- ([shift={(2.5mm,-2.5mm)}]R1);

  \node at (0,-1.2){(a)};
\end{scope}

\end{tikzpicture}
\caption{Part (a) of this figure shows the first embedding $\E_7\supset\SO(12)\times\SU(2)$, while part (b) depicts the second one $\E_7\supset\SU(4)\times \SU(4)\times\SU(2)$.}\label{fig:e7dyn-brok}
\end{figure}

The relation between fugacities of the different subgroups of $\E_7$ can be obtained analogously to the case of $\E_6$. The simple roots corresponding to each node of the affine Dynkin diagram for $\E_7$, depicted in figure \ref{fig:E7-Dynkin}, can be written in terms of orthonormal basis $\{ e_i \}$ as 
\begin{align}\label{eq:S-root-SO12-SU2}
    \alpha_0 &= -\sqrt{2}\,e_0, &
    \alpha_1 &= \frac{1}{2}\left(\sqrt{2}\,e_0+e_1-e_2-e_3-e_4-e_5-e_6\right), &\nonumber\\
    \alpha_2 &= e_1+e_2, &
    \alpha_3 &= -e_1+e_2, \nonumber\\
    \alpha_4 &= -e_2+e_3, &
    \alpha_5 &= -e_3+e_4, \nonumber\\
    \alpha_6 &= -e_4+e_5, &
    \alpha_7 &= -e_5+e_6.
\end{align}
This is an expression that makes manifest the structure of the $\SO(12) \times \SU(2)$ subgroup of $\E_7$. If we remove the node of $\alpha_1$ as in (a) of figure \ref{fig:e7dyn-brok}, we find the Dynkin diagram for $\SO(12) \times \SU(2)$ with its standard expression for the simple roots. 
The expression for $\alpha_1$ in \eqref{eq:S-root-SO12-SU2} has been determined by using the relation
\begin{align}\label{eq:E7-marks-id}
\alpha_0 + 2 \alpha_1 + 2 \alpha_2 + 3 \alpha_3 + 4 \alpha_4 + 3 \alpha_5 + 2 \alpha_6 + \alpha_7 = 0.
\end{align}

The simple roots identical to \eqref{eq:S-root-SO12-SU2} can be expressed differently by using another orthonormal basis $\{ f_i, g_i, h_j \}$, with $i=1,2,3,4$ and $j=1,2$, as
\begin{align}\label{eq:S-root-SU4SU2}
\alpha_0 &= - g_3 + g_4,
\qquad
\alpha_1 = - g_2 + g_3,
\qquad
\alpha_2 = - h_1 + h_2,
\qquad
\alpha_3 = - g_1 + g_2
\cr
\alpha_4 &= \frac14 (3f_1 - f_2 - f_3 -f_4 + 3g_1 - g_2 - g_3 - g_4 + 2h_1 - 2h_2 )
\cr
\alpha_5 &= -f_1 + f_2,
\qquad
\alpha_6 = - f_2 + f_3,
\qquad
\alpha_7 = - f_3 + f_4.
\end{align} 
In this expression, the structure of the $\SU(4)^2 \times \SU(2)$ subgroup of $E_7$ is manifest. If we remove the node of $\alpha_4$ as in (b) of figure \ref{fig:e7dyn-brok}, we obtain the Dynkin diagram for $\SU(4)^2 \times \SU(2)$ and their standard expression for their simple roots. Again, the expression of $\alpha_4$ in \eqref{eq:S-root-SU4SU2} can be determined from the others by using \eqref{eq:E7-marks-id}.

The relation between the fugacities of $\SO(12) \times \SU(2)$ and the fugacities of $\SU(4)^2 \times \SU(2)$ is obtained by equating the eigenvalues corresponding to the identical simple roots expressed in different bases \eqref{eq:S-root-SO12-SU2} and \eqref{eq:S-root-SU4SU2} with the replacement 
\begin{align}\label{eq:replace-em-E7}
e_0 \to \sqrt{2} m_0, \quad e_i \to m_i \quad (i=1,\dots 6),
\cr
f_i \to \tilde{m}_i, 
\qquad
g_i \to \tilde{\ell}_i, 
\qquad
h_j \to \tilde{n}_j.
\end{align}
Here, again, $m_0$ and $m_i$ ($i=1,\dots,6$) are the instanton mass and the flavor masses, respectively, and related to the fugacity of $\SO(12) \times \SU(2)$ as \eqref{eq:M=expm}. The $\E_7$ counterpart of \eqref{eq:cartan-E6-LC} in the $E_6$ case is
\begin{align}
\sqrt{2}m_0H_{e_0}+\sum_{i=1}^6m_iH_{e_i}\in\mathfrak {h}\ .
\end{align}
Also, $\tilde{\ell}_i, \tilde{m}_i, \tilde{n}_j$ satisfy the traceless condition 
\begin{align}
\sum_{i=1}^4 \tilde{\ell}_i 
= 
\sum_{i=1}^4 \tilde{m}_i 
= 
\sum_{j=1}^2 \tilde{n}_j 
= 0
\end{align}
and are related to the fugacity of $\SU(4)^2 \times \SU(2)$ as in \eqref{eq:LMNlmn}. 
Solving these relations for $m_i$ in terms of $\tilde{\ell}_i, \tilde{m}_i, \tilde{n}_j$, we find
\begin{alignat}{3}\label{eq:SO12SU2-vs-SU4SU2}
    m_0 &= \frac12(\tilde{\ell}_3-\tilde{\ell}_4), \qquad\qquad&
    m_1 &= \frac12(\tilde{\ell}_1-\tilde{\ell}_2-\tilde{n}_1+\tilde{n}_2), \qquad&
    m_2 &= \frac12(-\tilde{\ell}_1+\tilde{\ell}_2-\tilde{n}_1+\tilde{n}_2), \nonumber\\
    m_3 &= \tilde{m}_1+\frac12(\tilde{\ell}_1+\tilde{\ell}_2), &
    m_4 &= \tilde{m}_2+\frac12(\tilde{\ell}_1+\tilde{\ell}_2), &
    m_5 &= \tilde{m}_3+\frac12(\tilde{\ell}_1+\tilde{\ell}_2), \nonumber\\
    m_6 &= \tilde{m}_4+\frac12(\tilde{\ell}_1+\tilde{\ell}_2).
\end{alignat}

Now we focus on the tuning
\begin{alignat}{4}
    \tilde{\ell}_1 &= -\frac{3\pi i}{4}, \qquad&
    \tilde{\ell}_2 &= -\frac{\pi i}{4}, \qquad&
    \tilde{\ell}_3 &= \frac{\pi i}{4}, \qquad&
    \tilde{\ell}_4 &= \frac{3\pi i}{4}, \nonumber\\
    \tilde{m}_1 &= -\frac{3\pi i}{4}, &
    \tilde{m}_2 &= -\frac{\pi i}{4}, &
    \tilde{m}_3 &= \frac{\pi i}{4}, &
    \tilde{m}_4 &= \frac{3\pi i}{4}, \nonumber\\
    \tilde{n}_1 &= -\frac{\pi i}{2}, &
    \tilde{n}_2 &= \frac{\pi i}{2}.
\end{alignat}
corresponding to \eqref{eq:E7LMN-tuning}. Using the relation \eqref{eq:SO12SU2-vs-SU4SU2}, we obtain 
\begin{alignat}{4}\label{eq:E7-tuning-mi}
    m_0 &= -\frac{\pi i}{4}, \qquad&
    m_1 &= \frac{\pi i}{4}, \qquad&
    m_2 &= \frac{3\pi i}{4}, \qquad&
    m_3 &= -\frac{5\pi i}{4}, \nonumber\\
    m_4 &= -\frac{3\pi i}{4}, &
    m_5 &= -\frac{\pi i}{4}, &
    m_6 &= \frac{\pi i}{4}.
\end{alignat}

Again, various equivalent tunings can be obtained by applying Weyl transformations of $\E_7$ to \eqref{eq:E7-tuning-mi}. Translating the Weyl reflection \eqref{eq:Weyl-refl} into the corresponding Weyl group action $W_{\alpha_i},i=0,1,\dots,7$, on $m_j$ with the replacement \eqref{eq:replace-em-E7}, we obtain
\begin{align}
&W_{\alpha_0}: 
\left\{
\begin{array}{ll}
m_0 \to -m_0 \\
m_j \to m_j  & (j = 1,2, \dots, 6)
\end{array}
\right.
\cr
&W_{\alpha_1}: 
\left\{
\begin{array}{ll}
m_0 \to m_0 - \frac14 (2m_0+m_1-m_2-m_3-m_4-m_5-m_6) & \\
m_1 \to m_1 - \frac14 (2m_0+m_1-m_2-m_3-m_4-m_5-m_6) & \\
m_j \to m_j + \frac14 (2m_0+m_1-m_2-m_3-m_4-m_5-m_6) & (j = 2, \dots, 6)
\end{array}
\right.
\cr
&W_{\alpha_2}:
\left\{
\begin{array}{ll}
m_1 \to - m_2 & \\
m_2 \to - m_1 & \\
m_j \to m_j & (j = 0, 3 , 4, 5, 6)
\end{array}
\right.
\qquad
W_{\alpha_3}:
\left\{
\begin{array}{ll}
m_1 \to m_2 & \\
m_2 \to m_1 & \\
m_j \to m_j & (j = 0, 3 , 4, 5,6)
\end{array}
\right.
\cr
&W_{\alpha_4}:
\left\{
\begin{array}{ll}
m_2 \to m_3 & \\
m_3 \to m_2 & \\
m_j \to m_j & (j = 0, 1 , 4, 5,6)
\end{array}
\right.
\qquad \quad
W_{\alpha_5}:
\left\{
\begin{array}{ll}
m_3 \to m_4 & \\
m_4 \to m_3 & \\
m_j \to m_j & (j = 0, 1, 2 , 5,6)
\end{array}
\right.
\cr
&W_{\alpha_6}:
\left\{
\begin{array}{ll}
m_4 \to m_5 & \\
m_5 \to m_4 & \\
m_j \to m_j & (j = 0,1,2,3,6)
\end{array}
\right.
\qquad \quad
W_{\alpha_7}:
\left\{
\begin{array}{ll}
m_5 \to m_6 & \\
m_6 \to m_5 & \\
m_j \to m_j & (j = 0,1,2,3,4)
\end{array}
\right.
\end{align}
These are eight generators of the $\E_7$ Weyl group, one of which is redundant. 
Combining these actions, we can derive the following $\E_7$ Weyl group actions on $m_k$
\begin{itemize}
\item $
\sigma_{i,j}: \left\{
\begin{array}{ll}
m_i \to m_j  \\
m_j \to m_i  \\
m_k \to m_k & (k \neq i,j) 
\end{array}
\right.
$ $\quad \,\,$ for $i,j=1,\dots, 6$ with $i \neq j$
\item $
\lambda_{i,j}: \left\{
\begin{array}{ll}
m_i \to - m_j  \\
m_j \to - m_i  \\
m_k \to m_k & (k \neq i,j) 
\end{array}
\right.
$ $\quad $ for $i,j=1,\dots, 6$ with $i \neq j$
\item 
$
\rho: \left\{
\begin{array}{ll}
m_k \to m_k - \frac14 \left( 2m_0 + m_1 - \sum_{j=2}^6 m_j \right)  & (k=0,1) \\
m_k \to m_k + \frac14 \left( 2m_0 + m_1 - \sum_{j=2}^6 m_j \right) & (k=2,\dots 6) 
\end{array}
\right.
$
\end{itemize}
Applying $\sigma_{3,6} \circ \sigma_{2,5} \circ \lambda_{2,3} \circ \rho$ to \eqref{eq:E7-tuning-mi}, we have
\begin{align}\label{eq:tun-e7-exp}
m_0 = -\frac{\pi i}{2}, \qquad
m_1 = m_2 = 0, \qquad 
m_3 =-m_4 = \frac{\pi i}{2}, \qquad
m_5 = m_6 = \pi i.
\end{align}
which leads to \eqref{eq:E7-tuning-M}. 

\subsection{\texorpdfstring{$\E_8$}{E8} group}\label{sec:rel-fug-3}

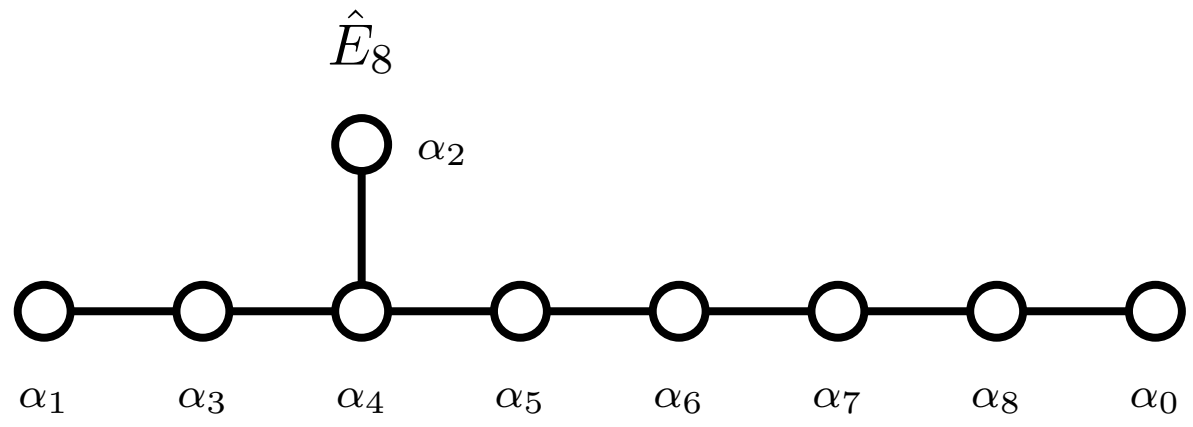
\begin{figure}
    \centering
\begin{tikzpicture}[
  x=10.5mm,
  y=10.5mm,
  dynkin edge/.style={
    draw,
    line width=1.5pt,
    line cap=round,
    line join=round
  },
  dynkin node/.style={
    circle,
    draw,
    fill=white,
    minimum size=3.5mm,
    inner sep=0pt,
    outer sep=0pt,
    line width=1.5pt
  },
  root label/.style={
    font=\scriptsize,
    inner sep=1pt,
    label distance=1.5mm
  },
  diagram title/.style={
    font=\normalsize,
    inner sep=1pt
  }
]

\coordinate (a0) at (-3.5,0);
\coordinate (a1) at (-2.5,0);
\coordinate (a2) at (-1.5,0);
\coordinate (a3) at (-0.5,0);
\coordinate (a4) at (0.5,0);
\coordinate (a5) at (1.5,0);
\coordinate (a6) at (2.5,0);
\coordinate (a7) at (3.5,0);
\coordinate (a8) at (-1.5,1.05);

\draw[dynkin edge] (a0) -- (a1) -- (a2) -- (a3) -- (a4) -- (a5) -- (a6) -- (a7);
\draw[dynkin edge] (a2) -- (a8);

\node[dynkin node, label={[root label]below:$\alpha_1$}] at (a0) {};
\node[dynkin node, label={[root label]below:$\alpha_3$}] at (a1) {};
\node[dynkin node, label={[root label]below:$\alpha_4$}] at (a2) {};
\node[dynkin node, label={[root label]below:$\alpha_5$}] at (a3) {};
\node[dynkin node, label={[root label]below:$\alpha_6$}] at (a4) {};
\node[dynkin node, label={[root label]below:$\alpha_7$}] at (a5) {};
\node[dynkin node, label={[root label]below:$\alpha_8$}] at (a6) {};
\node[dynkin node, label={[root label]below:$\alpha_0$}] at (a7) {};
\node[dynkin node, label={[root label]right:$\alpha_2$}] at (a8) {};

\node[diagram title] at (-1.5,1.70) {$\hat E_8$};

\end{tikzpicture}
    \caption{Affine Dynkin diagram for $E_8$.}
\label{fig:E8-Dynkin}
\end{figure}

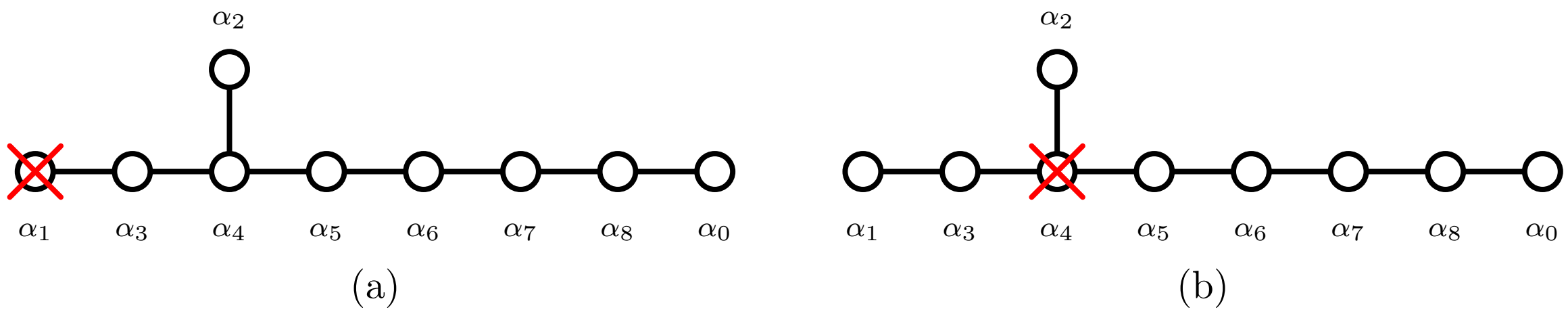
\begin{figure}
    \centering
\begin{tikzpicture}[
  x=9.5mm,
  y=9.5mm,
  dynkin edge/.style={
    draw,
    line width=1.5pt,
    line cap=round,
    line join=round
  },
  dynkin node/.style={
    circle,
    draw,
    fill=white,
    minimum size=3.5mm,
    inner sep=0pt,
    outer sep=0pt,
    line width=1.5pt
  },
  root label/.style={
    font=\scriptsize,
    inner sep=1pt,
    label distance=1.5mm
  },
  diagram title/.style={
    font=\normalsize,
    inner sep=1pt
  }
]
\coordinate (L0) at (-3.5,0);
\coordinate (L1) at (-2.5,0);
\coordinate (L2) at (-1.5,0);
\coordinate (L3) at (-0.5,0);
\coordinate (L4) at (0.5,0);
\coordinate (L5) at (1.5,0);
\coordinate (L6) at (2.5,0);
\coordinate (L7) at (3.5,0);
\coordinate (L8) at (-1.5,1.05);

\draw[dynkin edge] (L0) -- (L1) -- (L2) -- (L3) -- (L4) -- (L5) -- (L6) -- (L7);
\draw[dynkin edge] (L2) -- (L8);

\node[dynkin node, label={[root label]below:$\alpha_1$}] at (L0) {};
\node[dynkin node, label={[root label]below:$\alpha_3$}] at (L1) {};
\node[dynkin node, label={[root label]below:$\alpha_4$}] at (L2) {};
\node[dynkin node, label={[root label]below:$\alpha_5$}] at (L3) {};
\node[dynkin node, label={[root label]below:$\alpha_6$}] at (L4) {};
\node[dynkin node, label={[root label]below:$\alpha_7$}] at (L5) {};
\node[dynkin node, label={[root label]below:$\alpha_8$}] at (L6) {};
\node[dynkin node, label={[root label]below:$\alpha_0$}] at (L7) {};
\node[dynkin node, label={[root label]above:$\alpha_2$}] at (L8) {};

\draw[dynkin edge, red] ([shift={(-2.5mm,-2.5mm)}]L2) -- ([shift={(2.5mm,2.5mm)}]L2);
\draw[dynkin edge, red] ([shift={(-2.5mm,2.5mm)}]L2) -- ([shift={(2.5mm,-2.5mm)}]L2);
\textbf{}
\node at (0,-1.2){(b)};

\begin{scope}[xshift=-8.1cm]
  \coordinate (R0) at (-3.5,0);
  \coordinate (R1) at (-2.5,0);
  \coordinate (R2) at (-1.5,0);
  \coordinate (R3) at (-0.5,0);
  \coordinate (R4) at (0.5,0);
  \coordinate (R5) at (1.5,0);
  \coordinate (R6) at (2.5,0);
  \coordinate (R7) at (3.5,0);
  \coordinate (R8) at (-1.5,1.05);

  \draw[dynkin edge] (R0) -- (R1) -- (R2) -- (R3) -- (R4) -- (R5) -- (R6) -- (R7);
  \draw[dynkin edge] (R2) -- (R8);

  \node[dynkin node, label={[root label]below:$\alpha_1$}] at (R0) {};
  \node[dynkin node, label={[root label]below:$\alpha_3$}] at (R1) {};
  \node[dynkin node, label={[root label]below:$\alpha_4$}] at (R2) {};
  \node[dynkin node, label={[root label]below:$\alpha_5$}] at (R3) {};
  \node[dynkin node, label={[root label]below:$\alpha_6$}] at (R4) {};
  \node[dynkin node, label={[root label]below:$\alpha_7$}] at (R5) {};
  \node[dynkin node, label={[root label]below:$\alpha_8$}] at (R6) {};
  \node[dynkin node, label={[root label]below:$\alpha_0$}] at (R7) {};
  \node[dynkin node, label={[root label]above:$\alpha_2$}] at (R8) {};

  \draw[dynkin edge,red] ([shift={(-2.5mm,-2.5mm)}]R0) -- ([shift={(2.5mm,2.5mm)}]R0);
  \draw[dynkin edge,red] ([shift={(-2.5mm,2.5mm)}]R0) -- ([shift={(2.5mm,-2.5mm)}]R0);

  \node at (0,-1.2){(a)};
\end{scope}
\end{tikzpicture}
    \caption{Part (a) of this figure shows the first embedding $\E_8\supset\SO(16)$, while part (b) depicts the second one $\E_8\supset\SU(6)\times \SU(3)\times\SU(2)$.}\label{fig:e8dyn-brok}
\end{figure}

The relation between fugacities of the different subgroups of $\E_8$ can be obtained analogously to the case of $\E_6$ and $\E_7$. The simple roots corresponding to each node of the affine Dynkin diagram for $\E_8$, depicted in figure \ref{fig:E8-Dynkin}, can be written in terms of orthonormal basis $\{ e_i \}$ as
\begin{align}\label{eq:S-root-SO16}
\alpha_0  &= - e_0 - e_7 ,
\quad
\alpha_1 = \frac12 (e_0 + e_1 - e_2 - e_3 - e_4 - e_5 - e_6 - e_7)
\cr
\alpha_2 &= e_1 + e_2,
\quad
\alpha_3 = - e_1 + e_2,
\quad
\alpha_4 = - e_2 + e_3,
\quad
\alpha_5 = - e_3 + e_4,
\cr
\alpha_6 &= - e_4 + e_5,
\quad
\alpha_7 = - e_5 + e_6,
\quad
\alpha_8 = - e_6 + e_7.
\end{align}
This is an expression that makes manifest  the structure of the $\SO(16)$ subgroup of $\E_8$. If we remove the node of $\alpha_1$ as in (a) of figure \ref{fig:e8dyn-brok}, we find the Dynkin diagram for $\SO(16)$ with its standard expression for the simple roots. 
The expression for $\alpha_1$ in \eqref{eq:S-root-SO16} has been determined by using the relation
\begin{align}\label{eq:E8-marks-id}
\alpha_0 + 2 \alpha_1 + 3 \alpha_2 + 4 \alpha_3 + 6 \alpha_4 + 5 \alpha_5 + 4 \alpha_6 + 3 \alpha_7 + 2 \alpha_8 = 0.
\end{align}

The simple roots identical to \eqref{eq:S-root-SO16} can be expressed differently by using another orthonormal basis $\{ f_i, g_j, h_k \}$, with $i=1,2,\dots,6$, $j=1,2,3$ and $k=1,2$, as
\begin{align}\label{eq:S-root-SU6SU3SU2}
\alpha_0 &= - f_5 + f_6,
\qquad
\alpha_1 = - g_2 + g_3,
\qquad
\alpha_2 = - h_1 + h_2,
\qquad
\alpha_3 = - g_1 + g_2,
\cr
\alpha_4
&= \frac16(5 f_1 - f_2 - f_3 - f_4 - f_5 - f_6 + 4g_1 - 2 g_2 - 2 g_3 + 3 h_1 - 3 h_2 ),
\cr
\alpha_5 &= -f_1 + f_2,
\qquad
\alpha_6 = - f_2 + f_3, 
\qquad
\alpha_7 = - f_3 + f_4,
\qquad
\alpha_8 = - f_4 + f_5,
\end{align}
In this expression, the structure of the $\SU(6)\times \SU(3) \times \SU(2)$ subgroup of $E_8$ is manifest. If we remove the node of $\alpha_4$ as in (b) of figure \ref{fig:e8dyn-brok}, we obtain the Dynkin diagram for $\SU(6)\times \SU(3) \times \SU(2)$ and their standard expression for their simple roots. Again, the expression of $\alpha_4$ in \eqref{eq:S-root-SU6SU3SU2} can be determined from the others by using \eqref{eq:E8-marks-id}.

The relation between the fugacities of $\SO(16)$ and the fugacities of $\SU(6) \times \SU(3) \times \SU(2)$ is obtained by equating the eigenvalues corresponding to the identical simple roots expressed in different bases \eqref{eq:S-root-SO16} and \eqref{eq:S-root-SU6SU3SU2} with the replacement 
\begin{alignat}{3}\label{eq:replace-em-E8}
    e_0 &\to m_0, \qquad & e_i \to m_i \qquad & (i=1,\ldots,7), \nonumber\\
    f_i &\to \tilde{m}_i, \qquad &    g_j \to \tilde{\ell}_j, \qquad &    h_k \to \tilde{n}_k.
\end{alignat}
Here, again, $m_i$ and $m_0$ are the flavor masses and the instanton mass, respectively, and related to the fugacity of $\SO(16)$ as \eqref{eq:M=expm}. The $\E_8$ counterpart of \eqref{eq:cartan-E6-LC} in the $\E_6$ case is
\begin{align}
\sum_{i=0}^7m_iH_{e_i}\in\mathfrak {h}\ .
\end{align}
Also, $\tilde{\ell}_j, \tilde{m}_i, \tilde{n}_k$ satisfy the traceless condition 
\begin{align}
\sum_{i=1}^6 \tilde{m}_i 
=
\sum_{j=1}^3 \tilde{\ell}_j 
= 
\sum_{k=1}^2 \tilde{n}_k 
= 0
\end{align}
and are related to the fugacity of $\SU(6)\times \SU(3) \times \SU(2)$ as in \eqref{eq:LMNlmn}. 
Solving these relations for $m_i$ in terms of $\tilde{\ell}_j, \tilde{m}_i, \tilde{n}_k$, we find
\begin{alignat}{3}\label{eq:SO162-vs-SU6SU3SU2}
    m_0 &= -\tilde{m}_6+\frac12\tilde{\ell}_3, \qquad&
    m_1 &= \frac12(\tilde{\ell}_1-\tilde{\ell}_2-\tilde{n}_1+\tilde{n}_2), \qquad&
    m_2 &= \frac12(-\tilde{\ell}_1+\tilde{\ell}_2-\tilde{n}_1+\tilde{n}_2), \nonumber\\
    m_3 &= \tilde{m}_1-\frac12\tilde{\ell}_3, &
    m_4 &= \tilde{m}_2-\frac12\tilde{\ell}_3, &
    m_5 &= \tilde{m}_3-\frac12\tilde{\ell}_3, \nonumber\\
    m_6 &= \tilde{m}_4-\frac12\tilde{\ell}_3, &
    m_7 &= \tilde{m}_5-\frac12\tilde{\ell}_3.
\end{alignat}

Now we focus on the tuning
\begin{alignat}{2}
    \tilde{m}_k &= \frac{1}{6}(2k-7)\pi i, \qquad && k=1,2,\ldots,6, \nonumber\\
    \tilde{\ell}_k &= \frac{1}{3}(2k-4)\pi i, \qquad && k=1,2,3, \nonumber\\
    \tilde{n}_k &= \frac{1}{2}(2k-3)\pi i, \qquad && k=1,2.
\end{alignat}
corresponding to \eqref{eq:mass-tune-Z6}. 
Using the relation \eqref{eq:SO162-vs-SU6SU3SU2}, we obtain 
\begin{alignat}{4}\label{eq:E8-tuning-mi}
    m_0 &= -\frac{\pi i}{2}, \qquad&
    m_1 &= \frac{\pi i}{6}, \qquad&
    m_2 &= \frac{5\pi i}{6}, \qquad&
    m_3 &= -\frac{7\pi i}{6}, \nonumber\\
    m_4 &= -\frac{5\pi i}{6}, &
    m_5 &= -\frac{\pi i}{2}, &
    m_6 &= -\frac{\pi i}{6}, &
    m_7 &= \frac{\pi i}{6}.
\end{alignat}

Again, various equivalent tunings can be obtained by applying Weyl transformations of $E_8$ on \eqref{eq:E8-tuning-mi}. 
Translating the Weyl reflection \eqref{eq:Weyl-refl} into the corresponding Weyl group action $W_{\alpha_i},i=0,1,\dots,7$, on $m_j$ with the replacement \eqref{eq:replace-em-E8}, we obtain
\begin{align}
&W_{\alpha_0}: 
\left\{
\begin{array}{ll}
m_0 \to -m_7 \\
m_7\to-m_0\\
m_j \to m_j  & (j = 1,2, \dots, 6)
\end{array}
\right.
\cr
&W_{\alpha_1}: 
\left\{
\begin{array}{ll}
m_0 \to m_0 - \frac14 (m_0+m_1-m_2-m_3-m_4-m_5-m_6-m_7) & \\
m_1 \to m_1 - \frac14 (m_0+m_1-m_2-m_3-m_4-m_5-m_6-m_7) & \\
m_j \to m_j + \frac14 (m_0+m_1-m_2-m_3-m_4-m_5-m_6-m_7) & (j = 2,\dots, 7)
\end{array}
\right.
\cr
&W_{\alpha_2}:
\left\{
\begin{array}{ll}
m_1 \to - m_2 & \\
m_2 \to - m_1 & \\
m_j \to m_j & (j = 0, 3,\dots,7)
\end{array}
\right.
\qquad
W_{\alpha_3}:
\left\{
\begin{array}{ll}
m_1 \to m_2 & \\
m_2 \to m_1 & \\
m_j \to m_j & (j = 0, 3,\dots,7)
\end{array}
\right.
\cr
&W_{\alpha_4}:
\left\{
\begin{array}{ll}
m_2 \to m_3 & \\
m_3 \to m_2 & \\
m_j \to m_j & (j = 0, 1 , 4,\dots,7)
\end{array}
\right.
\qquad \quad
W_{\alpha_5}:
\left\{
\begin{array}{ll}
m_3 \to m_4 & \\
m_4 \to m_3 & \\
m_j \to m_j & (j = 0, 1, 2 , 5,6,7)
\end{array}
\right.
\cr
&W_{\alpha_6}:
\left\{
\begin{array}{ll}
m_4 \to m_5 & \\
m_5 \to m_4 & \\
m_j \to m_j & (j = 0,\dots,3,6,7)
\end{array}
\right.
\qquad \quad
W_{\alpha_7}:
\left\{
\begin{array}{ll}
m_5 \to m_6 & \\
m_6 \to m_5 & \\
m_j \to m_j & (j = 0,\dots,4,7)
\end{array}
\right.
\cr
&W_{\alpha_8}:
\left\{
\begin{array}{ll}
m_6 \to m_7 & \\
m_7 \to m_6 & \\
m_j \to m_j & (j = 0,\dots,5)
\end{array}
\right.
\end{align}
These are nine generators of the $\E_8$ Weyl group.
Combining these actions, we can derive the following $\E_8$ Weyl group actions on $m_k$
\begin{itemize}
\item $
\sigma_{i,j}: \left\{
\begin{array}{ll}
m_i \to m_j  \\
m_j \to m_i  \\
m_k \to m_k & (k \neq i,j) 
\end{array}
\right.
$ $\quad \,\,$ for $i,j=0,1,\dots, 7$ with $i \neq j$ 
\item $
\lambda_{i,j}: \left\{
\begin{array}{ll}
m_i \to - m_j  \\
m_j \to - m_i  \\
m_k \to m_k & (k \neq i,j) 
\end{array}
\right.
$ $\quad $ for $i,j=0,1,\dots, 7$ with $i \neq j$ 
\item 
$
\rho: 
m_k \to m_k - \frac14 \sum_{j=0}^7 m_j  \quad (k=0,1,\dots 7)
$
\end{itemize}
Applying
$\lambda_{4,7}\circ\rho\circ\lambda_{3,6}
\circ\sigma_{2,7}\circ\sigma_{2,4}\circ\lambda_{3,5}$ to \eqref{eq:E8-tuning-mi}, we obtain
\begin{align}\label{eq:e8-mi-tun}
    m_1 = m_2 = m_3 = 0, \quad 
m_4 = m_5 = \pi i, \quad
m_6 = m_7 = m_0 = -\frac{2\pi i}{3}\ .
\end{align}
Here, $m_0$ is introduced to extend the $E_6$ and $E_7$ Weyl group actions by adding one node, with the chemical potential $m_0$ identified with the physical instanton mass. In the $E_8$ case, however, $m_0$ is not the instanton mass itself, but should instead be identified with half the instanton mass. Consequently, the $E_8$
fugacities used in the main text are given by $\{M_0^2, M_1, \dots, M_7\}$. This identification naturally reproduces the $E_8$-tuning listed in \eqref{eq:E8-tuning-M}.

\subsection{\texorpdfstring{Characters of $\E_6$, $\E_7$, and $\E_8$}{Characters of E6, E7 and E8}}\label{app:characters}

For ease of reference, we collect here the characters of $\E_6,\E_7$ and $\E_8$ that appear in the Seiberg–Witten curves presented in the main text. In particular, in this Appendix, we used a Mathematica package called
Susyno \cite{Fonseca:2011sy} to re-express the characters of $\E_6,\E_7,\E_8$ groups in terms of the following subgroup embeddings:
\begin{align}
    \E_6&\supset \SO(10)\times \U(1)\ ,\notag\\
    \E_7&\supset\SO(12)\times \SU(2)\ ,\notag\\
    \E_8&\supset\SO(16)\ .
\end{align}

\subsection*{$\E_6$ characters}
Under the embedding shown in figure \ref{fig:e6dyn-brok}(a), i.e., $\E_6\supset \SO(10)\times\U(1)$,
the $\E_6$ characters decompose as follows
\begin{align}\label{eq:E6cha}
    \chi_1^{\E_6}&=M_0^{4/3}+\chi_1M_0^{-2/3}+\chi_5M_0^{1/3}\ ,\notag\\
\chi_2^{\E_6}&=\chi_2M_0^{-1}+\chi_5M_0+\chi_3+1\ ,\notag\\
\chi_3^{\E_6}&=\chi_3M_0^{-4/3}+\chi_1\chi_5M_0^{-1/3}
+\chi_1M_0^{2/3}+\chi_4M_0^{2/3}+\chi_5M_0^{5/3}\ ,\notag\\
\chi_4^{\E_6}&=\chi_4M_0^{-2}+\chi_4M_0^2
+\chi_5\chi_3M_0^{-1}+\chi_3\chi_2M_0
+\chi_3+\chi_1\chi_4\ ,\notag\\
\chi_5^{\E_6}&=\chi_2M_0^{-5/3}+\chi_3M_0^{4/3}
+\chi_1M_0^{-2/3}+\chi_4M_0^{-2/3}
+\chi_1\chi_2M_0^{1/3}\ ,\notag\\
\chi_6^{\E_6}&=M_0^{-4/3}+\chi_1M_0^{2/3}
+\chi_2M_0^{-1/3}\ .
\end{align}
Here, $M_0$ denotes the U(1)$_I$ fugacity. The characters $\chi_i^{\E_6},i=1,2,\dots,6$, correspond respectively to  the  $27$-, $78$-, $\overline{351}$-,  $2925$-, $351$-, and $\overline{27}$-dimensional representations of $\E_6$, while $\chi_i,i=1,2,\dots,5$, denote the characters of the 10-, $\overline{16}$-, 45-, 120-, $16$-dimensional representations of $\SO(10)$, respectively.

\subsection*{$\E_7$ characters}

For $\E_7$, we denote by $\chi_i^{\E_7}$, $i=1,2,\dots,7$, the characters  of the representations of dimensions 133, 912, 8645, 365750, 27664, 1539,  and 56, respectively.
Under the embedding $\E_7\supset\SO(12)\times \SU(2)$ , their decompositions into $\SO(12)\times\SU(2)$  characters are given as follow:
\begin{align}\label{eq:E7cha}
    \chi_1^{\E_7}&=\chi_0^2+\chi_5 \chi_0+\chi_2-1\ ,\notag\\
    \chi_2^{\E_7}&=\chi_6 \chi_0^2+\chi_1 \chi_0+\chi_3 \chi_0+\chi_1 \chi_5-2 \chi_6\ ,
    \notag\\
   \chi_3^{\E_7}&=\chi_5 \chi_0^3+\chi_2 \chi_0^2+\chi_4 \chi_0^2+\chi_0^2+\chi_2 \chi_5 \chi_0-2 \chi_5 \chi_0+\chi_5^2-2 \chi_2+\chi_1 \chi_3-3 \chi_4-2\ ,\notag\\
   \chi_4^{\E_7}&=\chi_4 \chi_0^4+\chi_3 \chi_6 \chi_0^3+\chi_1 \chi_3 \chi_0^2+\chi_2 \chi_4
   \chi_0^2-4 \chi_4 \chi_0^2+\chi_1 \chi_3 \chi_5 \chi_0-3 \chi_3 \chi_6 \chi_0+\chi_1^2-\chi_2^2 \notag\\
   &\quad +\chi_3^2+\chi_2 \chi_5^2+\chi_6^2-2 \chi_2+\chi_1^2 \chi_4-4 \chi_2 \chi_4+\chi_4-\chi_1 \chi_5 \chi_6-1 \ ,\notag\\
   \chi_5^{\E_7}&=\chi_3 \chi_0^3+\chi_2 \chi_6 \chi_0^2+\chi_1 \chi_2 \chi_0-3 \chi_3 \chi_0+\chi_1 \chi_4 \chi_0+\chi_3 \chi_5+\chi_1^2 \chi_6-3 \chi_2 \chi_6 \ ,\notag\\
   \chi_6^{\E_7}&=\chi_1^2+\chi_6 \chi_0 \chi_1+\chi_2 \chi_0^2-2 \chi_2+\chi_4 \ ,\notag\\
   \chi_7^{\E_7}&=\chi_6+\chi_1 \chi_0\ ,
\end{align}
where $\chi_0$ is the character of the fundamental representation of $\SU(2)$ group defined by
\begin{align}
    \chi_0=M_0+\frac1{M_0}
\end{align}
and $\chi_i,i=1,2,\dots,6$, are respectively the characters of $12$-, $66$-, $220$-,  $495$-, $\overline{32}$-, and $32$-dimensional representations of the $\SO(12)$ group.

\subsection*{$\E_8$ characters}
For $\E_8$, the dimensions of the representations associated with  the characters of  $\chi^{\E_8}_i$, $i=1,2,\dots,8$ are given by \eqref{eq:cha-dim-e8}.
We here use the embedding $\E_8\supset \SO(16)$, as shown in figure \ref{fig:e8dyn-brok}(a). 
Explicitly,  the eight characters of $\E_8$ decompose as follows
\begin{align}\label{eq:chie8}
    \chi_1^{\E_8}&=\chi_1^2+\chi_8 \chi_1-\chi_2+\chi_4-\chi_7-1\nonumber\\
    \chi_2^{\E_8}&=\chi_7 \chi_1^2+\chi_1^2+\chi_3 \chi_1+\chi_5 \chi_1+\chi_8^2-3 \chi_2-3 \chi
   _4-2 \chi_6-2 \chi_2 \chi_7-\chi_7+\chi_3 \chi_8-2 \nonumber\\ 
    \chi_3^{\E_8}&=\chi_8 \chi_1^3+\chi_2 \chi_1^2+\chi_4 \chi_1^2+\chi_6 \chi_1^2-\chi_7 \chi
   _1^2-\chi_3 \chi_1-2 \chi_2 \chi_8 \chi_1+\chi_4 \chi_8 \chi_1-\chi_7 \chi
   _8 \chi_1-2 \chi_2^2+\chi_7^2\nonumber\\
   &\quad+\chi_2 \chi_8^2-\chi_2-2 \chi_2 \chi_4+\chi_3
   \chi_5-3 \chi_2 \chi_6-\chi_6+\chi_2 \chi_7-2 \chi_4 \chi_7+\chi_3 \chi_8\nonumber\\
    \chi_4^{\E_8}&=\chi_6 \chi_1^4+\chi_1^4+\chi_3 \chi_1^3+\chi_5 \chi_1^3+\chi_5 \chi_7 \chi
   _1^3-\chi_7 \chi_8 \chi_1^3+\chi_7^2 \chi_1^2+\chi_8^2 \chi_1^2-5 \chi_2 \chi
   _1^2-2 \chi_4 \chi_1^2+\chi_3 \chi_5 \chi_1^2\nonumber\\
   &\quad-4 \chi_2 \chi_6 \chi_1^2+\chi
   _4 \chi_6 \chi_1^2-2 \chi_6 \chi_1^2-\chi_4 \chi_7 \chi_1^2-\chi_6 \chi_7
   \chi_1^2-2 \chi_7 \chi_1^2-\chi_1^2-\chi_3 \chi_7^2 \chi_1-\chi_2 \chi_3
   \chi_1\nonumber\\
   &\quad+2 \chi_3 \chi_1-3 \chi_2 \chi_5 \chi_1-2 \chi_4 \chi_5 \chi_1+\chi_5
   \chi_1+3 \chi_3 \chi_6 \chi_1-\chi_5 \chi_6 \chi_1-\chi_3 \chi_7 \chi_1-3
   \chi_2 \chi_5 \chi_7 \chi_1\nonumber\\
   &\quad-2 \chi_5 \chi_7 \chi_1+2 \chi_7^2 \chi_8 \chi
   _1+\chi_2 \chi_8 \chi_1+\chi_4 \chi_8 \chi_1+\chi_3 \chi_5 \chi_8 \chi
   _1-\chi_6 \chi_8 \chi_1+3 \chi_2 \chi_7 \chi_8 \chi_1-\chi_7 \chi_8 \chi
   _1\nonumber\\
   &\quad-\chi_8 \chi_1-\chi_2^3-2 \chi_7^3+\chi_3 \chi_8^3+4 \chi_2^2+\chi_2 \chi
   _3^2+\chi_3^2-\chi_2 \chi_4^2+\chi_2 \chi_5^2+2 \chi_5^2+\chi_6^2+\chi_2^2
   \chi_7^2-2 \chi_2 \chi_7^2\nonumber\\
   &\quad+2 \chi_4 \chi_7^2-\chi_7^2-2 \chi_2 \chi_8^2+\chi
   _2 \chi_4 \chi_8^2-\chi_2 \chi_7 \chi_8^2-\chi_7 \chi_8^2+5 \chi_2-2 \chi
   _2^2 \chi_4+2 \chi_2 \chi_4+\chi_3 \chi_5+\chi_2^2 \chi_6\nonumber\\
   &\quad+\chi_3^2 \chi_6+6
   \chi_2 \chi_6-4 \chi_2 \chi_4 \chi_6-2 \chi_4 \chi_6+\chi_6+\chi_2^2 \chi
   _7+\chi_3^2 \chi_7-\chi_4^2 \chi_7+3 \chi_2 \chi_7+2 \chi_2 \chi_4 \chi_7\nonumber\\
   &\quad+4
   \chi_4 \chi_7+2 \chi_3 \chi_5 \chi_7+3 \chi_2 \chi_6 \chi_7+4 \chi_6 \chi
   _7+4 \chi_7-2 \chi_2 \chi_3 \chi_8-3 \chi_3 \chi_8-2 \chi_3 \chi_4 \chi
   _8+\chi_2 \chi_5 \chi_8\nonumber\\
   &\quad-\chi_5 \chi_8-3 \chi_3 \chi_6 \chi_8-\chi_2 \chi_3
   \chi_7 \chi_8-\chi_3 \chi_7 \chi_8-\chi_5 \chi_7 \chi_8\nonumber\\
    \chi_5^{\E_8}&=\chi_5 \chi_1^3-\chi_6 \chi_1^2+\chi_4 \chi_7 \chi_1^2-\chi_1^2+\chi_3 \chi_4
   \chi_1-3 \chi_2 \chi_5 \chi_1-\chi_5 \chi_1+\chi_3 \chi_6 \chi_1-\chi_3
   \chi_7 \chi_1-\chi_5 \chi_7 \chi_1\nonumber\\
   &\quad+\chi_7 \chi_8 \chi_1+\chi_3^2-3 \chi
   _4^2+\chi_5^2-\chi_7^2+\chi_4 \chi_8^2-\chi_8^2+2 \chi_2-2 \chi_2 \chi_4+\chi
   _4+3 \chi_3 \chi_5+\chi_2 \chi_6-3 \chi_4 \chi_6\nonumber\\
   &\quad+3 \chi_6+\chi_3^2 \chi
   _7+\chi_2 \chi_7-3 \chi_2 \chi_4 \chi_7+\chi_4 \chi_7-\chi_7+\chi_2 \chi_5
   \chi_8-\chi_3 \chi_7 \chi_8+2\nonumber\\
    \chi_6^{E_8}&=\chi_4 \chi_1^2-\chi_5 \chi_1+\chi_3 \chi_7 \chi_1+\chi_3^2-\chi_7^2+\chi_2-2
   \chi_2 \chi_4+\chi_2 \chi_6+\chi_6-\chi_2 \chi_7-2 \chi_4 \chi_7+\chi_5
   \chi_8\nonumber\\
    \chi_7^{\E_8}&=\chi_1 \chi_3-\chi_4+\chi_6+\chi_2 \chi_7-\chi_7\nonumber\\ 
    \chi_8^{\E_8}&=\chi_2+\chi_7\ ,
\end{align}
where $\chi_i$~ $(i=1,2,\dots,8)$, denote the characters of SO(16), whose dimensions are respectively given by
\begin{align}
   \{16,120,560,1820,4368,8008,\overline{128},128\}\ .
\end{align}
Under our physical mass parameterization, these are the SO(16)-basis fugacities \begin{align}
    \{M_0^2,M_1,M_2,M_3,M_4,M_5,M_6,M_7\}\ ,
\end{align}
which are used to express the $E_8$ characters. Here, $M_0$ is the instanton parameter of the $\E_8$ theory.
 
\bigskip
\bibliographystyle{JHEP}
\bibliography{ref}

@article{Velu1971,
  author  = {J. V{\'e}lu},
  title   = {Isog{\'e}nies entre courbes elliptiques},
  journal = {\href{https://aghitza.github.io/publications/translation-velu.pdf}{C. R. Acad. Sci. Paris S{\'e}r. A}},
  volume  = {273},
  pages   = {238},
  year    = {1971}
}

@article{Eguchi:2000fv,
    author = "Eguchi, Tohru and Kanno, Hiroaki",
    title = "{Five-dimensional gauge theories and local mirror symmetry}",
    eprint = "hep-th/0005008",
    archivePrefix = "arXiv",
    reportNumber = "UT-882",
    doi = "10.1016/S0550-3213(00)00375-8",
    journal = "Nucl. Phys. B",
    volume = "586",
    pages = "331--345",
    year = "2000"
}

@article{Nekrasov:1996cz,
    author = "Nekrasov, Nikita",
    title = "{Five dimensional gauge theories and relativistic integrable systems}",
    eprint = "hep-th/9609219",
    archivePrefix = "arXiv",
    reportNumber = "ITEP-TH-26-96, HUTP-96-A023",
    doi = "10.1016/S0550-3213(98)00436-2",
    journal = "Nucl. Phys. B",
    volume = "531",
    pages = "323--344",
    year = "1998"
}

@book{schuttshioda,
  title={Mordell--Weil Lattices},
  author={Sch{\"u}tt, M. and Shioda, T.},
  series={Ergebnisse der Mathematik und ihrer Grenzgebiete. 3. Folge / A Series of Modern Surveys in Mathematics},
  year={2019},
  publisher={Springer Singapore},
doi={10.1007/978-981-32-9301-4},
}

@article{Aharony:1997ju,
	title = "{Branes, superpotentials and superconformal fixed points}",
	author = "Aharony, Ofer and Hanany, Amihay",
	year = "1997",
	journal = "Nucl. Phys. B",
	volume = "504",
	pages = "239--271",
	doi = "10.1016/S0550-3213(97)00472-0",
	eprint = "hep-th/9704170",
	archivePrefix = "arXiv",
	reportNumber = "RU-97-25, IASSNS-HEP-97-38"
}

@article{Aharony:1997bh,
	title = "{Webs of (p,q) five-branes, five-dimensional field theories and grid diagrams}",
	author = "Aharony, Ofer and Hanany, Amihay and Kol, Barak",
	year = "1998",
	journal = "JHEP",
	volume = "01",
	pages = "002",
	doi = "10.1088/1126-6708/1998/01/002",
	eprint = "hep-th/9710116",
	archivePrefix = "arXiv",
	reportNumber = "IASSNS-HEP-97-113, RU-97-81, SU-ITP-97-40"
}

@article{Intriligator:1997pq,
	title = "{Five-dimensional supersymmetric gauge theories and degenerations of Calabi-Yau spaces}",
	author = "Intriligator, Kenneth A. and Morrison, David R. and Seiberg, Nathan",
	year = "1997",
	journal = "Nucl. Phys. B",
	volume = "497",
	pages = "56--100",
	doi = "10.1016/S0550-3213(97)00279-4",
	eprint = "hep-th/9702198",
	archivePrefix = "arXiv",
	reportNumber = "RU-96-99, IASSNS-HEP-96-112"
}

@article{Seiberg:1996bd,
	title = "{Five-dimensional SUSY field theories, nontrivial fixed points and string dynamics}",
	author = "Seiberg, Nathan",
	year = "1996",
	journal = "Phys. Lett. B",
	volume = "388",
	pages = "753--760",
	doi = "10.1016/S0370-2693(96)01215-4",
	eprint = "hep-th/9608111",
	archivePrefix = "arXiv",
	reportNumber = "RU-96-69"
}

@article{Leung:1997tw,
	title = "{Branes and toric geometry}",
	author = "Leung, Naichung Conan and Vafa, Cumrun",
	year = "1998",
	journal = "Adv. Theor. Math. Phys.",
	volume = "2",
	pages = "91--118",
	doi = "10.4310/ATMP.1998.v2.n1.a4",
	eprint = "hep-th/9711013",
	archivePrefix = "arXiv",
	reportNumber = "HUTP-97-A090"
}

@article{Kol:1997fv,
	title = "{5-D field theories and M theory}",
	author = "Kol, Barak",
	year = "1999",
	journal = "JHEP",
	volume = "11",
	pages = "026",
	doi = "10.1088/1126-6708/1999/11/026",
	eprint = "hep-th/9705031",
	archivePrefix = "arXiv",
	reportNumber = "SU-ITP-97-22"
}

@article{Witten:1997sc,
	title = "{Solutions of four-dimensional field theories via M-theory}",
	author = "Witten, Edward",
	year = "1997",
	journal = "Nucl. Phys. B",
	volume = "500",
	pages = "3--42",
	doi = "10.1016/S0550-3213(97)00416-1",
	eprint = "hep-th/9703166",
	archivePrefix = "arXiv",
	reportNumber = "IASSNS-HEP-97-19"
}

@article{Brandhuber:1997ua,
	title = "{On the M theory approach to (compactified) 5-D field theories}",
	author = "Brandhuber, A. and Itzhaki, N. and Sonnenschein, J. and Theisen, S. and Yankielowicz, S.",
	year = "1997",
	journal = "Phys. Lett. B",
	volume = "415",
	pages = "127--134",
	doi = "10.1016/S0370-2693(97)01249-5",
	eprint = "hep-th/9709010",
	archivePrefix = "arXiv",
	reportNumber = "TAUP-2449-97, LMU-TPW-97-22"
}

@article{Benini:2009gi,
	title = "{Webs of five-branes and N=2 superconformal field theories}",
	author = "Benini, Francesco and Benvenuti, Sergio and Tachikawa, Yuji",
	year = "2009",
	journal = "JHEP",
	volume = "09",
	pages = "052",
	doi = "10.1088/1126-6708/2009/09/052",
	eprint = "0906.0359",
	archivePrefix = "arXiv",
	primaryClass = "hep-th",
	reportNumber = "PUPT-2303"
}

@article{Kim:2014nqa,
	title = "{5d E$_{n}$ Seiberg-Witten curve via toric-like diagram}",
	author = "Kim, Sung-Soo and Yagi, Futoshi",
	year = "2015",
	journal = "JHEP",
	volume = "06",
	pages = "082",
	doi = "10.1007/JHEP06(2015)082",
	eprint = "1411.7903",
	archivePrefix = "arXiv",
	primaryClass = "hep-th",
	reportNumber = "KIAS-P14064"
}

@article{Eguchi:2002fc,
	title = "{Seiberg-Witten curve for the E string theory}",
	author = "Eguchi, Tohru and Sakai, Kazuhiro",
	year = "2002",
	journal = "JHEP",
	volume = "05",
	pages = "058",
	doi = "10.1088/1126-6708/2002/05/058",
	eprint = "hep-th/0203025",
	archivePrefix = "arXiv",
	reportNumber = "UT-975"
}

@article{Conrad:2006LangNeron,
    author = "Conrad, Brian",
    title = "{Chow's $K/k$-image and $K/k$-trace, and the Lang--N\'eron theorem}",
    journal = "Enseign. Math. (2)",
    volume = "52",
    number = "1--2",
    pages = "37--108",
    year = "2006",
    doi = "10.5169/seals-2226"
}

@article{Kim:2026yeh,
    author = "Kim, Sung-Soo and Liu, Song and Sabir, Mudassar and Yagi, Futoshi",
    title = "{5d $\mathbb{Z}_2$ S-folds and Seiberg--Witten curves}",
    eprint = "2611.09804",
    archivePrefix = "arXiv",
    primaryClass = "hep-th",
    month = "11",
    year = "2026"
}

@article{Acharya:2021jsp,
	title = "{Gauging discrete symmetries of T$_{N}$-theories in five dimensions}",
	author = "Acharya, Bobby and Lambert, Neil and Najjar, Marwan and Svanes, Eirik Eik and Tian, Jiahua",
	year = "2022",
	journal = "JHEP",
	volume = "04",
	pages = "114",
	doi = "10.1007/JHEP04(2022)114",
	eprint = "2110.14441",
	archivePrefix = "arXiv",
	primaryClass = "hep-th",
	reportNumber = "KCL-PH-TH/2021-78"
}

@article{Kim:2021fxx,
	title = "{S-foldings of 5d SCFTs}",
	author = "Kim, Hee-Cheol and Kim, Sung-Soo and Lee, Kimyeong",
	year = "2022",
	journal = "JHEP",
	volume = "05",
	pages = "178",
	doi = "10.1007/JHEP05(2022)178",
	eprint = "2112.14550",
	archivePrefix = "arXiv",
	primaryClass = "hep-th",
	reportNumber = "KIAS-21060"
}

@article{Xie:2022lcm,
	title = "{Classification of rank one 5d $\mathcal{N}=1$ and 6d $(1,0)$ SCFTs}",
	author = "Xie, Dan",
	year = "2022",
	month = "10",
	eprint = "2210.17324",
	archivePrefix = "arXiv",
	primaryClass = "hep-th"
}

@article{Apruzzi:2022nax,
	title = "{Generalized quotients and holographic duals for 5d S-fold SCFTs}",
	author = "Apruzzi, Fabio and Bergman, Oren and Kim, Hee-Cheol and Uhlemann, Christoph F.",
	year = "2023",
	journal = "JHEP",
	volume = "04",
	pages = "027",
	doi = "10.1007/JHEP04(2023)027",
	eprint = "2211.13243",
	archivePrefix = "arXiv",
	primaryClass = "hep-th",
	reportNumber = "LCTP-22-15"
}

@article{Closset:2023pmc,
	title = "{Reading between the rational sections: Global structures of 4d $\mathcal{N}=2$ KK theories}",
	author = "Closset, Cyril and Magureanu, Horia",
	year = "2024",
	journal = "SciPost Phys.",
	volume = "16",
	number = "5",
	pages = "137",
	doi = "10.21468/SciPostPhys.16.5.137",
	eprint = "2308.10225",
	archivePrefix = "arXiv",
	primaryClass = "hep-th"
}

@article{Furrer:2024zzu,
	title = "{Coulomb branch surgery: Holonomy saddles, S-folds and discrete symmetry gaugings}",
	author = "Furrer, Elias and Magureanu, Horia",
	year = "2024",
	journal = "SciPost Phys.",
	volume = "17",
	number = "3",
	pages = "073",
	doi = "10.21468/SciPostPhys.17.3.073",
	eprint = "2404.02955",
	archivePrefix = "arXiv",
	primaryClass = "hep-th"
}

@article{Arias-Tamargo:2026rrf,
	title = "{Discrete Gauging of 5d $\mathcal{N}=1$$E_n$ SCFTs}",
	author = "Arias-Tamargo, Guillermo and De Marco, Mario and Lawrie, Craig and Meynet, Shani Nadir and Mininno, Alessandro",
	year = "2026",
	month = "9",
	eprint = "2609.20917",
	archivePrefix = "arXiv",
	primaryClass = "hep-th",
	reportNumber = "DESY-26-032"
}

@article{Caorsi:2018ahl,
	title = "{Special Arithmetic of Flavor}",
	author = "Caorsi, Matteo and Cecotti, Sergio",
	year = "2018",
	journal = "JHEP",
	volume = "08",
	pages = "057",
	doi = "10.1007/JHEP08(2018)057",
	eprint = "1803.00531",
	archivePrefix = "arXiv",
	primaryClass = "hep-th"
}

@article{Huang:2013yta,
	title = "{Refined stable pair invariants for E-, M- and $[p, q]$-strings}",
	author = "Huang, Min-Xin and Klemm, Albrecht and Poretschkin, Maximilian",
	year = "2013",
	journal = "JHEP",
	volume = "11",
	pages = "112",
	doi = "10.1007/JHEP11(2013)112",
	eprint = "1308.0619",
	archivePrefix = "arXiv",
	primaryClass = "hep-th",
	reportNumber = "USTC-ICTS-13-10, BONN-TH-2013-10"
}

@article{Minahan:1997ch,
	title = "{Investigating the BPS spectrum of noncritical E(n) strings}",
	author = "Minahan, J. A. and Nemeschansky, D. and Warner, N. P.",
	year = "1997",
	journal = "Nucl. Phys. B",
	volume = "508",
	pages = "64--106",
	doi = "10.1016/S0550-3213(97)00527-0",
	eprint = "hep-th/9705237",
	archivePrefix = "arXiv",
	reportNumber = "USC-97-006, NSF-ITP-97-055"
}

@article{Eguchi:2002nx,
	title = "{Seiberg-Witten curve for E string theory revisited}",
	author = "Eguchi, Tohru and Sakai, Kazuhiro",
	year = "2003",
	journal = "Adv. Theor. Math. Phys.",
	volume = "7",
	number = "3",
	pages = "419--455",
	doi = "10.4310/ATMP.2003.v7.n3.a3",
	eprint = "hep-th/0211213",
	archivePrefix = "arXiv",
	reportNumber = "UT-02-46"
}

@article{Closset:2021lhd,
	title = "{The $U$-plane of rank-one 4d $\mathcal{N}=2$ KK theories}",
	author = "Closset, Cyril and Magureanu, Horia",
	year = "2022",
	journal = "SciPost Phys.",
	volume = "12",
	number = "2",
	pages = "065",
	doi = "10.21468/SciPostPhys.12.2.065",
	eprint = "2107.03509",
	archivePrefix = "arXiv",
	primaryClass = "hep-th"
}

@article{Rosen:2016extensions,
    author = "Rosen, Julian and Shnidman, Ariel",
    title = "{Extensions of CM elliptic curves and orbit counting on the projective line}",
    eprint = "1608.01390",
    archivePrefix = "arXiv",
    primaryClass = "math.NT",
    doi = "10.1007/s40993-017-0073-y",
    journal = "Res. Number Theory",
    volume = "3",
    pages = "9",
    year = "2017"
}

@article{Bao:2011rc,
	title = "{M5-Branes, Toric Diagrams and Gauge Theory Duality}",
	author = "Bao, Ling and Pomoni, Elli and Taki, Masato and Yagi, Futoshi",
	year = "2012",
	journal = "JHEP",
	volume = "04",
	pages = "105",
	doi = "10.1007/JHEP04(2012)105",
	eprint = "1112.5228",
	archivePrefix = "arXiv",
	primaryClass = "hep-th"
}

@article{Bao:2013pwa,
	title = "{Non-Lagrangian Theories from Brane Junctions}",
	author = "Bao, Ling and Mitev, Vladimir and Pomoni, Elli and Taki, Masato and Yagi, Futoshi",
	year = "2014",
	journal = "JHEP",
	volume = "01",
	pages = "175",
	doi = "10.1007/JHEP01(2014)175",
	eprint = "1310.3841",
	archivePrefix = "arXiv",
	primaryClass = "hep-th",
	reportNumber = "DESY-13-176, HU-Mathematik-17-13, HU-EP-13-50, KIAS-P-13058, RIKEN-MP-78"
}

@article{Oguiso:1991mw,
    author  = {Oguiso, Keiji and Shioda, Tetsuji},
    title   = {{The Mordell--Weil lattice of a rational elliptic surface}},
    journal = {Comment. Math. Univ. St. Pauli},
    volume  = {40},
    number  = {1},
    pages   = {83--99},
    year    = {1991},
    note    = {\href{https://www.researchgate.net/profile/Tetsuji-Shioda/publication/240320684_The_Mordell-Weil_lattice_of_a_rational_elliptic_surface/links/5704b7fe08ae44d70ee06790/The-Mordell-Weil-lattice-of-a-rational-elliptic-surface.pdf}{[PDF]}}
}

@article{Fonseca:2011sy,
	title = "{Calculating the renormalisation group equations of a SUSY model with Susyno}",
	author = "Fonseca, Renato M.",
	year = "2012",
	journal = "Comput. Phys. Commun.",
	volume = "183",
	pages = "2298--2306",
	doi = "10.1016/j.cpc.2012.05.017",
	eprint = "1106.5016",
	archivePrefix = "arXiv",
	primaryClass = "hep-ph",
	reportNumber = "CFTP-11-011"
}

@article{Gaiotto:2009we,
    author = "Gaiotto, Davide",
    title = "{N=2 dualities}",
    eprint = "0904.2715",
    archivePrefix = "arXiv",
    primaryClass = "hep-th",
    doi = "10.1007/JHEP08(2012)034",
    journal = "JHEP",
    volume = "08",
    pages = "034",
    year = "2012"
}

@article{Schuett:2009pej,
    author = "Schuett, Matthias and Shioda, Tetsuji",
    title = "{Elliptic Surfaces}",
    eprint = "0907.0298",
    archivePrefix = "arXiv",
    primaryClass = "math.AG",
    month = "7",
    year = "2009"
}

@article{Billingsley:2021specialization,
title = {An algorithm for checking injectivity of specialization maps from elliptic surfaces},
journal = {Journal of Number Theory},
volume = {238},
pages = {1081-1106},
year = {2022},
issn = {0022-314X},
doi = {https://doi.org/10.1016/j.jnt.2021.11.002},
url = {https://www.sciencedirect.com/science/article/pii/S0022314X21003772},
author = {Tyler Raven Billingsley} 
}

@book{Silverman2009arithmetic,
  author    = {Joseph H. Silverman},
  title     = {Arithmetic of Elliptic Curves},
  year      = {2009},
  publisher = {Springer London, Limited},
  isbn      = {9780387094946}
}

@article{Lyness1942,
  author  = {Lyness, R. C.},
  title   = {1581. {C}ycles},
  journal = {The Mathematical Gazette},
  volume  = {26},
  number  = {268},
  pages   = {62},
  year    = {1942},
  doi     = {10.2307/3606036}
}

@article{Hone:2021,
  author        = {Hone, Andrew N. W. and Kouloukas, Theodoros E.},
  title         = {Deformations of cluster mutations and invariant
                   presymplectic forms},
  journal       = {J. Algebraic Combin.},
  volume        = {57},
  pages         = {763--791},
  year          = {2023},
  doi           = {10.1007/s10801-022-01203-5},
  eprint        = {2107.11866},
  archivePrefix = {arXiv},
  primaryClass  = {math-ph}
}

@article{Ducat:2021,
  author        = {Ducat, Tom},
  title         = {The 3-dimensional {Lyness} map and a self-mirror
                   log {Calabi--Yau} 3-fold},
  journal       = {Manuscripta Math.},
  volume        = {174},
  pages         = {87--140},
  year          = {2024},
  doi           = {10.1007/s00229-023-01497-0},
  eprint        = {2105.07843},
  archivePrefix = {arXiv}, 
  primaryClass  = {math.AG},
url={https://arxiv.org/abs/2105.07843}, 
  }
\end{document}